\documentclass{amsart}
\usepackage{amssymb}
\usepackage{subfigure}
\usepackage{amsmath}
\usepackage{amsthm}
\usepackage{mathrsfs}
\usepackage{graphicx}
\usepackage{newtxtext}
\usepackage{newtxmath}
\usepackage[ruled,vlined]{algorithm2e}
\usepackage{bm}
\usepackage{xcolor}

\usepackage{fullpage}

\usepackage{float}

\newcommand{\bx}{\mathbf{x}}
\newcommand{\bn}{\mathbf{n}}
\newcommand{\bu}{\mathbf{u}}
\newcommand{\bbu}{\bar{\mathbf{u}}}

\begin{document}

\title[Deep learning of inverse water waves problems using multi-fidelity data]{Deep learning of inverse water waves problems using multi-fidelity data:  Application to Serre-Green-Naghdi equations}


\author{Ameya D. Jagtap}
\address{\textbf{A.~Jagtap:} Division of Applied Mathematics, Brown University, 182 George Street, Providence, RI, 02912, USA}
\email{ameya$\_$jagtap@brown.edu}

\author{Dimitrios Mitsotakis}
\address{\textbf{D.~Mitsotakis:} Victoria University of Wellington, School of Mathematics and Statistics, PO Box 600, Wellington 6140, New Zealand}
\email{dimitrios.mitsotakis@vuw.ac.nz}

\author{George Em Karniadakis}
\address{\textbf{G.~Karniadakis:} Division of Applied Mathematics, Brown University, 182 George Street, Providence, RI, 02912, USA}
\email{george$\_$karniadakis@brown.edu}

\begin{abstract}
We consider strongly-nonlinear and weakly-dispersive surface water waves governed by equations of Boussinesq type, known as the Serre-Green-Naghdi system; it describes future states of the free water surface and depth averaged horizontal velocity, given their initial state. The lack of knowledge of the velocity field as well as the initial states provided by measurements lead to an ill-posed problem that cannot be solved by traditional techniques. To this end, we employ physics-informed neural networks (PINNs) to generate solutions to such ill-posed problems using only data of the free surface elevation and depth of the water. PINNs can readily incorporate the physical laws and the observational data, thereby enabling inference of the physical quantities of interest. In the present study, both experimental and synthetic (generated by numerical methods) training data are used to train PINNs. Furthermore, multi-fidelity data are used to solve the inverse water wave problem by leveraging both high- and low-fidelity data sets. The applicability of the PINN methodology for the estimation of the impact of water waves onto solid obstacles is demonstrated after deriving the corresponding equations. The present methodology can be employed to efficiently design offshore structures such as oil platforms, wind turbines, etc. by solving the corresponding ill-posed inverse water waves problem.
\end{abstract}



\maketitle

\section{Introduction}
The propagation of surface water waves is usually described by deterministic partial differential  equations, which  are nonlinear and dispersive equations derived asymptotically, and approximate the equations of fluid mechanics known as Euler's equations. The complete description of a deterministic water waves problem requires the knowledge of both the depth averaged horizontal velocity and the free surface elevation of the water as functions of space and time. Usually, these data are not available since the laboratory and {\em in situ measurements} of water waves consist of recorded data of the free surface elevation of the water at sparsely chosen locations only. Although sometimes it is feasible to reconstruct approximations of the unknown components using asymptotic expansions, these are not very accurate. The lack of knowledge of one of these components of water waves leads to ill-posed inverse problems. Therefore, reconstructing the complete solution or measure important quantities of the fluid from the recorded experimental measurements is impossible with deterministic modeling.

Traditional numerical methods such as finite element methods or  finite difference methods have undergone remarkable progress over the years, but they still operate under very stringent requirements such as requirement of precise knowledge of the physical model and boundary as well as initial conditions to simulate future states of physical systems. Moreover, they often need time consuming workflows such as mesh generation and long time simulations.  Along with these requirements, the simulations of real observational data for ill-posed problems become very time consuming due to the multiple simulations required to calibrate the free parameters of the model or the missing physics in the system.  In such scenarios a very interesting regime arises where some of the physics may be unknown and at the same time some sparse, multi-fidelity or multi-modal  observational data can be made available through the experimental techniques and/or high-resolution numerical methods (synthetic data). The sparse and multi-fidelity nature of the data with incomplete knowledge makes the use of traditional numerical methods ineffective. Thus, the \textit{physics-informed machine learning} based methods can be employed to solve such ill-posed problems.

Physics-informed machine learning is a relatively new paradigm for bridging the physical models and the observational data. In the earlier study, Lagaris et al. \cite{lagaris1998artificial} proposed artificial neural networks (ANN) for solving differential equations, which was further employed to solve the boundary value problems, see Lagaris et al. \cite{lagaris2000neural}. Later McFall \& Mahan \cite{mcfall2009artificial} proposed ANN for solving boundary value problems with exact satisfaction of boundary condition. Beidokhti \& Malek \cite{beidokhti2009solving} solved initial-boundary value problem for system of partial differential equations (PDEs) using neural networks. The ANN approach was further used to solve the Stokes problem by Baymani et al. \cite{baymani2010artificial}. In the recent years, Owhadi \cite{owhadi2015bayesian} used  systematically structured prior information about
the solution to construct the physics-informed
learning machines. Brunton et al. \cite{brunton2016discovering} proposed SINDy framework for dictionary learning of dynamical systems. Wang et al. \cite{wang2017comprehensive} proposed physics-informed learning machines for turbulence modeling. More recently, Raissi et al. \cite{raissi2019physics} proposed the \textit{physics-informed neural network} (PINN) for solving forward and inverse problems of PDEs using automatic differentiation to deal with all the differential operators, hence removing the tyranny of moving mesh generation. The PINN can smoothly integrate the sparse, noisy and multi-fidelity data along with the governing equations, and thereby recast the original PDE problem into an equivalent optimization problem. One of the major limitation of PINN methodology is its high computational cost associated with the training of the neural networks. This was first addressed by domain decomposition based PINN methodology namely, \textit{conservative PINN (cPINN)} methodology, see Jagtap et al. \cite{jagtap2020conservative}, and further by more general space-time domain decomposition based \textit{extended PINN (XPINN)} methodology, see Jagtap \& Karniadakis \cite{jagtap2020extended}, and for the theory of XPINN see \cite{hu2021extended}. PINNs have been successfully applied to solve many problems in the field of computational science, see \cite{mao2020physics, shukla2020physics,shukla2021physics, Shukla2022A, kharazmi2021hp, cai2021flow} for more details. In a recent study \cite{shin2020convergence} establishes the mathematical foundation of the PINNs, whereas \cite{mishra2020estimates} presented estimates on the generalization error of the PINN methodology.

The propagation of surface oceanic water waves is described by the incompressible Euler equations of water wave theory \cite{whitham2011}. Solving the Euler equations apparently is a very difficult task. For this reason, several model equations that approximate the Euler equations have been derived using asymptotic techniques. Such model equations are the Boussinesq equations that describe weakly nonlinear and weakly dispersive surface water waves while  strongly nonlinear and weakly dispersive surface water waves can be described by the Serre-Green-Naghdi equations, \cite{Serre,GN1976}.
The Serre-Green-Naghdi system of equations has most of the ingredients that characterise good approximations of the full Euler equations: It is mathematically and physically justified; it is  Galilean invariant; it is locally in time well-posed in $\mathbb{R}$; it admits classical solitary wave solutions; and its solutions agree very well with classical benchmark problems \cite{lannes2013,duran2013,mitsotakis2017,israwi2011,mitsotakis2017b}.

Mathematical and numerical modeling of water waves problems require theoretical and experimental justification. Theoretical justification ensures the deterministic properties of the mathematical equations while experimental validation is based on the comparison of mainly numerical solutions with laboratory data to explore its physical relevance. Usually, the experimental data consist of measurements of the free surface elevation of the ocean while the velocity profile is almost impossible to be measured. Furthermore, the available laboratory data are limited to a few recorded wave-gauge measurements.

In this work we employ the PINN approach to solve ill-posed inverse problems for nonlinear water waves, and in particular for the Serre-Green-Naghdi equations. The approach presented in the current work exploits the strengths of machine learning algorithms and of scientific computing for solving nonlinear inverse water wave problems. In the absence of the velocity profiles, an ill-posed problem can be formulated since all model water wave systems of equations have as minimum requirement the knowledge of the initial state of the free surface elevation and velocity.


The main contributions of this work are
\begin{itemize}
    \item  The PINN is employed to solve highly ill-posed inverse water wave problems governed by Serre-Green-Naghadi equations. In particular, the aim is to infer the unknown velocity field as well as the corresponding transient forces acting on the solid obstacles from the known free surface elevation data obtained from the wave gauges and total depth at the discrete spatial locations. Furthermore, we also discussed about how to choose the optimal locations for the gauges in order to obtained the best predictive accuracy.
    \item The number of high-fidelity sensors or gauges used is often limited due to high operational cost. On the other hand, plenty of low-fidelity data can be made available using low-fidelity cheap gauges. In this work, we have performed multi-fidelity simulations using PINNs for inverse water wave problems by leveraging both high- and low-fidelity data sets.
    \item Asymptotic derivation of the Serre-Green-Naghdi equations that describes the propagation of strongly nonlinear and weakly dispersive surface water waves is presented. Although the particular system is well-known, the new detailed derivation reveals terms significant for the determination of the energy and the instantaneous force of water waves. In addition, the equations can be used for the generation of water waves due to deformations of the ocean floor with high-accuracy.
\end{itemize}

The structure of this paper is as follows: A brief introduction to the Serre-Green-Naghdi equations is presented in Section 2. Section 3 presents the problem formulation for the PINN methodology, which is followed by Section 4, where the PINN methodology is discussed in detail. In Section 5, we solve various one- and two-dimensional inverse problems for the Serre-Green-Naghdi equations inferring the unknown velocity field and the corresponding transient force acting on the solid walls of the domain. Finally, we summarize our findings in Section 6.

\section{Governing equations}

Denoting the space independent variables by the tuple $\bx=(x,y)$ and the time independent variable by $t$, the free surface elevation from its rest position ($z=0$) is described by the function $\eta(\bx,t)$. In what follows the gradient operator is denoted by $\nabla=(\partial/\partial x,\partial/\partial y)=(\partial_x,\partial_y)$. Assume that the depth of the seafloor (distance of seafloor from the rest position of the free surface) is described by the function $D(\bx)>0$.  If $h(\bx,t)=D(\bx)+\eta(\bx,t)$ denotes the total depth and $\bbu(\bx,t)$ the depth averaged horizontal velocity of the water
$$\bbu(\bx,t)=\frac{1}{h}\int_0^hu(\bx,z,t)~dz\ ,$$
then, the Serre-Green-Naghdi equations in dimensional variables can be written in the form 
\begin{align}
&h_t+\nabla\cdot[h\bbu]=0\ , \label{eq:Serre1}\\
&\bbu_t+g\nabla \eta+(\bbu\cdot \nabla)\bbu-\frac{1}{h}\nabla\left[ h^2\left(\frac{1}{3}P+\frac{1}{2}Q\right)\right]+\left(\frac{1}{2}P+Q \right)\nabla D=0\ , \label{eq:Serre2}
\end{align}
where
\begin{align}
& P=h[\nabla\cdot(\bbu_t+ \bbu\nabla\cdot\bbu)-2(\nabla\cdot \bbu)^2]\ ,\label{eq:PQG1}\\
& Q=[\bbu_t+ (\bbu\cdot\nabla)\bbu+ \bbu(\bbu\cdot\nabla)]\cdot \nabla D\ , \label{eq:PQG2}
\end{align}
and $g$ is the acceleration due to gravity (usually taken $g\approx 9.81~{\rm m/s}^2$). In the special one-dimensional case, the Serre-Green-Naghdi system is reduced to
\begin{align}
& h_t+[h\bar{u}]_x=0\ , \label{eq:cserre1}\\
&\bar{u}_t+g\eta_x+\bar{u}\bar{u}_x-\frac{1}{h}\frac{\partial}{\partial x}\left\{h^2\left(\frac{1}{3}P+\frac{1}{2}Q \right)\right\}+\left(\frac{1}{2}P+Q\right)D_x=0 \ ,  \label{eq:cserre2}
\end{align}
where
\begin{align}
P&=h(\bar{u}_{xt}+\bar{u}\bar{u}_{xx}-\bar{u}_x^2)\ , \label{eq:cserre3} \\
Q&=D_x(\bar{u}_t+ \bar{u}\bar{u}_x)+ \bar{u}^2D_{xx}\ . \label{eq:cserre4}
\end{align}
In this paper we will consider these equations in a bounded domain $\Omega$ with wall boundary conditions on its boundary $\partial\Omega$. The wall boundary conditions in the interval $\Omega=(a,b)$ can be described by the condition $u(a,t)=u(b,t)=0$, while in two-dimensional domains this will take the form of the slip-wall boundary condition $\bu\cdot\bn=0$ on $\partial\Omega$ where $\bn$ is the unit outward normal vector on the boundary.

For the sake of completeness, we present a derivation of these equations in \ref{sec:derivation}. The particular derivation reveals asymptotic approximations of the horizontal and vertical components of the velocity field as well as an asymptotic formula for the instantaneous force applied by the water waves at a particular location. Alternative derivations of the Serre-Green-Naghdi equations can be found in \cite{CiCP2018} and references therein. In \ref{sec:force} we derive a formula for the instantaneous force applied at a point $\bx=\bx_w$ by the water column at the specific location. This formula in the case of flat bottom $D(\bx,t)=D_0$ is
\begin{equation}\label{eq:iforce2}
F_w(t)=\rho g\left[\frac{h^2}{2}-\frac{h^3}{3g}\left(\nabla\cdot\bar{\bu}_t+\bar{\bu}\cdot\nabla(\nabla\cdot\bar{\bu})-(\nabla\cdot\bar{\bu})^2\right) \right]_{\bx=\bx_w}\ .
\end{equation}
The particular formula can be used to measure the external forces on a submerged obstacle in the water as well as the computation of the wave energy released by the deformation of the bottom topography.

Although, Serre-Green-Naghdi equations have complicated structure, one can compute analytical formulas for the classical solitary wave solutions in the case of a horizontal bottom of depth $D(x)=D_0$. Specifically, a solitary wave traveling with constant speed $c_s$ and amplitude $A$ is given analytically by the formulas
\begin{equation}\label{eq:solitwav}
\begin{aligned}
& \eta(x,t)=\eta_s(x-c_st)\doteq A {\rm sech}^2[\lambda(x-x_0-c_st)]\ ,\\
& \bar{u}(x,t)=u_s(x-c_st)\doteq c_s\left(1-\frac{D_0}{D_0+\eta_s(x,t)} \right)\ ,
\end{aligned}
\end{equation}
where
$$\lambda=\sqrt{\frac{3A}{4D_0^2(D_0+A)}},\quad \text{and}\quad c_s=c_0\sqrt{1+\frac{A}{D_0}}\ ,$$
are the spread parameter and the phase speed of the solitary wave, respectively, while $c_0=\sqrt{gD_0}$ is known as the linear speed of propagation and $x_0$ a translation parameter.

\section{Problem Formulation}

The general form of a system of partial differential equations can be expressed as
\begin{align}\label{ProSet}
\mathcal{L}_{\bm{x}}(\mathcal{U}) & = f(\bm{x}),~~~~~ \bm{x}=(\bx,t) \in \bar{\Omega}= \Omega\times [0,T) \subset \mathbb{R}^3\ ,
\\ \mathcal{B}_k(\mathcal{U}) & = g_k(\bm{x}),~~~~~ \bm{x} \in \Gamma_k \subset \partial \bar{\Omega},\ \nonumber
\end{align}
for $k = 1, 2, \cdots, n_b$, where $\mathcal{L}_{\bm{x}}(\cdot)$ is a differential operator, $\mathcal{U}$ is the solution, $\mathcal{B}_k(\cdot)$ any type of boundary conditions including Dirichlet, Neumann and Robin boundary conditions, and the forcing term is denoted by $f(\bm{x})$. The components of $\bm{x} = (\bx,t)$ where, for brevity, we consider time $t$ as one of the components of $\bm{x}$, and
the initial conditions can be treated as a particular type of boundary condition on the given computational domain. The above setup encapsulates a wide range of problems in scientific computations. We define the residual $\mathcal{F}(\mathcal{U})$ for equation (\ref{ProSet}) as 
\begin{equation}\label{Resi}
\mathcal{F}(\mathcal{U}) \coloneqq \mathcal{L}_{\bm{x}}(\mathcal{U}) - f(\bm{x}).
\end{equation}
In this work we employ simple feed forward neural networks (FFNNs) to solve the ill-posed inverse PDE problems. The computational examples we present in Section \ref{sec:results} provide details of the neural network architecture and other hyperparameters like learning rate, activation function, optimization procedure, etc.

\section{Methodology} 

In this section we describe the new PINN methodology for solving ill-posed problems of the Serre-Green-Naghdi equations. Details on the implementation of the neural network and the optimization methods used are provided below.

\subsection{Feed-forward neural network: An abstract mathematical setup}

Let $\mathcal{N}^L: \mathbb{R}^{D_i} \rightarrow \mathbb{R}^{D_o}$ be a FFNN of $L$ layers and $N_k$ neurons in $k^{th}$ layer, where the input layer has $N_0 = D_i$ and the output layer $N_L = D_o$ neurons. The $\bm{W}^k \in \mathbb{R}^{N_k \times N_{k-1}}$ and $\bm{b}^k \in \mathbb{R}^{N_k}$ represents the weight matrix and bias vector in the $k^{th}$ layer ($1 \leq k \leq L$), respectively. If $\bm{z}\in \mathbb{R}^{D_i}$ is the input vector at the $k^{th}$ layer, then the output vector at $k^{th}$ layer is denoted by $\mathcal{N}^k(\bm{z})$. Thus, the input vector can be expressed as $\mathcal{N}^0(\bm{z}) = \bm{z}$. The layer-wise adaptive activation function is denoted by $\Phi(n a^k)$, where $a^k$ are the trainable parameters, and $n$ is the predefined scaling factor. The introduction of the additional slope parameters $a^k$ dynamically changes the slope of activation function, thereby increases the training speed of the neural networks, see Jagtap et al. \cite{jagtap2020adaptive, jagtap2020locally} for more details. Recently, Jagtap et al. \cite{jagtap2022deep} proposed the \textit{Rowdy activation function}  that is designed to eliminate any saturation region by injecting sinusoidal
noise like effects, which further increases the training speed. With these adaptive activation functions, the learning speed of the network can be enhanced significantly, especially during the early training. In this paper, we employ layer-wise locally adaptive activation functions with scaling factor $n = 10$ for all hidden-layers and initialize $n a^k = 1, ~\forall k$. The $(L-1)$-hidden layer FFNN is defined as
\begin{equation}
\begin{aligned}
& \mathcal{N}^1(\bm{z}) = \bm{W}^1 \bm{z} + \bm{b}^1, \quad \text{ for }\quad k=1,\\
& \mathcal{N}^k(\bm{z}) = \bm{W}^k \Phi(a^{k-1} \mathcal{N}^{k-1}(\bm{z})) + \bm{b}^k \in \mathbb{R}^{N_k},\quad \text{ for }\quad 2\leq k \leq L\ ,
\end{aligned}
\end{equation} where in the last layer, the activation function is the identity.
By letting $\tilde{\boldsymbol{\Theta}} = \{ \bm{W}^k, \bm{b}^k, a^k \}_{k=1}^L \in \mathcal{V}$ as the collection of all trainable parameters including weights, biases, and activation slopes, and taking $\mathcal{V}$ as the parameter space, the output of the neural network is written as
$$\mathcal{U}_{\tilde{\boldsymbol{\Theta}}}(\bm{z}) = \mathcal{N}^L(\bm{z}; \tilde{\boldsymbol{\Theta}})\ ,$$
where $\mathcal{N}^L(\bm{z}; \tilde{\boldsymbol{\Theta}})$ emphasizes the dependence of the neural network output $\mathcal{N}^L(\bm{z})$ on $ \tilde{\boldsymbol{\Theta}}$. In general, weights and biases are initialized from known probability distributions. In this work we used the Xavier initialization \cite{glorot2010understanding}. 

\subsection{Physics-informed neural networks}

PINN is the surrogate model, which incorporates the governing physical laws in the loss function along with the sparse and noisy data obtained experimentally or with numerical simulations. In the PINN algorithm,  minimizing the violation of governing PDEs by the solution of NN as well as data mismatch terms, can drastically reduce the space of admissible solutions. One of the main advantages of the PINN methodology is its ability to solve highly ill-posed inverse problems arising in various settings. The simulation of observation data for the inverse problems can become very costly where simulations are performed multiple times to either discover missing physics in the model or calibrate the free parameters in the model. Due to such sparse, noisy data or incomplete knowledge of the underlying physical laws the traditional methods are not appropriate, but the PINN can be easily employed. In the literature, various inverse problems are solved by PINN, see for example, \cite{raissi2018deep, mao2020physics, raissi2020hidden, shukla2021physics, jagtap2020extended, jagtap2020conservative, Shukla2022A, shukla2021parallel}.  

Let $\{\bm{x}^{(i)}_\mathcal{U}\}_{i=1}^{N_\mathcal{U}}$ and $\{\bm{x}^{(i)}_F\}_{i=1}^{N_F}$ be the set of randomly selected training data and residual points, respectively. These points are drawn from a distribution, which is usually not known a priori and often need to be chosen from the given input training data. The PINN algorithm aims to learn a surrogate $\mathcal{U}= \mathcal{U}_{\tilde{\boldsymbol{\Theta}}}$ for predicting the solution $\mathcal{U}$ of the given PDE. The PINN loss function is given as
\begin{equation}\label{Loss}
 \mathcal{J}(\tilde{\boldsymbol{\Theta}})  =   W_{\mathcal{U}} ~ \text{MSE}_{\mathcal{U}}(\tilde{\boldsymbol{\Theta}}; \{\bm{x}^{(i)}_\mathcal{U}\}_{i=1}^{N_\mathcal{U}} ) + W_{\mathcal{F}} ~ \text{MSE}_{\mathcal{F}}(\tilde{\boldsymbol{\Theta}}; \{\bm{x}^{(i)}_F\}_{i=1}^{N_F} )\ ,
\end{equation}
 where $W_{\mathcal{U}}$ and $W_{\mathcal{F}}$ are the weights for the data mismatch and residual terms, respectively. The mean squared error (MSE) is given by
\begin{align*}
\text{MSE}_{\mathcal{U}}(\tilde{\boldsymbol{\Theta}}; \{\bm{x}^{(i)}_\mathcal{U}\}_{i=1}^{N_\mathcal{U}} ) & = \frac{1}{N_{\mathcal{U}}} \sum_{i=1}^{N_{\mathcal{U}}}\left|\mathcal{U}^{(i)} - \mathcal{U}_{{\tilde{\boldsymbol{\Theta}}}}(\bm{x}^{(i)}_{\mathcal{U}})\right|^2\ , \ \\  \text{MSE}_{\mathcal{F}}(\tilde{\boldsymbol{\Theta}}; \{\bm{x}^{(i)}_F\}_{i=1}^{N_F} )  &= \frac{1}{N_{F}} \sum_{i=1}^{N_{F}}\left|\mathcal{F}_{{\tilde{\boldsymbol{\Theta}}}}(\bm{x}^{(i)}_{F})\right|^2\ .
\end{align*}
 The term $\text{MSE}_{\mathcal{U}}$ is the MSE for data mismatch term, which enforces the given experimental or synthetic data. The term $\text{MSE}_{\mathcal{F}}$ is the MSE for the  residue of the governing physical laws where $\mathcal{F}_{{\tilde{\boldsymbol{\Theta}}}}(\bm{x}_{F}) \coloneqq \mathcal{F}(\mathcal{U}_{{\tilde{\boldsymbol{\Theta}}}}(\bm{x}_{F}))$ represents the residual of the governing PDEs given by equation \eqref{Resi}. The notations $N_\mathcal{U}$ and $N_F$ are the number of training data and residual points, respectively. The trainable parameters of the neural networks $\mathcal{U}_{\tilde{\boldsymbol{\Theta}}}$ can be estimated by minimizing the loss function given by Eq. \eqref{Loss}.

The PDE residual $\mathcal{F}_{{\tilde{\boldsymbol{\Theta}}}}$ in the loss function is constructed using automatic differentiation \cite{baydin2018automatic}. Automatic differentiation is a graph-based differentiation method, which can calculate derivatives accurately in a computational graph compared to numerical differentiation, since, they
do not suffer from truncation and round-off errors. Thus, evaluation of PDE operator $\mathcal{L}_{\bm{x}}(\cdot)$ acting on $\mathcal{U}_{\tilde{\boldsymbol{\Theta}}}$ is achieved easily with  automatic differentiation, which can be incorporated in the loss function along with the data mismatch term. Hence, the PINN method does not require any grid. This makes them easily implemented even in case of complicated computational domains.

\subsection{Optimization Methods}

In the parameter space, we search for the optimal parameters of the network $\tilde{\boldsymbol{\Theta}}^*$ that minimizes the loss function $\mathcal{J}(\tilde{\boldsymbol{\Theta}})$.  
In the literature, there are several optimization algorithms available that can be used to minimize the loss function. The gradient-based methods is a popular class of optimization methods that can be employed for the minimization of loss functions. In the basic form, given an initial value of tunable parameters $\tilde{\boldsymbol{\Theta}}$, these parameters are updated as
$ \tilde{\boldsymbol{\Theta}}^{m+1} = \tilde{\boldsymbol{\Theta}}^m - \eta_l \left. \frac{\partial \mathcal{J}(\tilde{\boldsymbol{\Theta}}) }{\partial \tilde{\boldsymbol{\Theta}}} \right|_{\tilde{\boldsymbol{\Theta}} = \tilde{\boldsymbol{\Theta}}^m}$,
here, the learning rate is denoted by $\eta_l$.
Such gradient-based methods are among the most popular optimization methods. In particular, we shall use the Adam optimizer \cite{kingma2014adam} followed by L-BFGS optimizer \cite{byrd1995limited}. For smooth and regular solutions L-BFGS optimizer can find a better solution with small number of iterations than Adam optimizer, because L-BFGS is second order accurate method as opposed to Adam, which converges linearly.

\subsection{PINN algorithm for Serre-Green-Naghdi equations }

\begin{figure} 
\centering 
\includegraphics[trim=0cm 0cm 0cm 0cm, clip=true, scale=0.8, angle = 0]{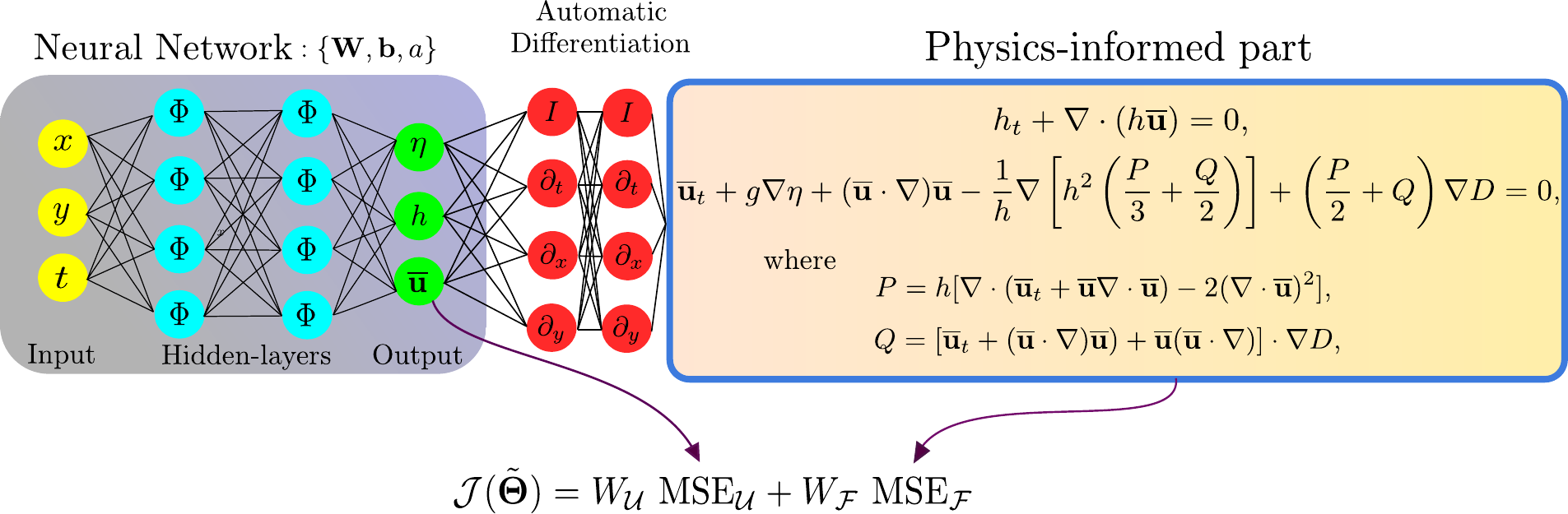}
\caption{Schematic representation of the physics-informed neural network for Serre-Green-Naghdi equations}
\label{fig:sketch_PINN}
\end{figure}

In this work we are solving an inverse problem of Serre-Green-Naghdi equations, which aims to infer the velocity of the water waves, and the instantaneous force on the wall by supplying the value of $\eta$ from the gauges as time series data. We further assume that the bottom topology, and therefore $h$ is also known at the discrete points. The data on the velocity is available only on the wall where normal velocity is prescribed as zero (reflection boundary conditions).
In the computational experiments both experimental as well as synthetic data are used. 
Figure \ref{fig:sketch_PINN} shows the schematics representation of the physics-informed neural network for Serre-Green-Naghdi equations. The input of the network is $x, y$ and $t$ and the output of the network is $\bar{\mathbf{u}}, \eta$ and $h$.  For illustration purposes, the network
depicted in this figure comprises two hidden layers and four neurons per hidden layer. 
The total loss consists of the training data loss of $\eta$, $h$ and no penetration wall boundary conditions on the respective data points, and the physics loss imposed by the governing Serre-Green-Naghdi equations on the residual points in the spatio-temporal domain. Here, $I$ denotes the identity operator and the differential operators such as $\partial_t, \partial_x, \ldots $ are computed using automatic differentiation. The PINN algorithm for the inverse problem of the Serre-Green-Naghdi system is given in Algorithm \ref{alg:one}.

\begin{algorithm}
\SetAlgoLined
\textbf{Given Data} : Water surface elevation from the gauges as time series data and total depth at the fixed spatial locations, and wall boundary conditions.\\
 \textbf{Output} : $(\bbu(\bm{x}), \eta(\bm{x}), h(\bm{x}))$ everywhere in the spatio-temporal domain and the instantaneous transient force acting on the walls. \\
 \textit{Training data points} : $\mathcal{U}_{\tilde{\boldsymbol{\Theta}}}$ network $\{\bm{x}^{(i)}_{\mathcal{U}}\}_{i=1}^{N_{\mathcal{U}}}$ \\
\textit{Residual training points} :  $\mathcal{F}_{{\tilde{\boldsymbol{\Theta}}}}$ network $\{\bm{x}^{(i)}_{F}\}_{i=1}^{N_{F}}$.\\
 \textbf{Step 1} : Construct the fully connected neural network $\mathcal{U}_{\tilde{\boldsymbol{\Theta}}}$ with random initialization of parameters $\tilde{\boldsymbol{\Theta}}$ in the domain. \\
  \textbf{Step 2} : Construct the residuals $\mathcal{F}_{{\tilde{\boldsymbol{\Theta}}}}$ in  the domain by
  substituting surrogate $\mathcal{U}_{\tilde{\boldsymbol{\Theta}}}$ into the governing equations using 
automatic differentiation as well as other arithmetic operations.\\
\textbf{Step 3} : Specify the loss function $\mathcal{J}(\tilde{\boldsymbol{\Theta}})$.\\
\textbf{Step 4}: Find the best parameters using suitable optimization method for minimizing the loss function 
\begin{equation}\tilde{\boldsymbol{\Theta}}^* =  \text{arg min}_{\tilde{\boldsymbol{\Theta}} \in \mathcal{V}}~\mathcal{J}(\tilde{\boldsymbol{\Theta}}),
\end{equation}
where $\mathcal{V}$ is the parameter space.\\x
 \caption{PINN algorithm for inverse problem of Serre-Green-Naghdi equations}\label{alg:one}
\end{algorithm}

\section{Results and discussion}\label{sec:results}

In this section we consider four test cases of significant physical importance, inspired by standard experimental benchmark problems in the literature. In every case we consider the Serre-Green-Naghdi equations written in dimensionless and unscaled variables. For all the test cases, we used $ W_{\mathcal{U}} = 20$ and $ W_{\mathcal{F}} = 1$, and all the computations are performed in double-precision unless otherwise specified. 

\subsection{Shoaling of a solitary wave on a plain beach}

\begin{figure}
  \centering
  \includegraphics[scale=0.85]{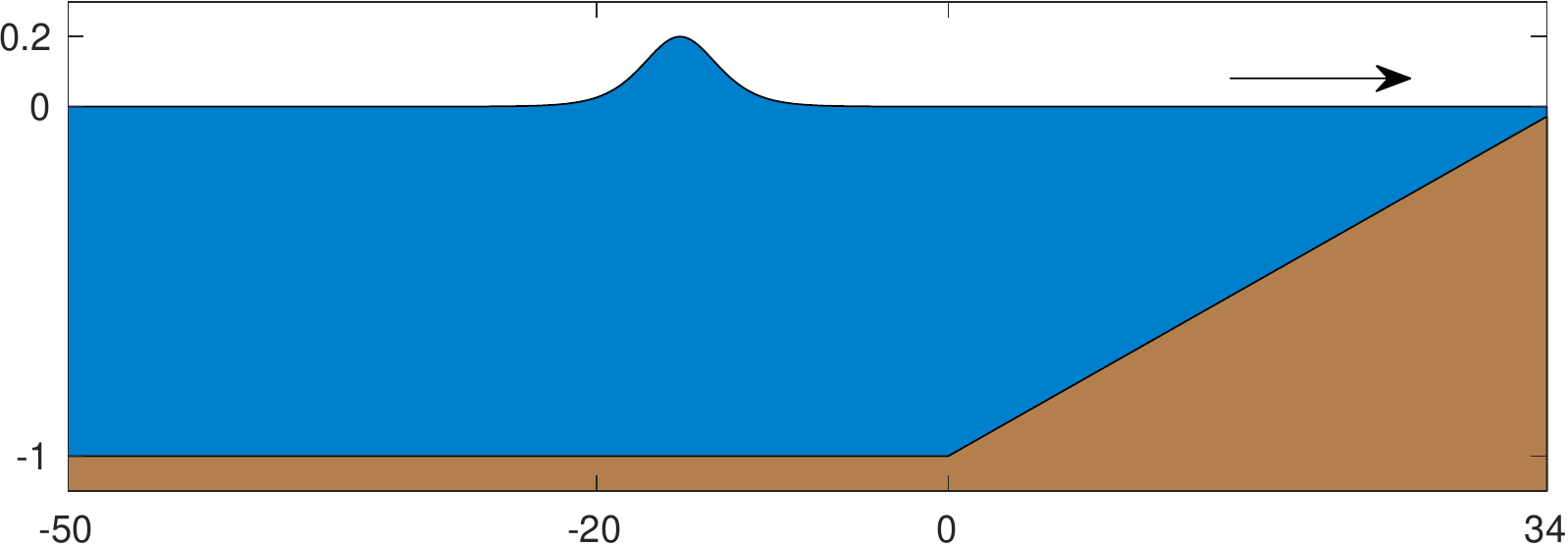}
  \caption{Sketch of the domain for the shoaling of solitary waves on a plain beach of slope $1:35$ in dimensionless and unscaled variables}
  \label{fig:slope}
\end{figure}

In this experiment we consider the shoaling of a solitary wave of the Serre-Green-Naghdi equations (\ref{eq:Serre1})--(\ref{eq:Serre2}). This particular experiment is a classical benchmark for testing models of weakly nonlinear and weakly dispersive equations, and proposed in \cite{grilli94}.
The initial location of the solitary wave (at $t=0$) is taken $x_0=-20.1171$ to fit with the experimental data of \cite{grilli94}. The topography of the bottom consists of a flat region with depth $D_0$ for $x\in[-50,0]$, and a plain beach of slope $1:35$ for $x\in[0,34]$. The initial profile of the solitary wave has amplitude $A/D_0=0.2$. A sketch of the physical domain is depicted in Figure \ref{fig:slope}. For the generation of synthetic data we used a spatial grid of size $\Delta x=0.1$ and $\Delta t=0.01$, while we computed the solution for up to time $T=50$.

\begin{figure}
\centering 
\begin{tabular}{cc}
\includegraphics[trim=0cm 0cm 0cm 0cm, clip=true, scale=0.28, angle = 0]{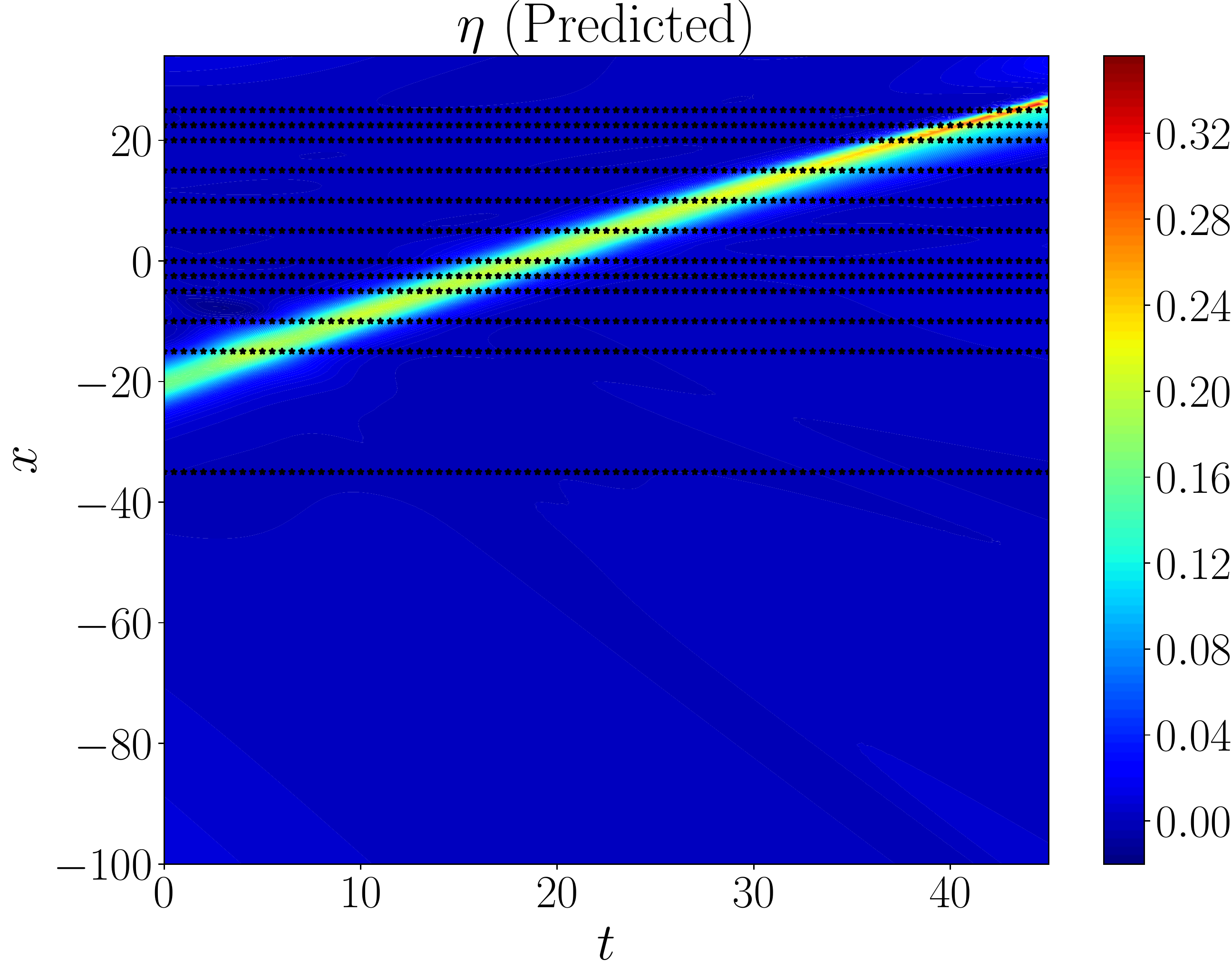}&
\includegraphics[trim=0cm 0cm 0cm 0cm, clip=true, scale=0.28, angle = 0]{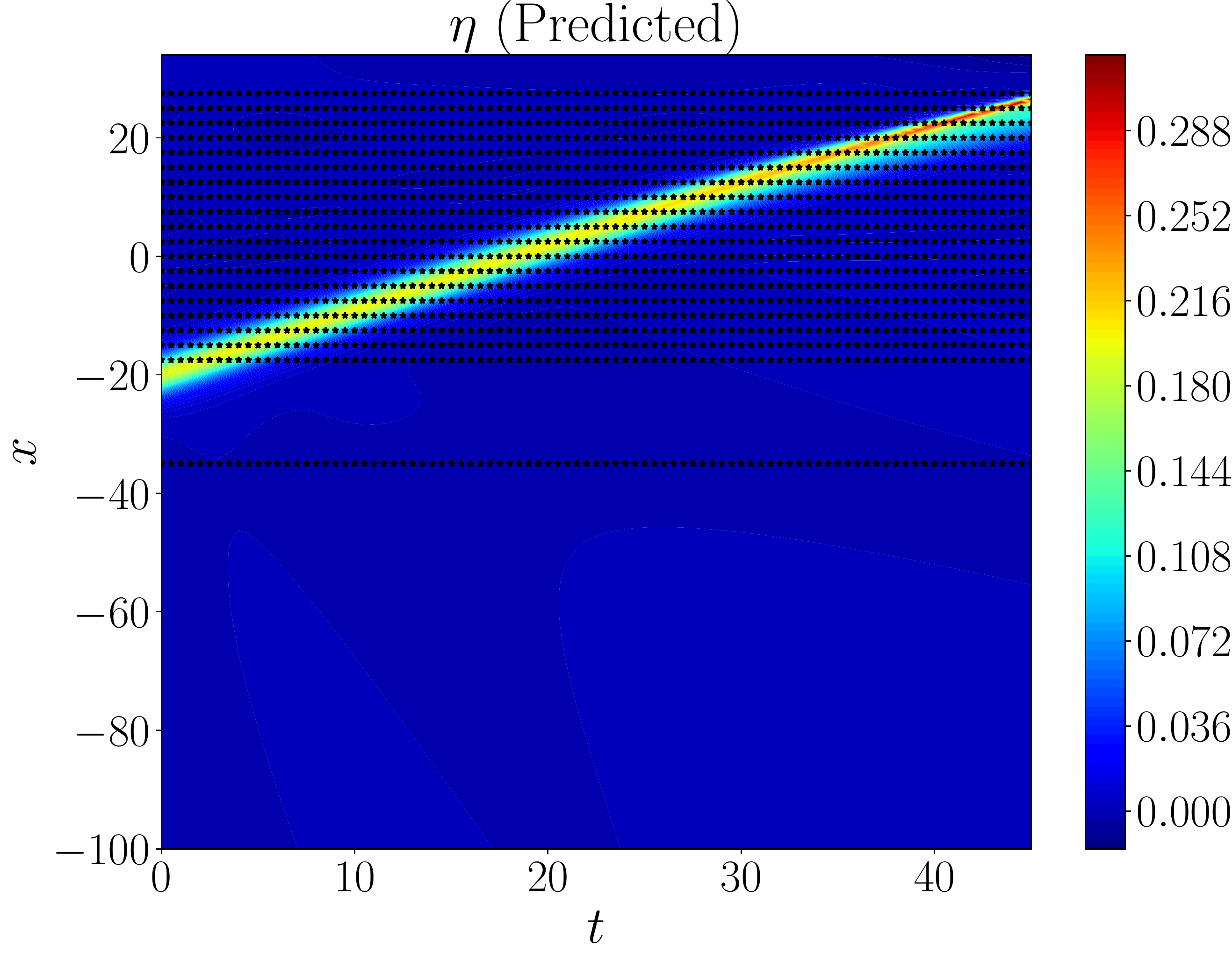}\\
\includegraphics[trim=0cm 0cm 0cm 0cm, clip=true, scale=0.28, angle = 0]{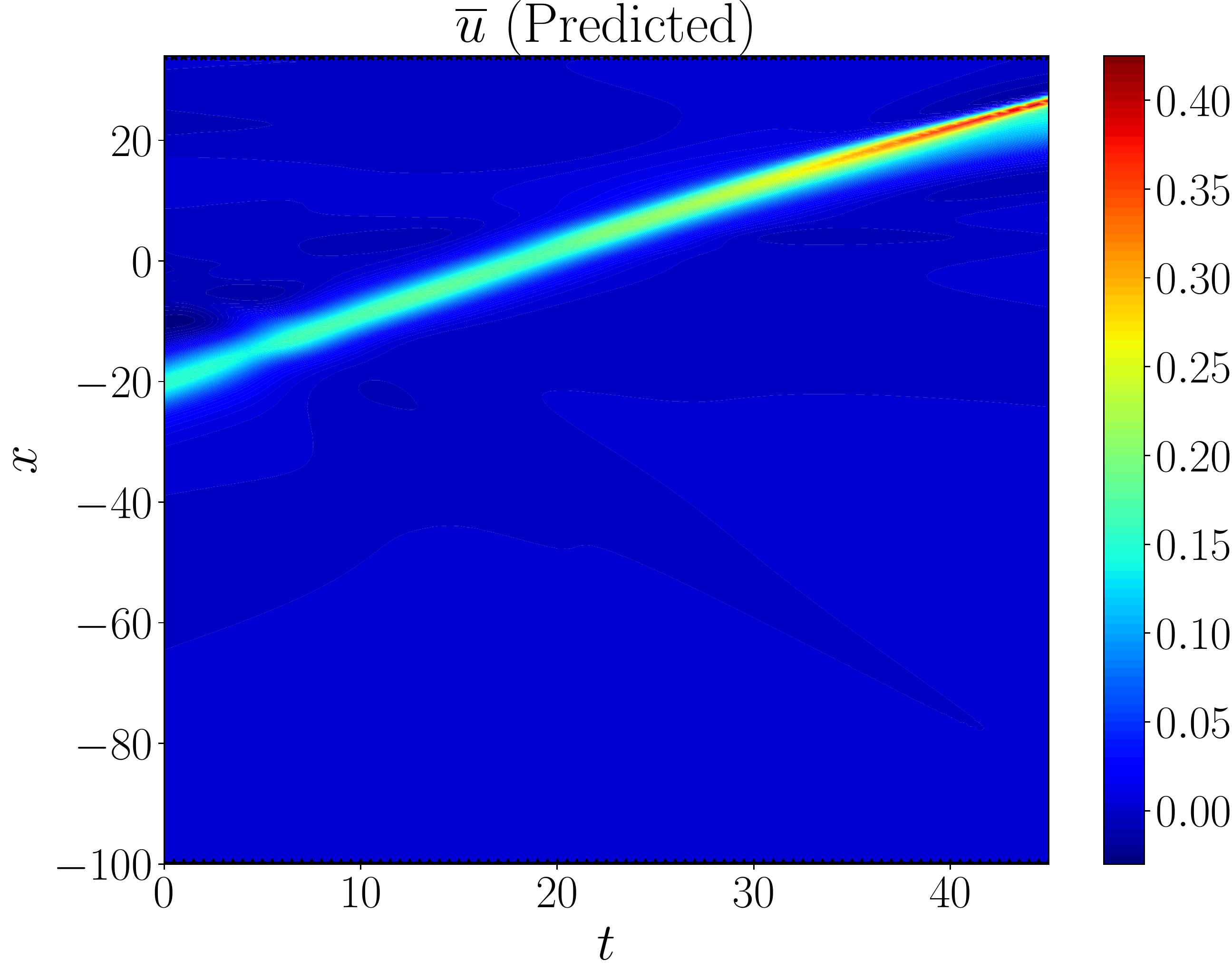}&
\includegraphics[trim=0cm 0cm 0cm 0cm, clip=true, scale=0.28, angle = 0]{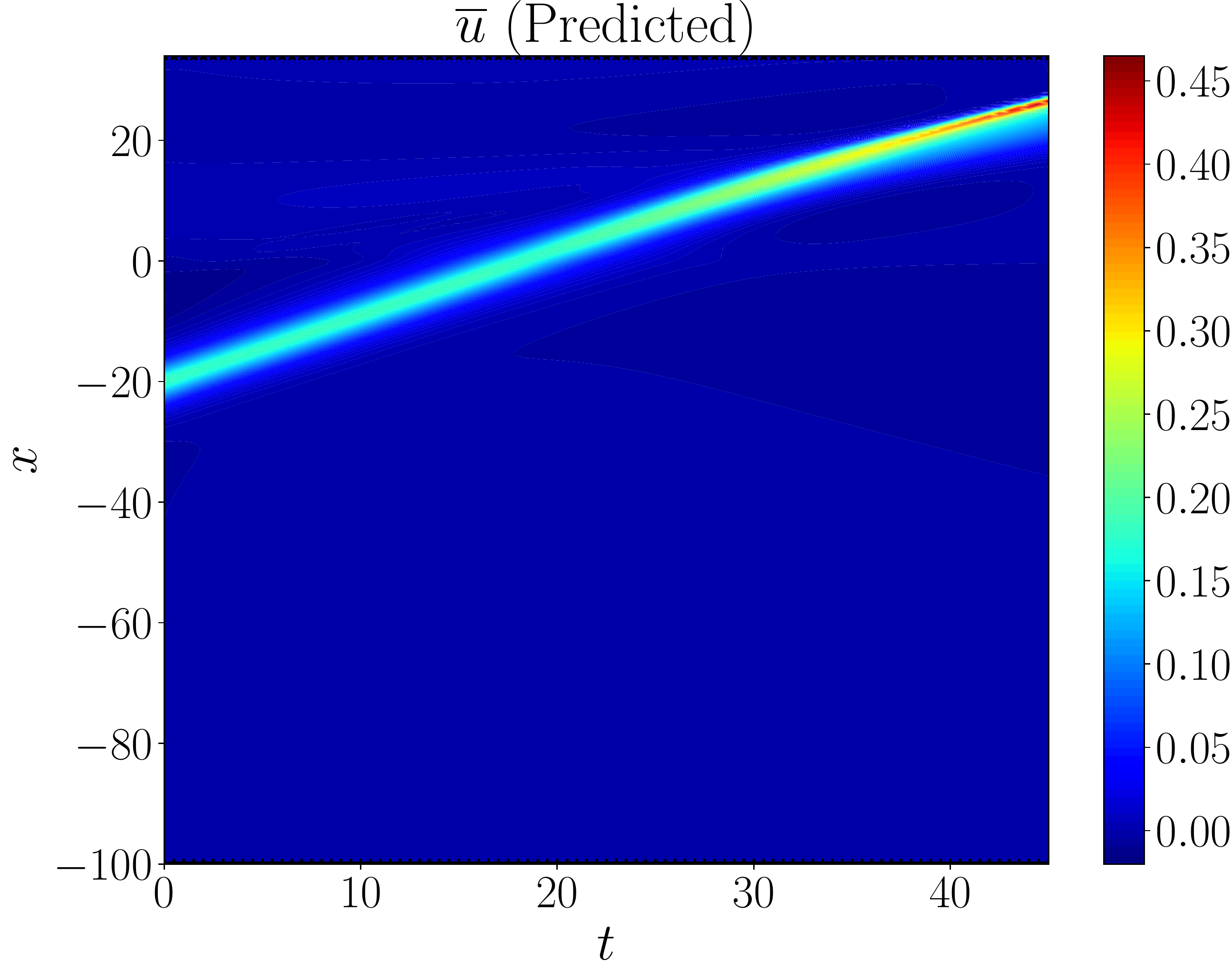}
\end{tabular}
\caption{Shoaling of a solitary wave: $\eta$ and $\bar{u}$ predictions using 12 (first column) and 20 (second column) gauges, respectively. The time series data obtained from the gauges is shown on the $\eta$ contour plots with '$\star$'}
\label{fig:TC1_1}
\end{figure}
\begin{figure} 
\centering 
\begin{tabular}{cc}
\includegraphics[trim=0cm 0cm 0cm 0cm, clip=true, scale=0.28, angle = 0]{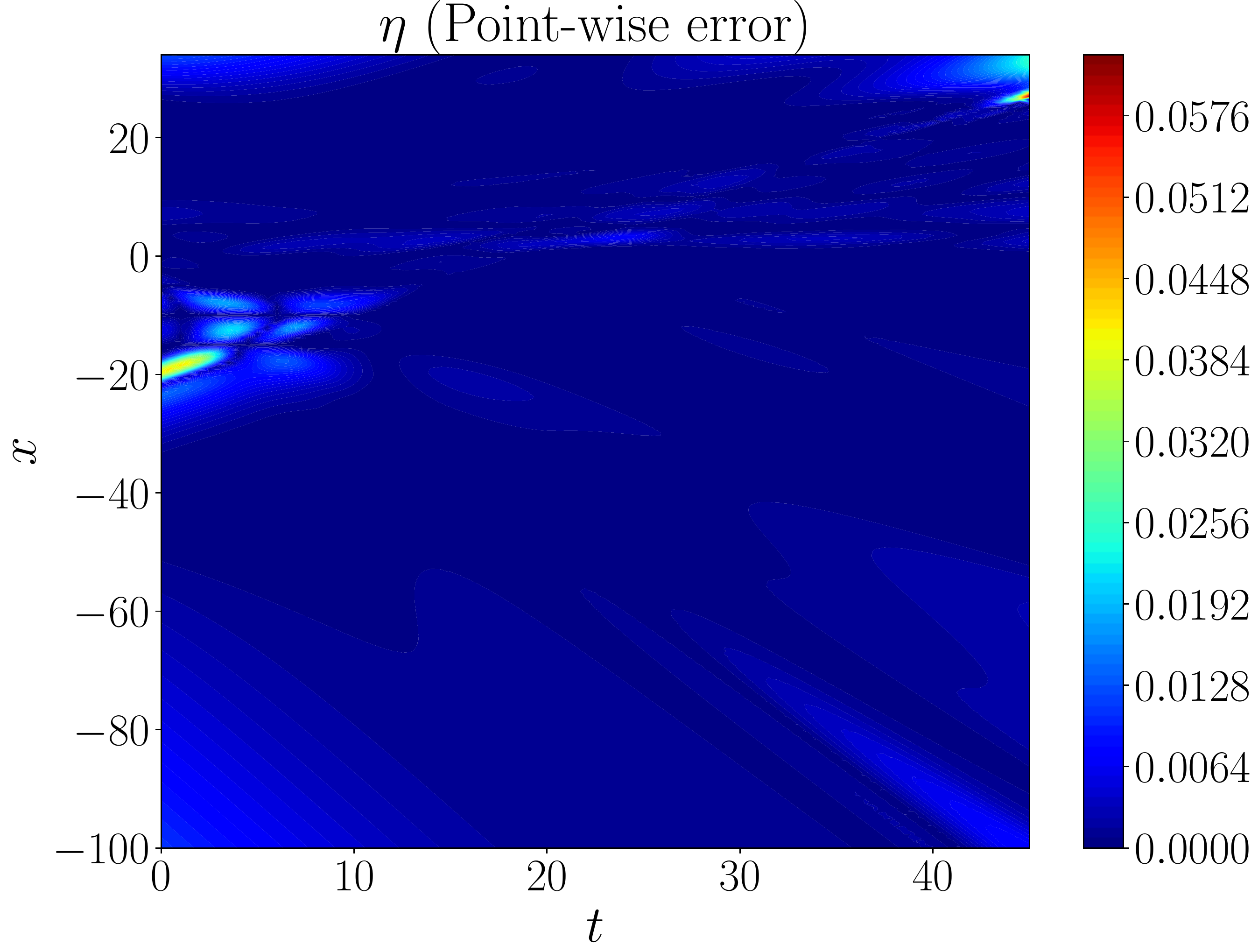} &
\includegraphics[trim=0cm 0cm 0cm 0cm, clip=true, scale=0.28, angle = 0]{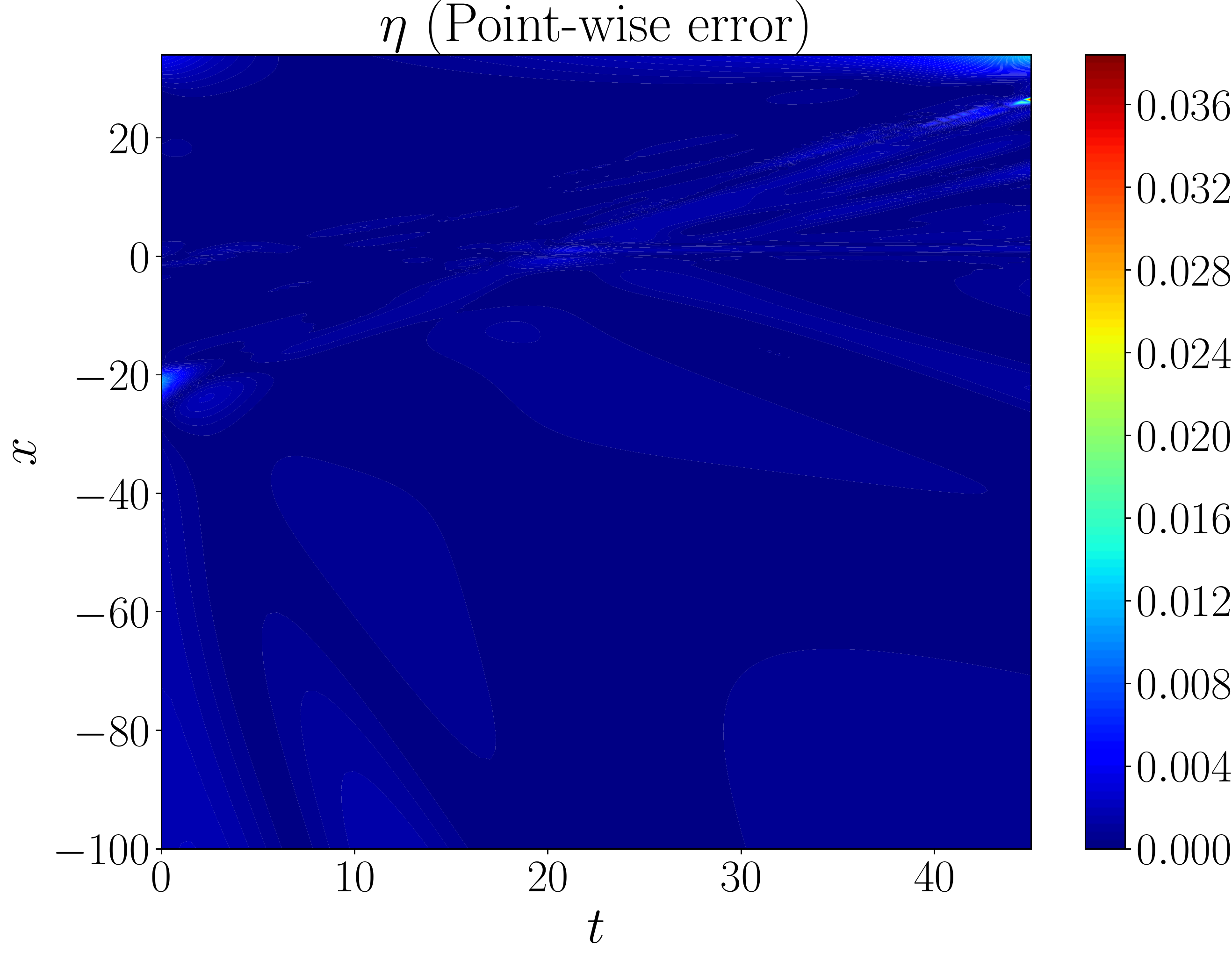}\\
\includegraphics[trim=0cm 0cm 0cm 0cm, clip=true, scale=0.28, angle = 0]{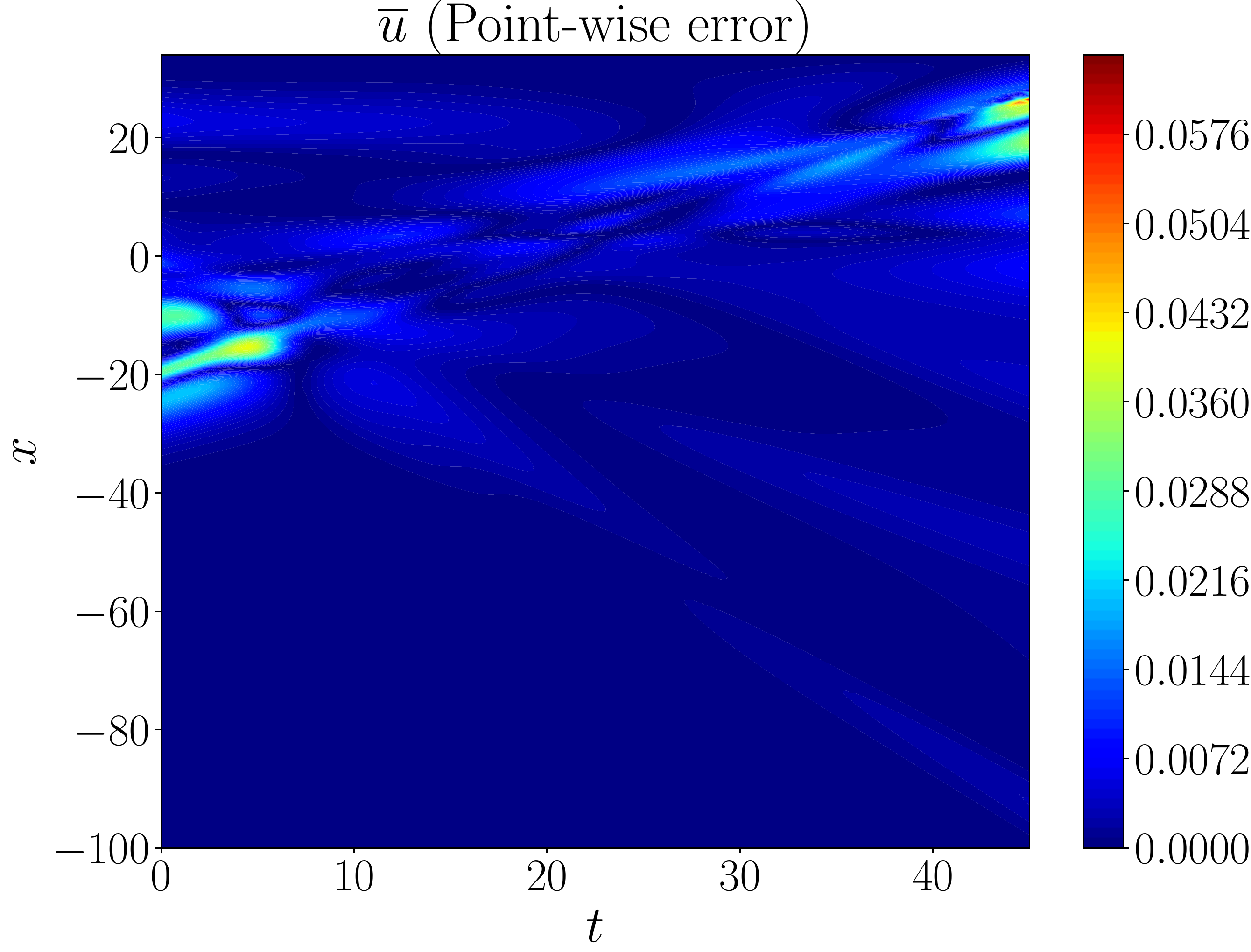}&
\includegraphics[trim=0cm 0cm 0cm 0cm, clip=true, scale=0.28, angle = 0]{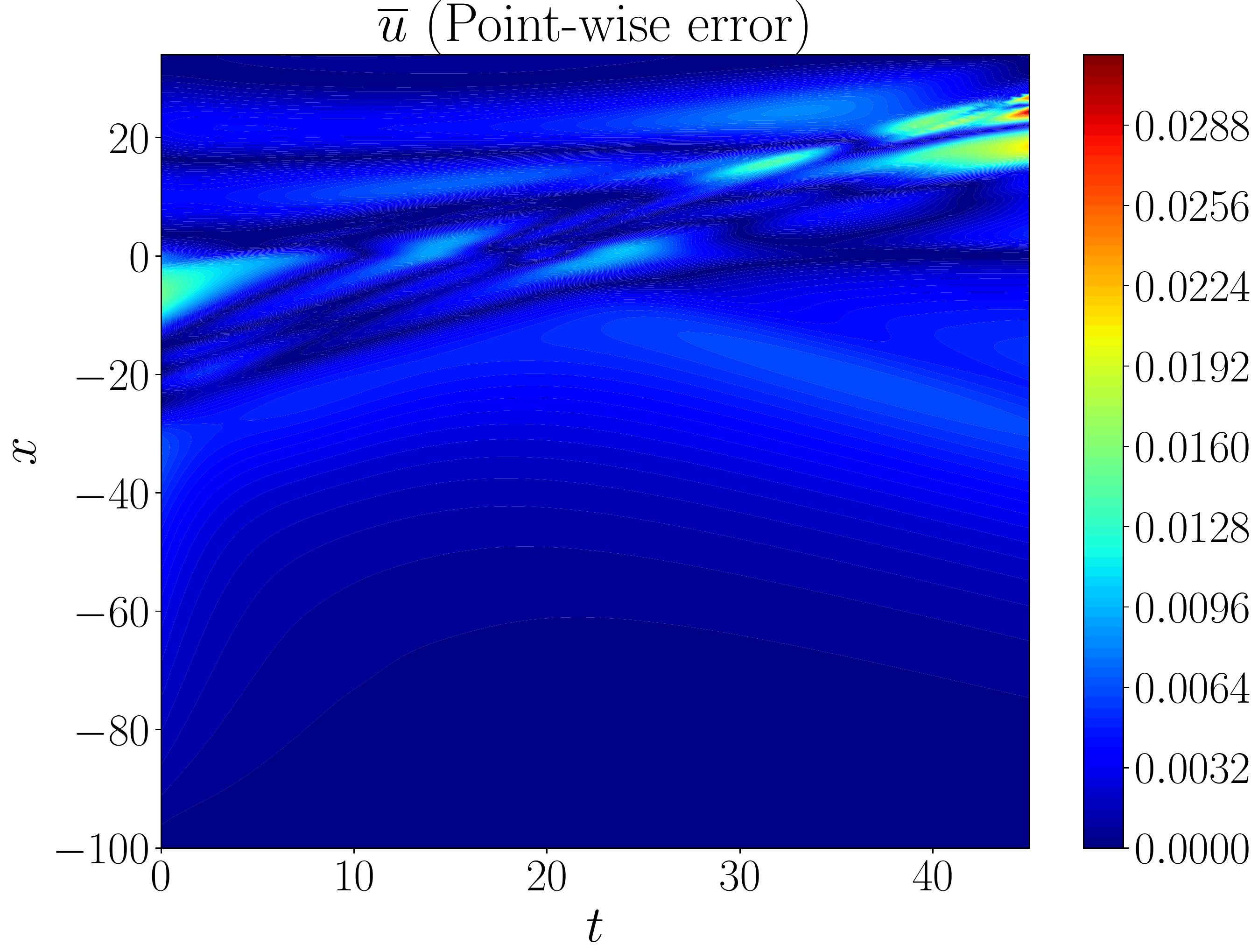}
\end{tabular}
\caption{Shoaling of a solitary wave : $\eta$ and $\bar{u}$ point-wise errors using 12 (first column) and 20 (second column) gauges, respectively}
\label{fig:TC1_2}
\end{figure}

In this test-case the aim is to infer the velocity field $\bar{u}$ from the given $\eta$ and $h$ data on a small subset of the domain. This data set is obtained from numerical simulations where synthetic data are recorded on wave gauges by solving the Serre-Green-Naghdi equations using the standard Galerkin/Finite element method accompanied with the Runge-Kutta method of order 4 \cite{mitsotakis2014,mitsotakis2017b}. Note that the available experimental data are very sparse to observe any form of convergence of our numerical algorithms \cite{grilli94}. The accuracy of the Serre-Green-Naghdi equations for this experiment has been tested in \cite{mitsotakis2014,antonopoulos2017,mitsotakis2017b}, and so we do not repeat it here but we only compare the solution of the aforementioned neural network with the synthetic data.

\begin{table}[h!]
\begin{center}
\begin{tabular}{|l||*{4}{c|}}\hline
&Width 4&Width 8&Width 16
& Width 32\\\hline\hline
Depth 1 &4.5829e-01& 4.2406e-01&1.9853e-01&9.9722e-02
\\\hline
Depth 2&3.0923e-01&3.9950e-01&1.9564e-01&9.8288e-02
\\\hline
Depth 4 &1.2095e-01&2.1495e-02&7.3498e-03&5.4968e-03
\\\hline
Depth 8 &1.9851e-01&2.4607e-02&6.0808e-03&6.1901e-03
\\\hline
\end{tabular}
\caption{Relative $L_2$ error in the velocity field for different width and depth of the fully connected feed forward neural networks. These errors represents the average of 10 different runs corresponding to different initialization.\label{TableNEW_TC1}}
\end{center}
\end{table}
Choosing the optimal hyperparameters of the neural networks such as depth, width, learning rate, optimizer, etc. are important for fast convergence, and the better predictive accuracy of the solution. 
Although, more sophisticated (but very expensive) meta-learning strategies are available in the literature, in this work, all the hyperparameters are chosen based on our past experience with PINN methodology, along with the complexity of the solution, for all test cases.
Table \ref{TableNEW_TC1} gives the mean relative $L_2$ error in the velocity field for different depth and width of the fully connected feed forward neural networks. In all cases the activation function is hyperbolic tangent, the learning rate is $8\times 10^{-4}$, and $12000$ residual points. It can be seen that with increase in depth as well as the width of the network the predictive accuracy increases.

For the computation, we used 5 hidden-layers with 20 neurons in each layer. The activation function, the learning rate, and the number of residual points used are same as before.
The predicted solution $\eta$ and $\bar{u}$ by using 12 (first column) and 20 (second column) gauges, respectively, as result of PINN is presented in Figure \ref{fig:TC1_1}. For the generation of these approximations we do not use velocity data except on the boundary where wall boundary conditions are applied. Figure \ref{fig:TC1_2} shows the corresponding absolute point-wise error in $\eta$ and $\bar{u}$. 

\begin{figure} 
\centering 
\includegraphics[trim=0cm 0cm 0cm 0cm, clip=true, scale=0.4, angle = 0]{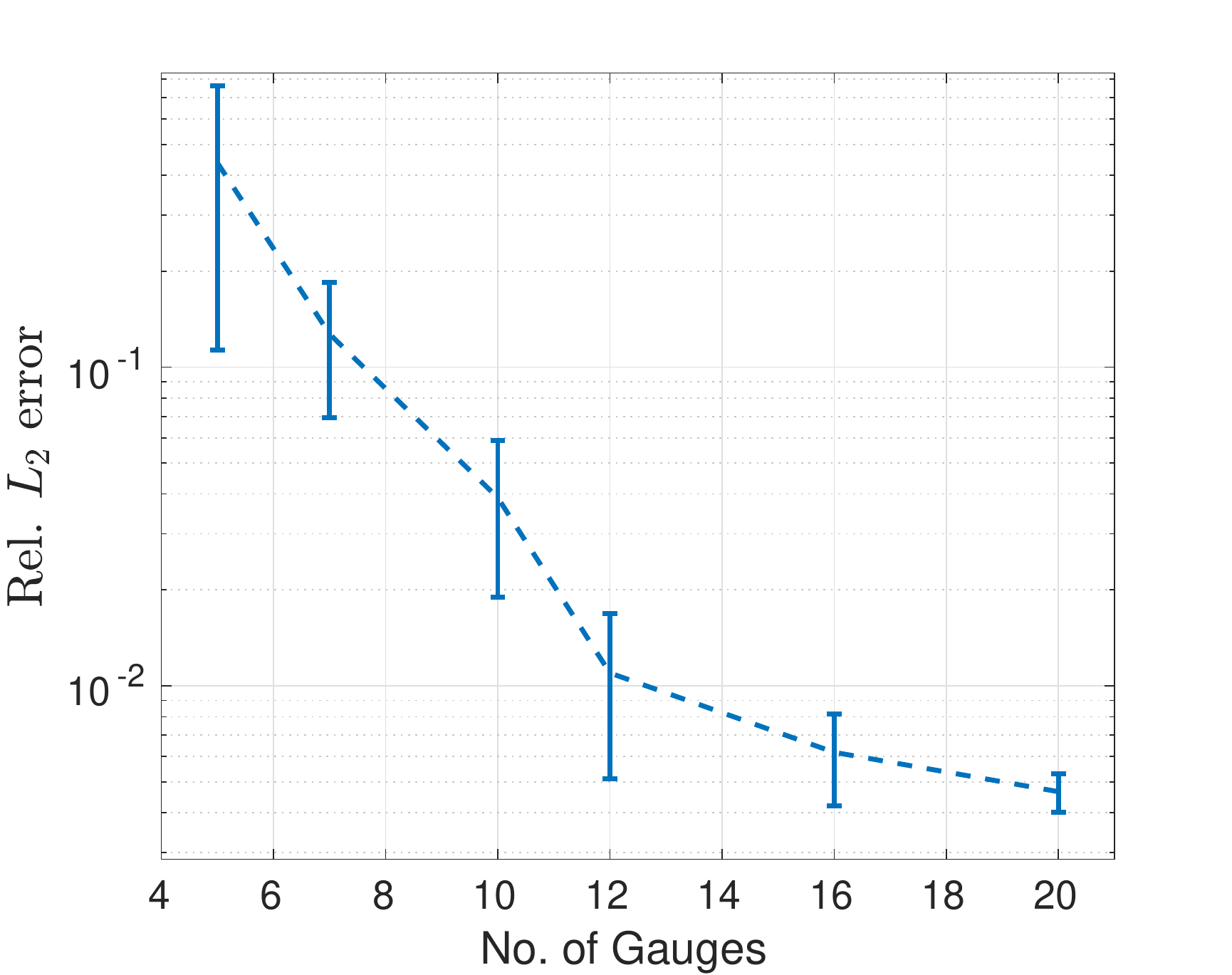}
\caption{Shoaling of a solitary wave on a plain beach. Mean and standard deviation of relative $L_2$ error in the velocity field versus the number of gauges for 10 different initialization of tunable parameters}
\label{fig:TC1_3}
\end{figure}
Increasing the  number of wave gauges as input to the PINN, we can infer the velocity field with more accuracy. This is presented in Figure \ref{fig:TC1_3}, where we plot the mean and standard deviation of the relative $L_2$ errors between the synthetic data and the current approximations, for different number of gauges, and for 10 different initializations of tunable parameters keeping the other hyperparameters of the network fixed. The gauges are uniformly distributed  for $-35 \leq x \leq 30$.
Note that the point-wise and relative $L_2$ errors are inversely proportional to the number of gauges we used. For example, when we used 12 gauges, the maximum point-wise error was approximately $0.06$ while increasing the number of gauges to almost its double, the error becomes approximately equal to $0.03$. Similar observations can be extracted from Figure \ref{fig:TC1_3}.

\subsubsection{Optimal gauge locations}
The choice for gauge locations is very important in order to extract the vital information from the time series data set with limited number of gauges. Instead of randomly choosing these locations, one can find their optimal locations. In this senario, we employed the procedure given in \cite{manohar2018data} for finding the optimal gauge locations. Assuming that the spatio-temporal evaluation of $\eta$ is known, denoted by $X$. This flow field can be efficiently represented using \textit{Singular Value Decomposition} in terms of the eigenbasis as $X = U\Sigma V^T$, where $U$ is an orthogonal matrix consist of eigenbasis. The aim is to sample a particular good rows of $U$ that corresponds to optimal gauge locations. For this purpose a pivoted QR factorization is employed, which turns out that the pivoted locations that essentially pivot the data to have better condition QR factorization, are the nearly optimal gauge locations.

\begin{figure} [htpb]
\centering 
\includegraphics[trim=7cm 3cm 7cm 0cm, clip=true, scale=0.084, angle = 0]{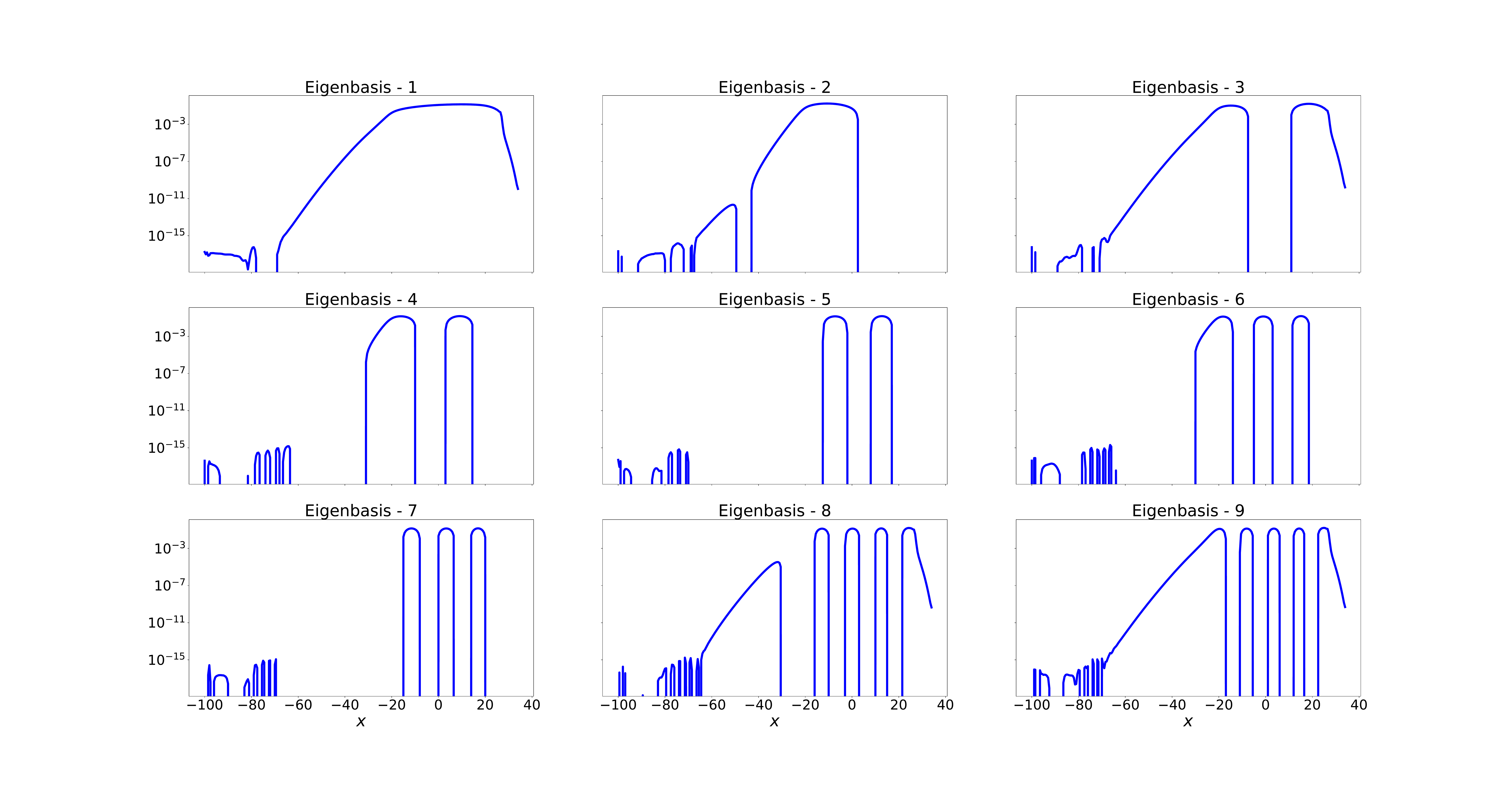}
\includegraphics[trim=0cm 0.4cm 0cm 0cm, clip=true, scale=0.345, angle = 0]{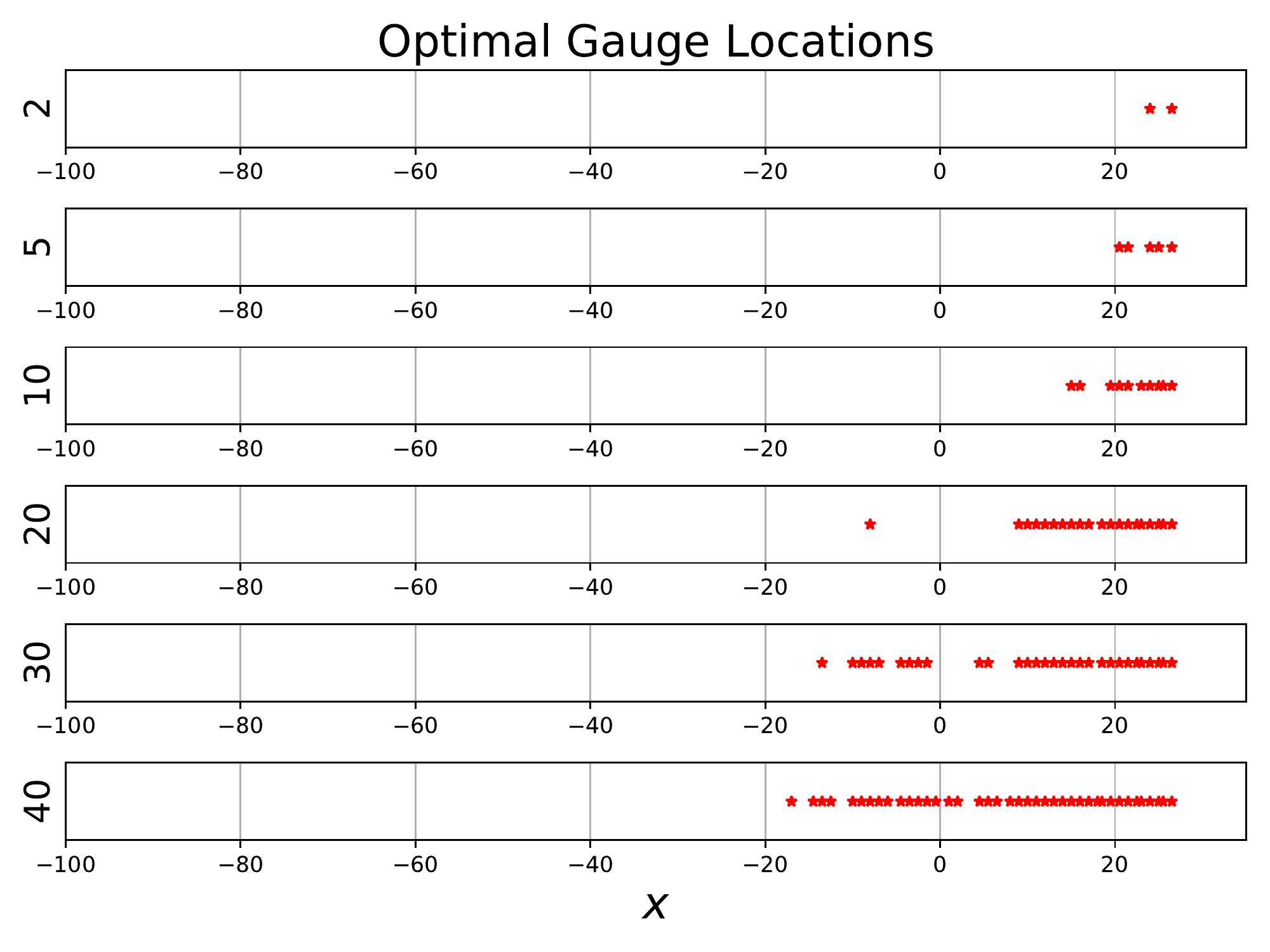}
\caption{First 9 eigenbasis for the $\eta$ field (left), and the optimal gauge locations (shown by the red star) for different number of gauges (right). The corresponding number of gauges are given on the vertical axis.}
\label{fig:TC1_4}
\end{figure}

For the present example, Figure \ref{fig:TC1_4} (left) shows the first nine eigenbasis for the $\eta$ field, and the right figure shows the optimal gauge locations (shown by the red star) for different number of gauges ranging from 2 (top) to 40 (bottom).
It can be observed that the optimal gauges are placed in the region $-20 \leq x \leq 30$ where all flow dynamics is present in spatio-temporal domain. This is consistent with our selection of location of gauges.

\subsection{Head-on collision of two solitary waves}

Another classical benchmark problem has been proposed in \cite{craig2006}. This is the head-on collision of two unequal solitary waves traveling in opposite directions over flat bottom topography of depth $D=D_0=5$. In this experiment two solitary waves with amplitudes $A_1 = 1.08$ and $A_2 = 1.20$ are located initially at $x_1 = 24.7$ and $x_2 = 134.8$, respectively. These solitary waves travel in opposite directions and collide in an inelastic way resulting in two new solitary waves. The inelastic head-on collision is characterized by a shift in the phase of the two solitary waves accompanied by an energy transfer to new small amplitude, trailing dispersive wave-trains and amplitude reduction of the two solitary pulses. For the generation of the synthetic data we solve the Serre-Green-Naghdi equations in the interval $[-500,500]$ up to $T=50$ with $\Delta x=0.1$ and $\Delta t=0.01$.

For the PINN we used 4 hidden-layers with 80 neurons in each layer. We chose the hyperbolic tangent as activation function, while we took the learning rate equal to $4\times 10^{-4}$ and $12000$ residual points.
We solved this problem with all the tunable parameters fixed using 32-bit as well as 64-bit precision. 
\begin{figure} 
\centering 
\includegraphics[trim=0cm 0cm 0cm 0cm, clip=true, scale=0.55, angle = 0]{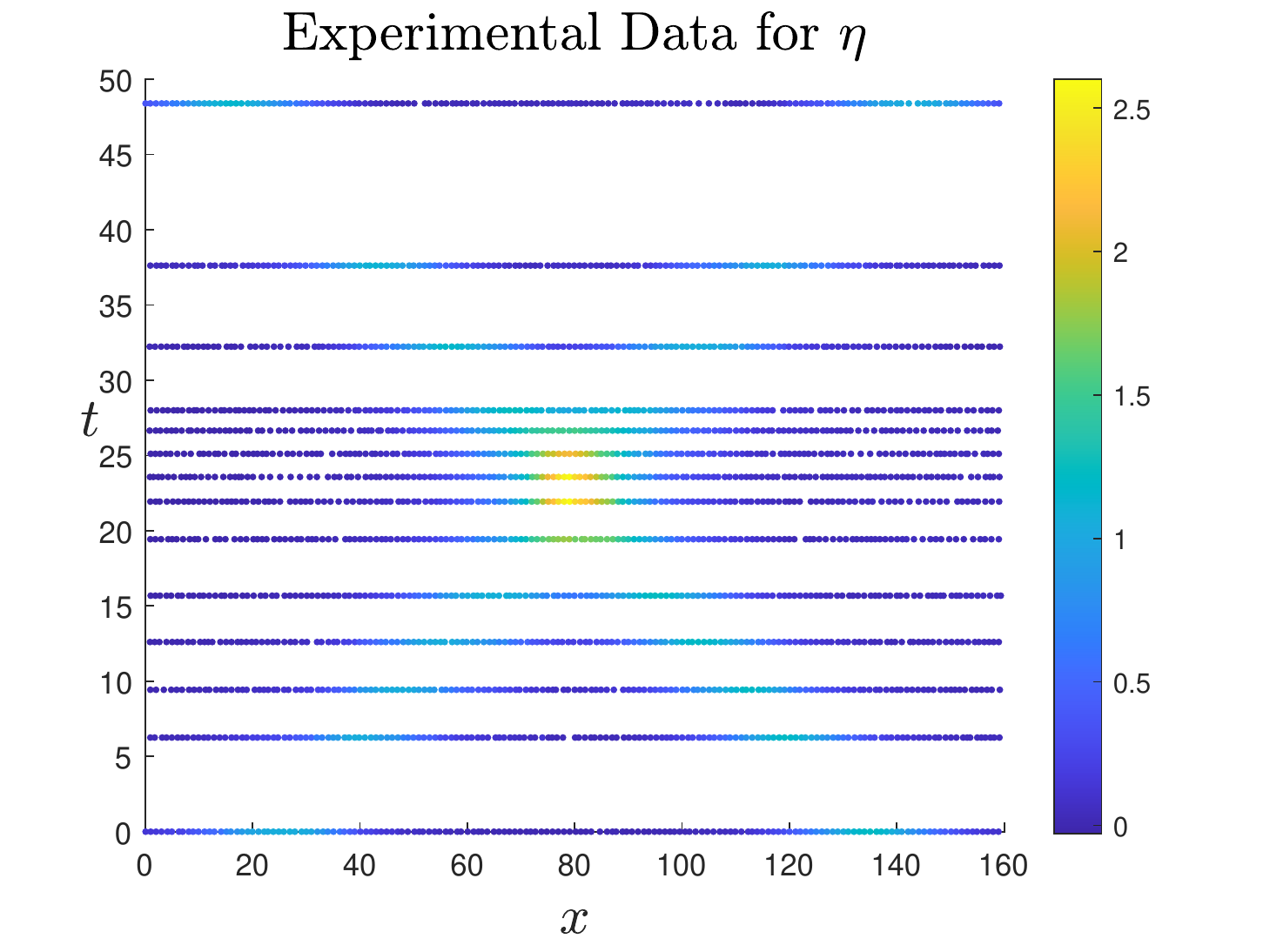}
\caption{{\color{blue}Head-on collision of two solitary waves: Experimental data for $\eta$ used in PINN.}}
\label{fig:TC2_4}
\end{figure}
\begin{figure} 
\centering 
\begin{tabular}{cc}
\includegraphics[trim=0cm 0cm 0cm 0cm, clip=true, scale=0.28, angle = 0]{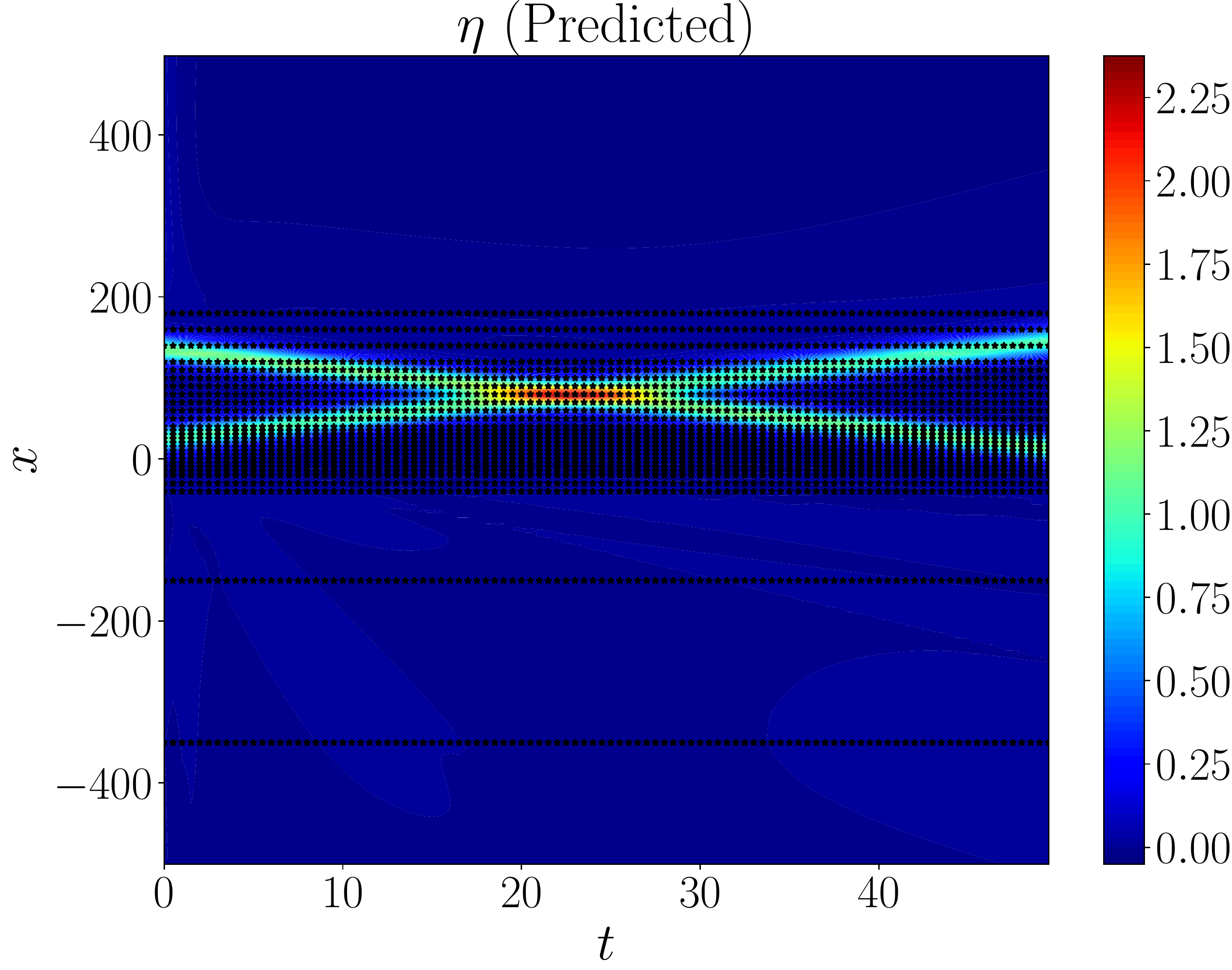} &
\includegraphics[trim=0cm 0cm 0cm 0cm, clip=true, scale=0.28, angle = 0]{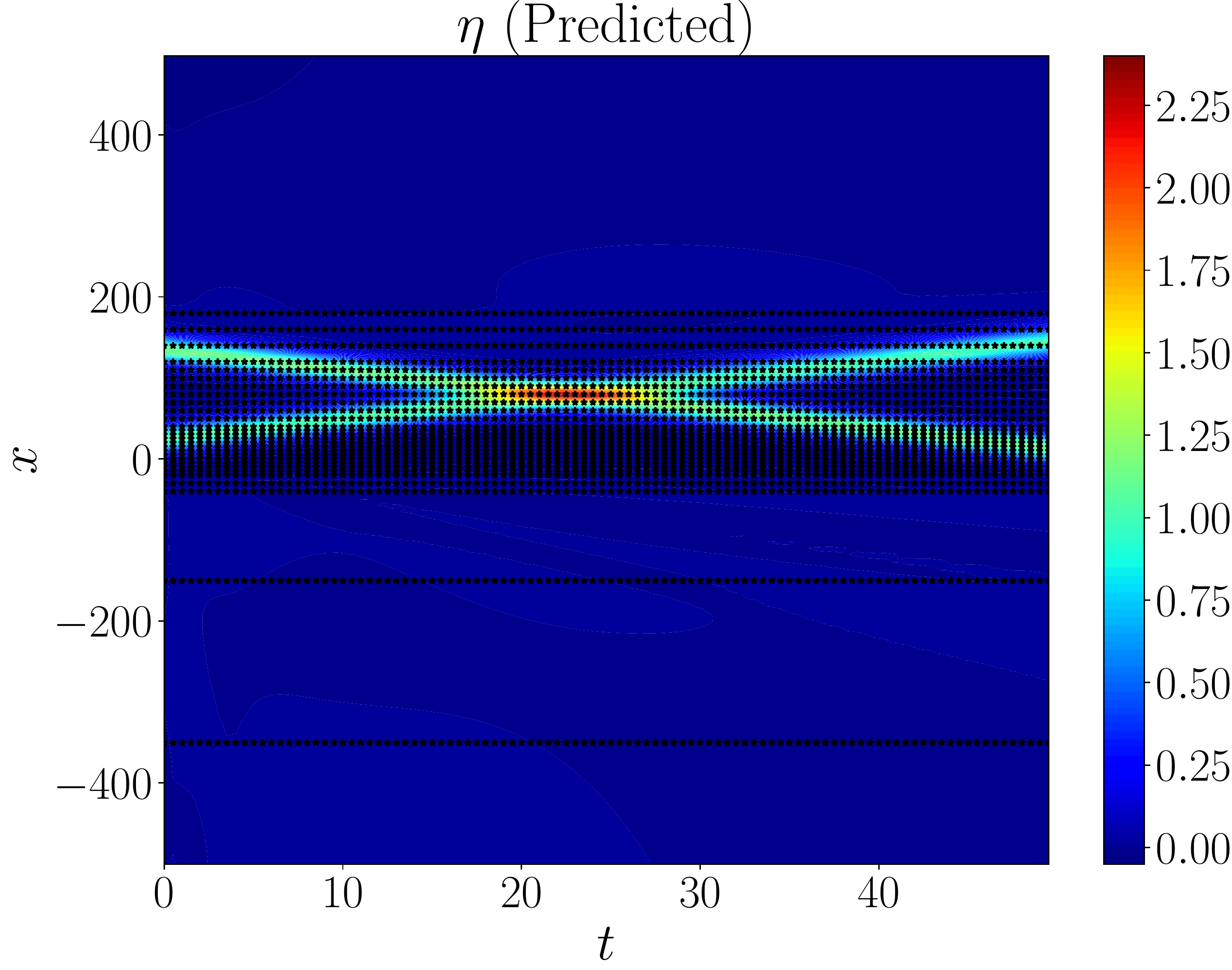} \\
\includegraphics[trim=0cm 0cm 0cm 0cm, clip=true, scale=0.28, angle = 0]{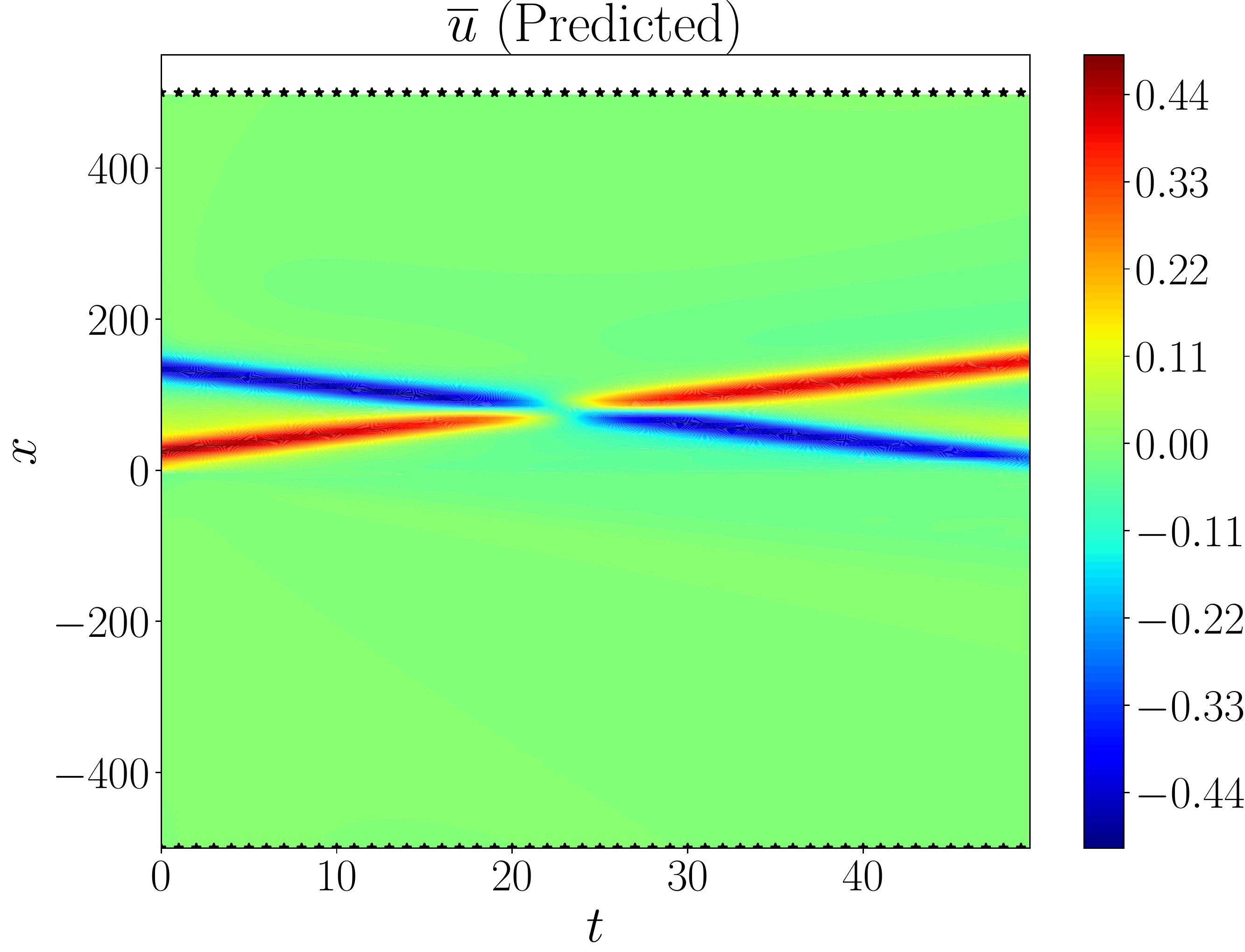} &
\includegraphics[trim=0cm 0cm 0cm 0cm, clip=true, scale=0.28, angle = 0]{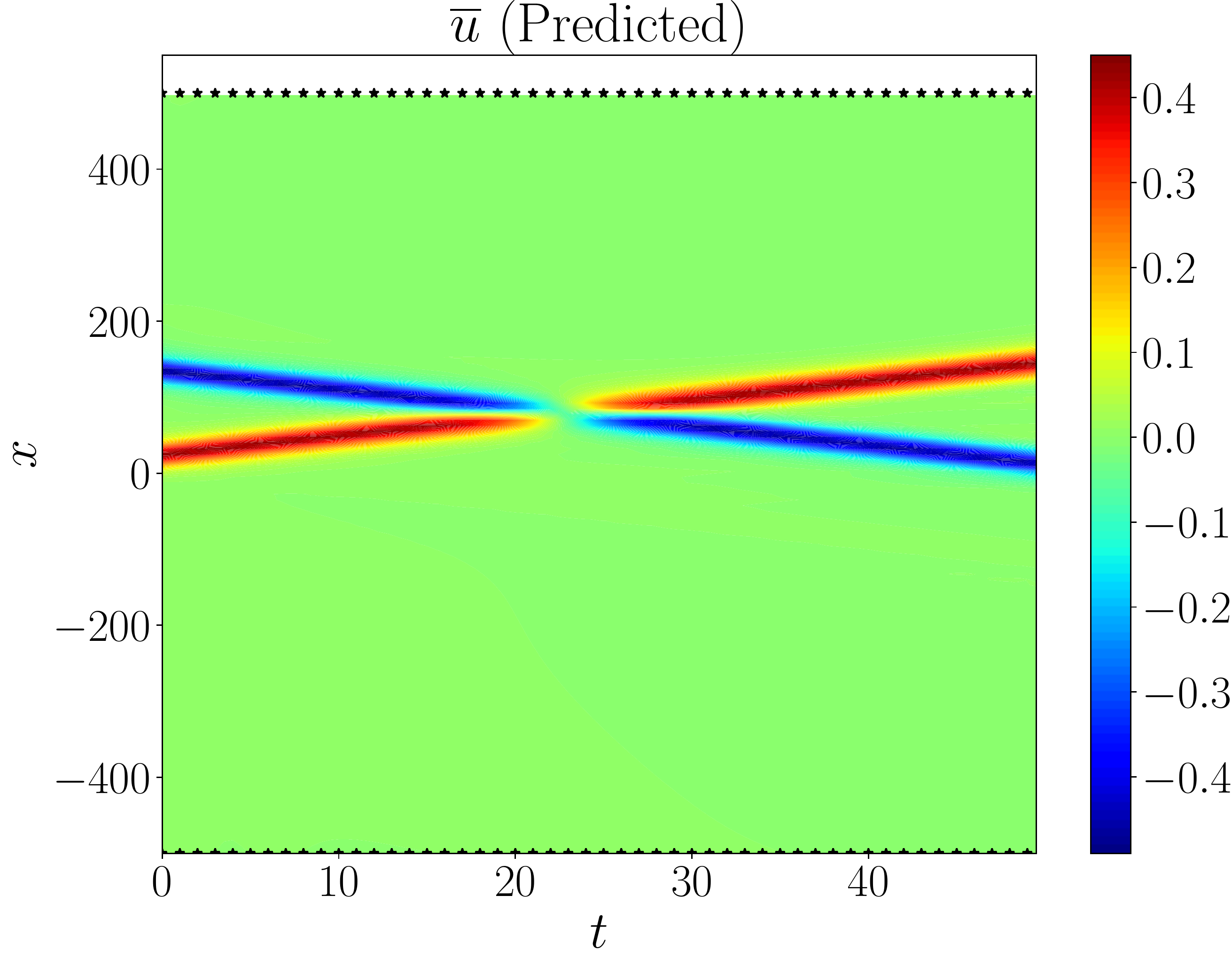}
\end{tabular}
\caption{Head-on collision of two solitary waves: $\eta$ and $\bar{u}$ predictions using 28 gauges. First column shows results in 32-bit precision and second column in 64-bit precision. The locations of gauge data are depicted with '$\star$'}
\label{fig:TC2_1}
\end{figure}

\begin{figure} 
\centering 
\begin{tabular}{cc}
\includegraphics[trim=0cm 0cm 0cm 0cm, clip=true, scale=0.28, angle = 0]{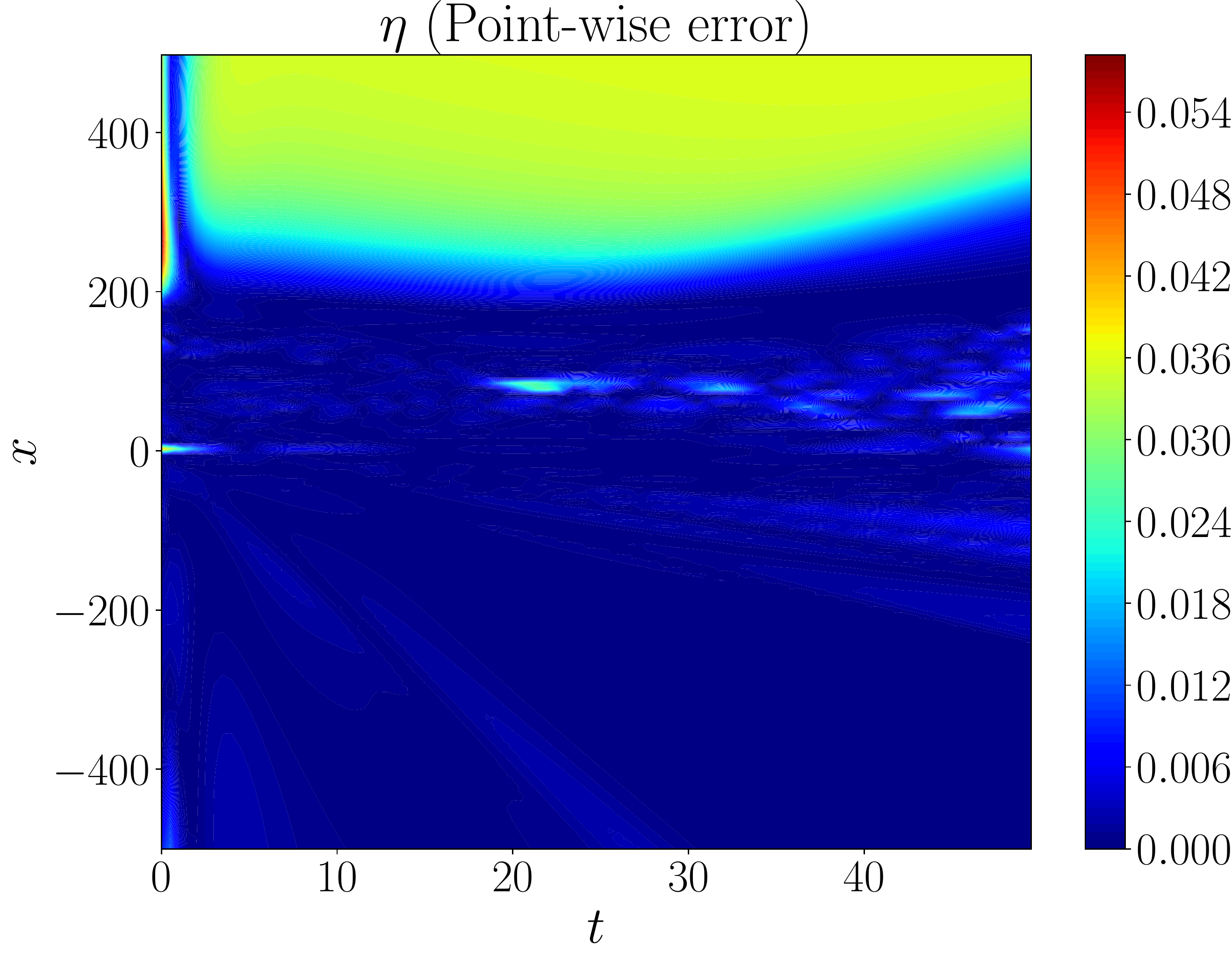} &
\includegraphics[trim=0cm 0cm 0cm 0cm, clip=true, scale=0.28, angle = 0]{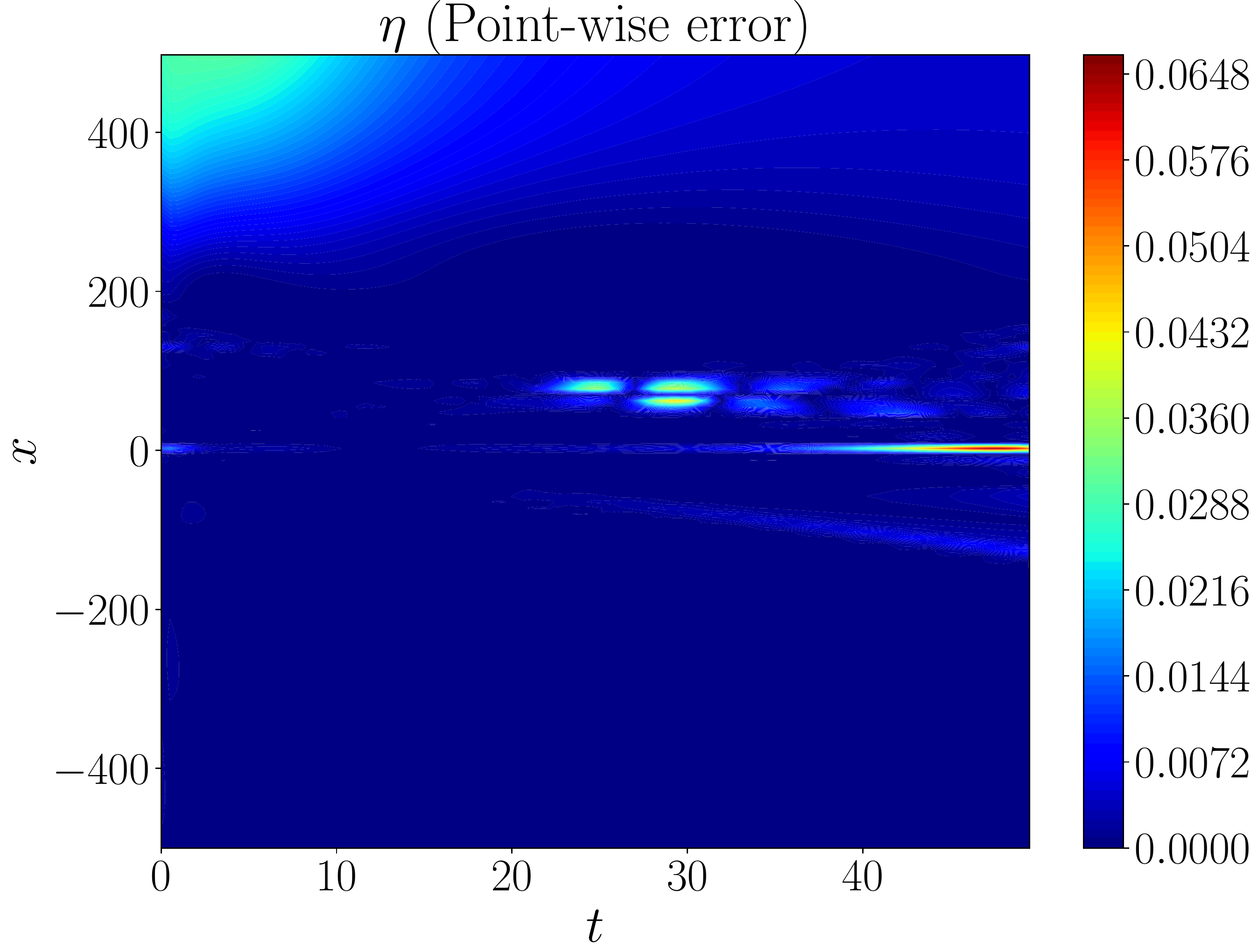} \\
\includegraphics[trim=0cm 0cm 0cm 0cm, clip=true, scale=0.28, angle = 0]{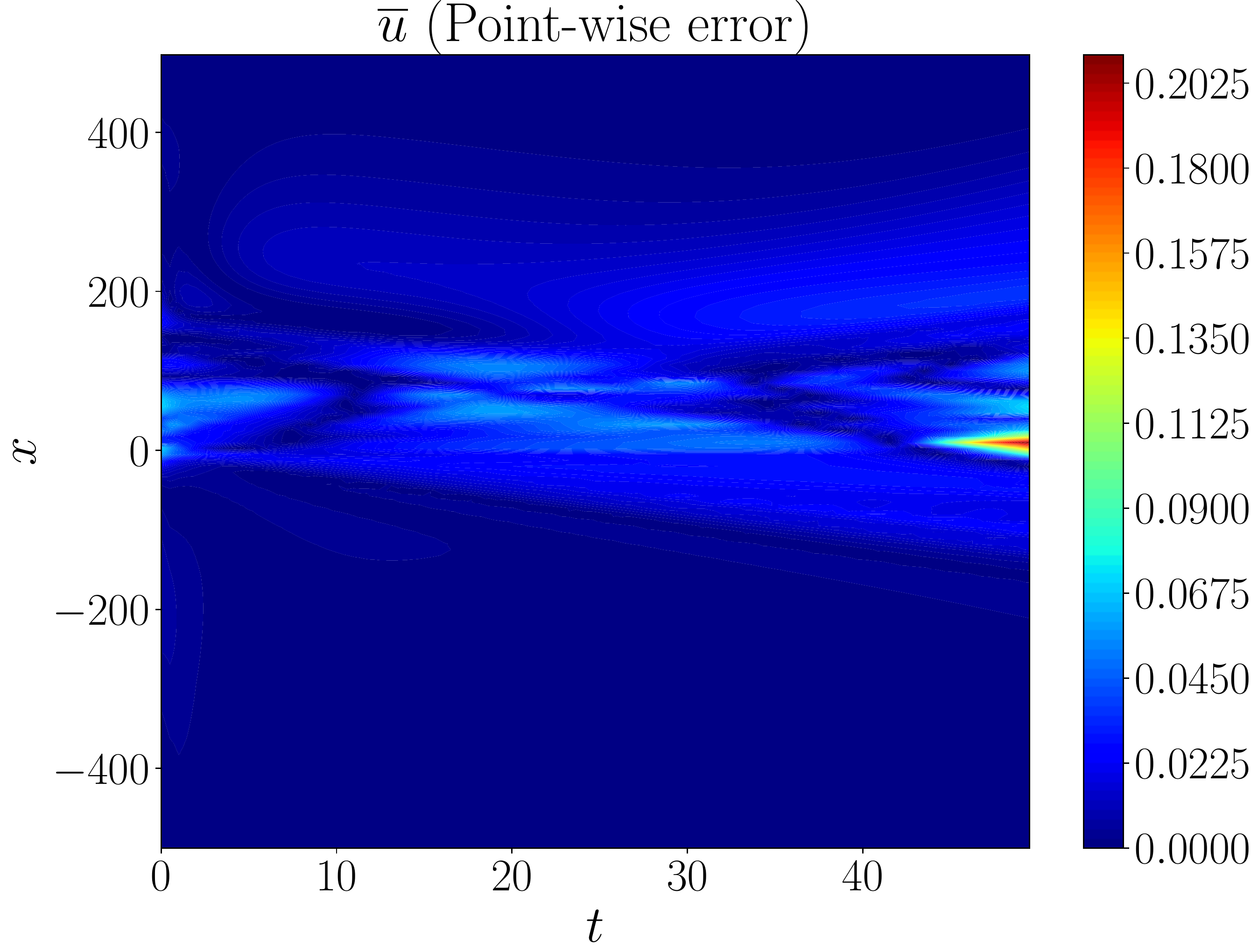} &
\includegraphics[trim=0cm 0cm 0cm 0cm, clip=true, scale=0.28, angle = 0]{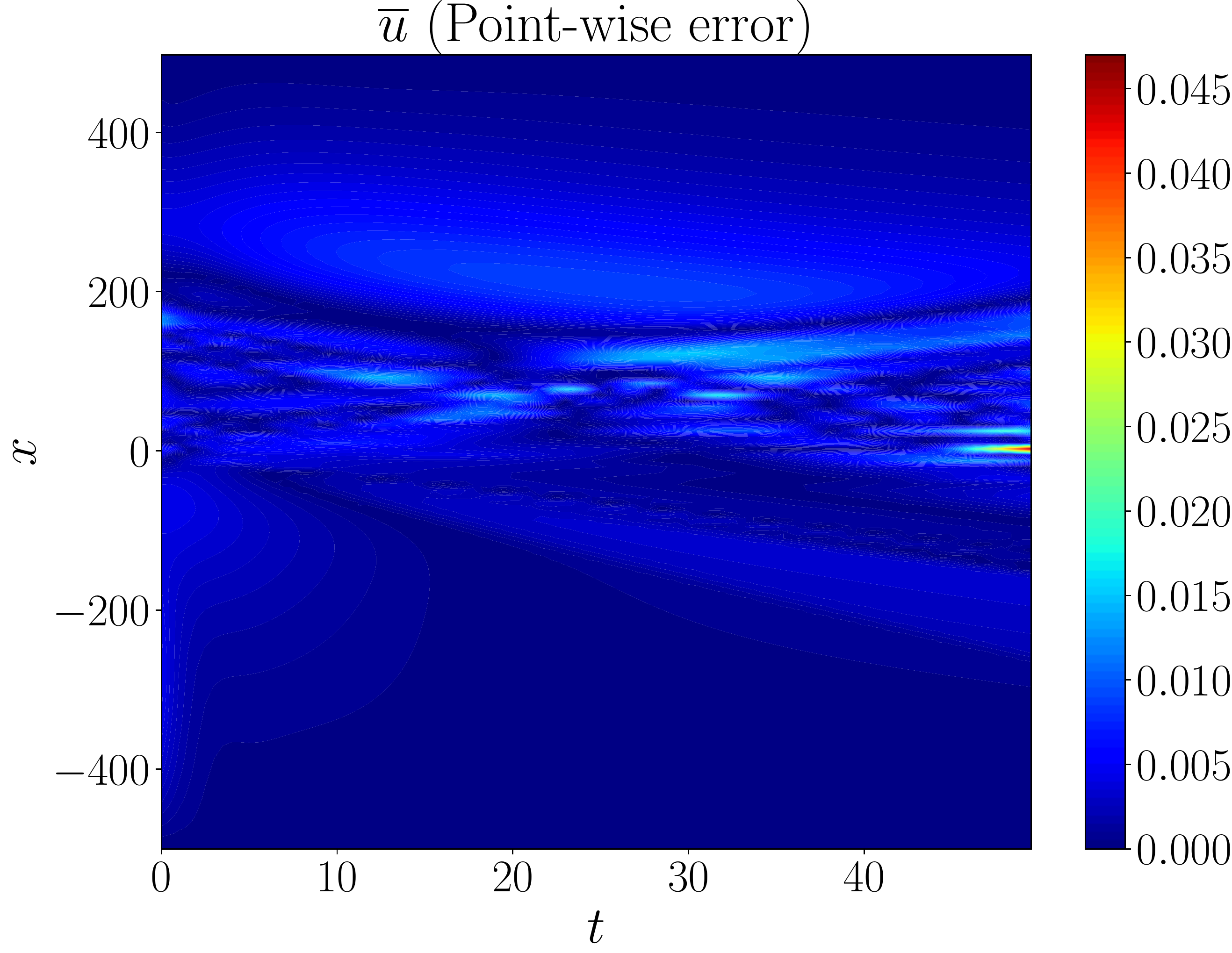}
\end{tabular}
\caption{Head-on collision of two solitary waves : $\eta$ and $\bar{u}$ point-wise errors using 28 gauges. First column shows results for 32 bit precision and second column is for 64 bit precision }
\label{fig:TC2_2}
\end{figure}
\begin{figure} 
\centering 
\includegraphics[trim=0cm 0cm 0cm 0cm, clip=true, scale=0.58, angle = 0]{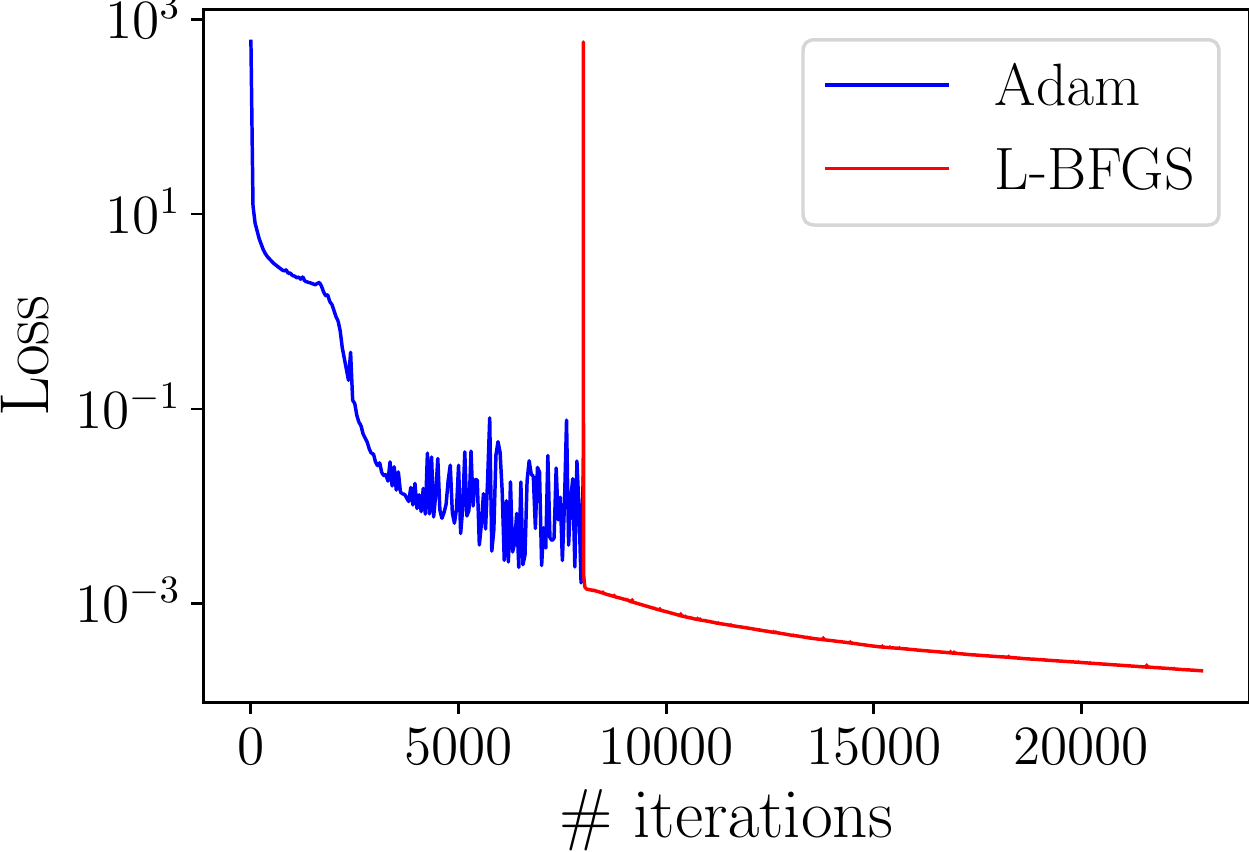}
\includegraphics[trim=0cm 0cm 0cm 0cm, clip=true, scale=0.58, angle = 0]{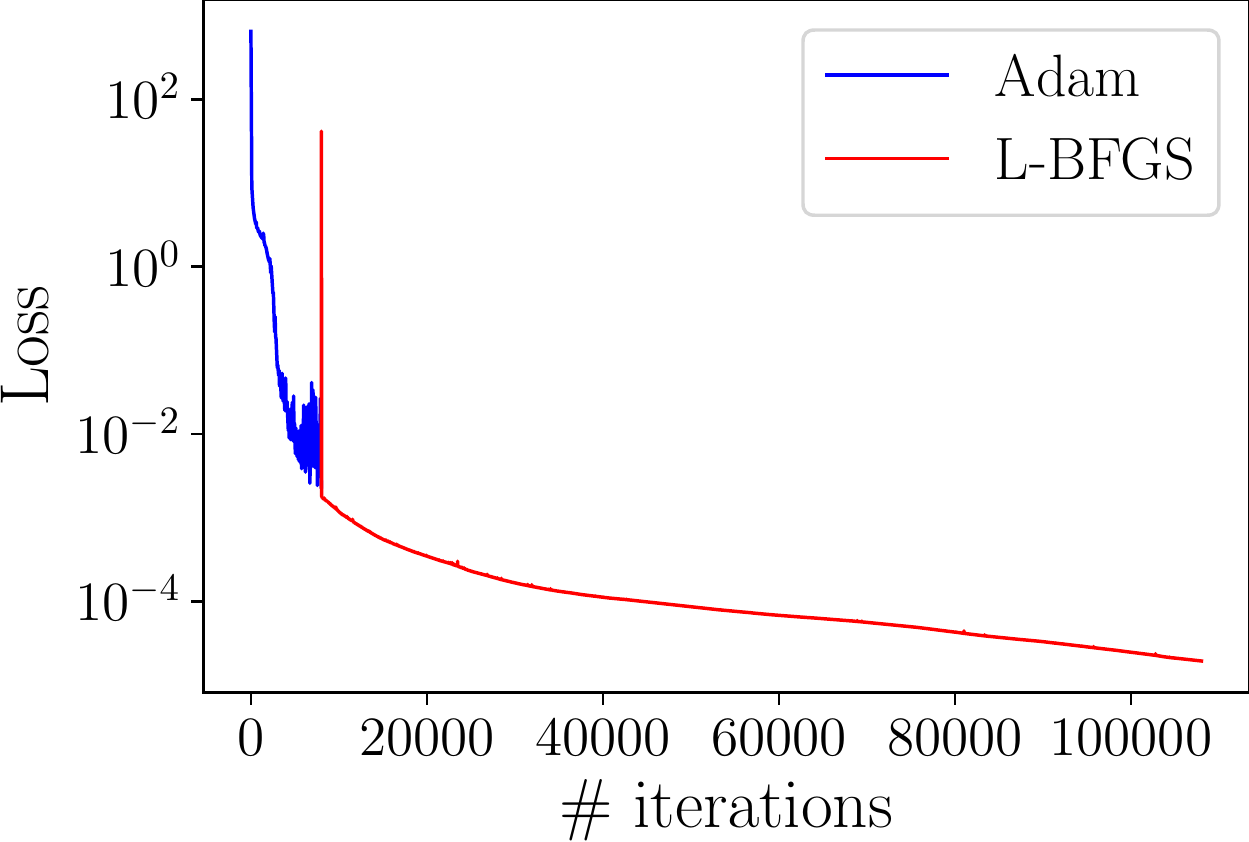}
\caption{Head-on collision of two solitary waves: Loss function versus number of iterations for 32 bit (left) and 64 bit (right) precision }
\label{fig:TC2_3}
\end{figure}
Figure \ref{fig:TC2_1} shows the $\eta$ (first row) and $\bar{u}$ (second row) predictions using 28 gauges. First column shows results for 32 bit precision and second column is for 64 bit precision. The experimental data of \cite{craig2006} that we used in PINN are presented in Figure \ref{fig:TC2_4}. The given experimental data is enriched with synthetic data. The locations of the synthetic gauge data that we used as the input in the PINN are depicted with asterisks. Note that we used only the reflective boundary conditions at $x=\pm 500$ as input for the velocity field. Figure \ref{fig:TC2_2} shows the corresponding point-wise errors for both $\eta$ and $\bar{u}$. The point-wise error is relatively more in the computational region where $\eta$ data is not available, but still the predicted velocity field is accurate. Moreover, with 64-bit precision the predictive accuracy of $\bar{u}$ was increased as expected. To this end, the loss function plots with number of iterations are shown in Figure \ref{fig:TC2_3}. It is worth mentioning that we employed the ADAM optimizer for the first 8000 iterations followed by the L-BFGS optimizer. The L-BFGS optimizer stops earlier for 32-bit precision compared to 64-bit case, giving more accurate results for the later one. In terms of computational cost, the computations in 32-bit precision machine lasted 2.04 hrs, whereas the computations in 64-bit precision required 10.36 hrs on the same computer. This is obvious because, the 64-bit case take up twice as much memory compared to 32-bit case, and doing operations on them can be much slower. Although such cost comes with accuracy, the predicted accuracy of 64-bit is more compared to 32-bit.

\subsection{Solitary waves reflection by a vertical wall}

One important application of strongly nonlinear and weakly dispersive water waves is the study of the impact of the collision of water waves on marine structures such as oil rigs and piers. In such cases the instantaneous force of the wave on the marine structure is the important quantity and its value can be approximated using the formula (\ref{eq:iforce2}). As one may observe in formula (\ref{eq:iforce2}), the computation of the force requires the knowledge of the horizontal velocity field. Here we test the accuracy of PINNs in the study of the force applied by solitary waves of the form (\ref{eq:solitwav}) with amplitudes $A/D=0.1,0.2,\ldots, 0.7$, in collisions with a vertical and impenetrable wall. Taking the computational domain to be $[0,100]$ in dimensionless and unscaled variables, we assume that the wall is located at $x=0$. For the generation of the synthetic data we used again $\Delta x=0.1$ and $\Delta t=0.01$.
Further, we used data from the 28 gauges placed at equal spatial interval of length 2.5 from the wall.
\begin{figure} 
\centering 
\includegraphics[trim=0cm 0cm 0cm 0cm, clip=true, scale=0.37, angle = 0]{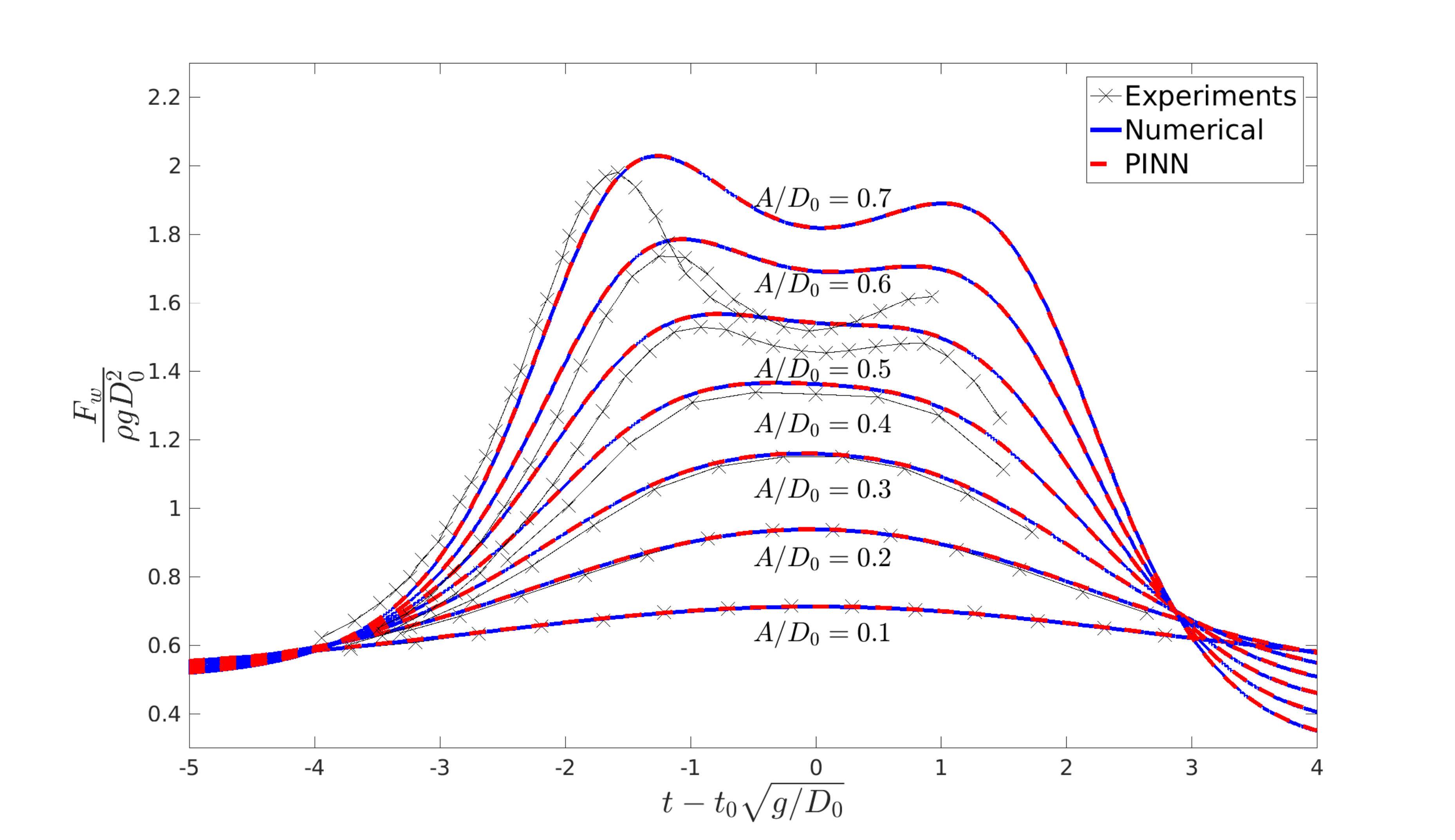}
\caption{Solitary waves reflection by a vertical wall : Experimental values, numerical solution and PINN solution comparison for various $A/D_0$ values}
\label{fig:TC3_1}
\end{figure}

In this test case we used 5 hidden-layers with 60 neurons in each layer of the PINN. The activation function was again the hyperbolic tangent, while we took the learning rate equal to $6\times 10^{-4}$, and the number of residual points used to be $12000$.
Figure \ref{fig:TC3_1} shows a comparison between the experimental data, PINN and numerical results for the computation of the instantaneous force on the wall during the reflection process. The quantities in the graph are scaled as follows: The force is scaled by the force applied to the mass of a water column of unit volume ($\rho g D_0^2$) while the time has been translated so as to fit with the time instance $t_0/\sqrt{g/D_0}$ where the maximum runup of the solitary wave on the wall was achieved. We observe that the PINN results are in complete agreement with the numerical data, while there is a disagreement between the experimental data of \cite{CWB1997} and numerical data for solitary waves of normalized amplitude greater than $0.5$. For solitary waves of larger amplitude the error was about $20-30\%$. On the other hand the maximum instantaneous force was predicted with satisfactory accuracy in all the cases. This shows that the Serre-Green-Naghdi system is able to describe accurately the propagation of waves of amplitude up to $0.5$. Note that the numerical results are plotted for the forward problem where initial velocity profile as well as wall boundary conditions on velocity are known, whereas PINN results are obtained from the inverse problem where velocity is known only on the wall boundary. 
 
\subsection{Interaction of solitary waves with an obstacle}

\begin{figure} 
\centering 
\includegraphics[trim=1cm 0cm 1.7cm 0cm, clip=true, scale=0.25, angle = 0]{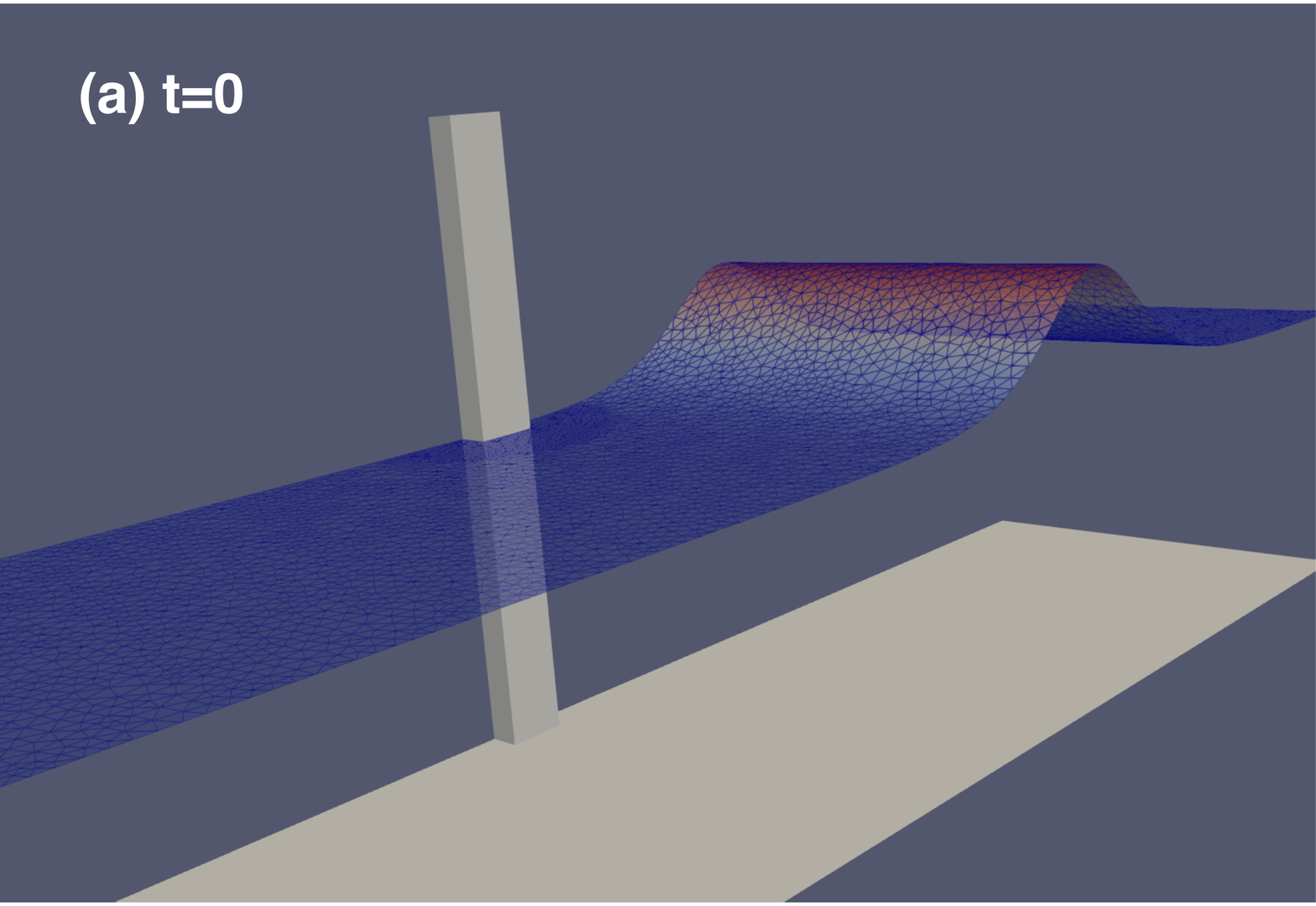}
\includegraphics[trim=1cm 0cm 1.7cm 0cm, clip=true, scale=0.25, angle = 0]{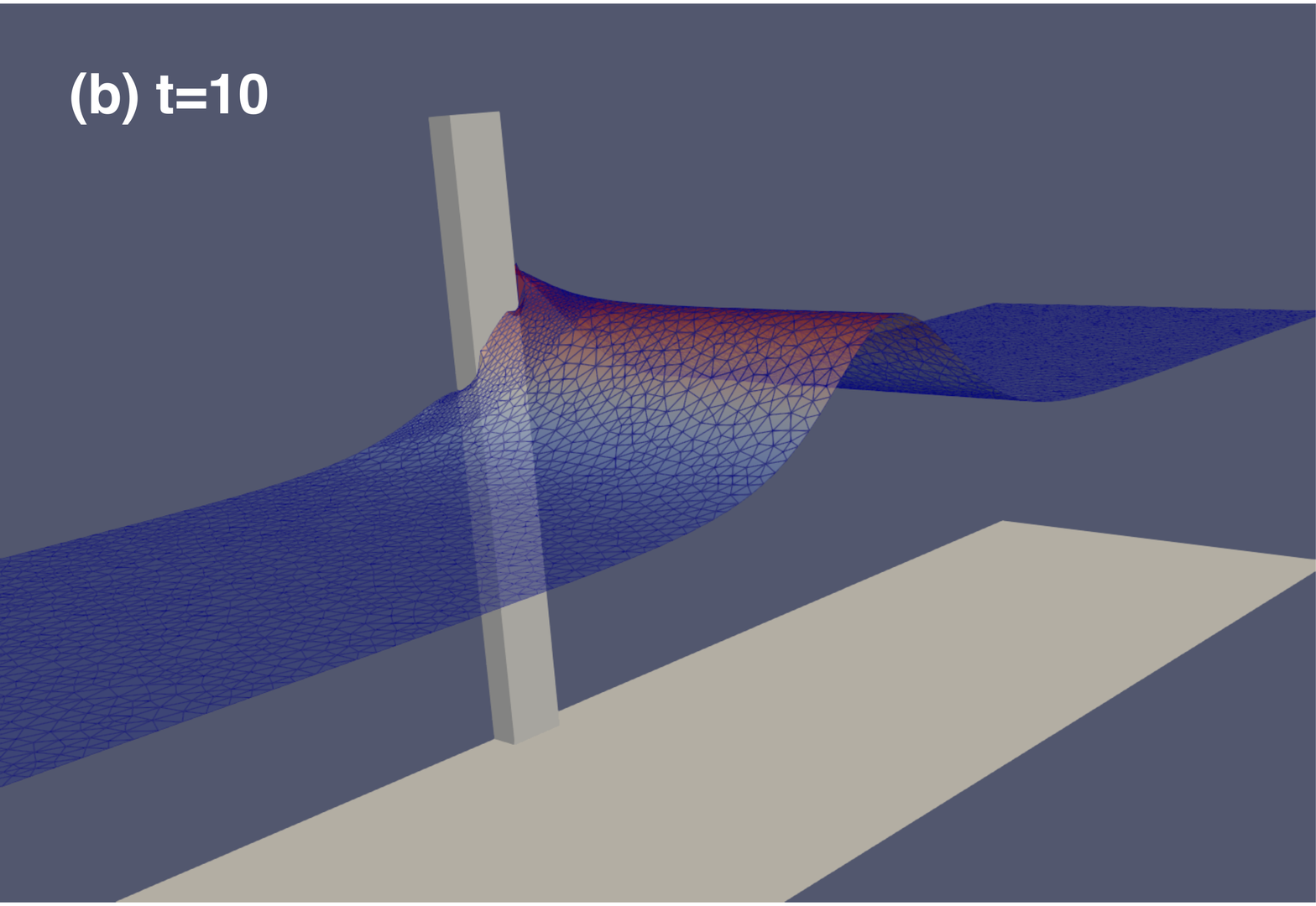}\\
\includegraphics[trim=1cm 0cm 1.7cm 0cm, clip=true, scale=0.25, angle = 0]{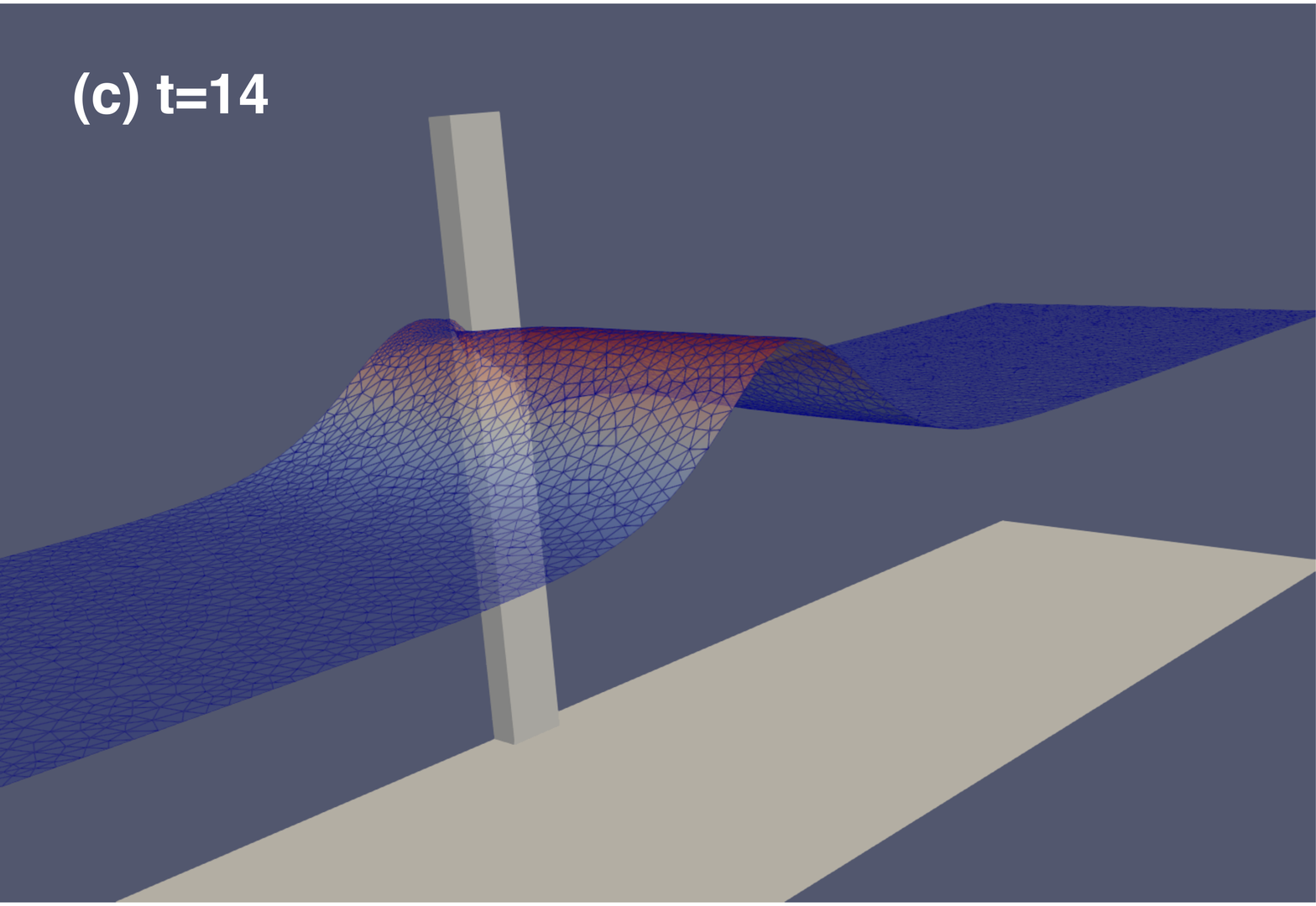}
\includegraphics[trim=1cm 0cm 1.7cm 0cm, clip=true, scale=0.25, angle = 0]{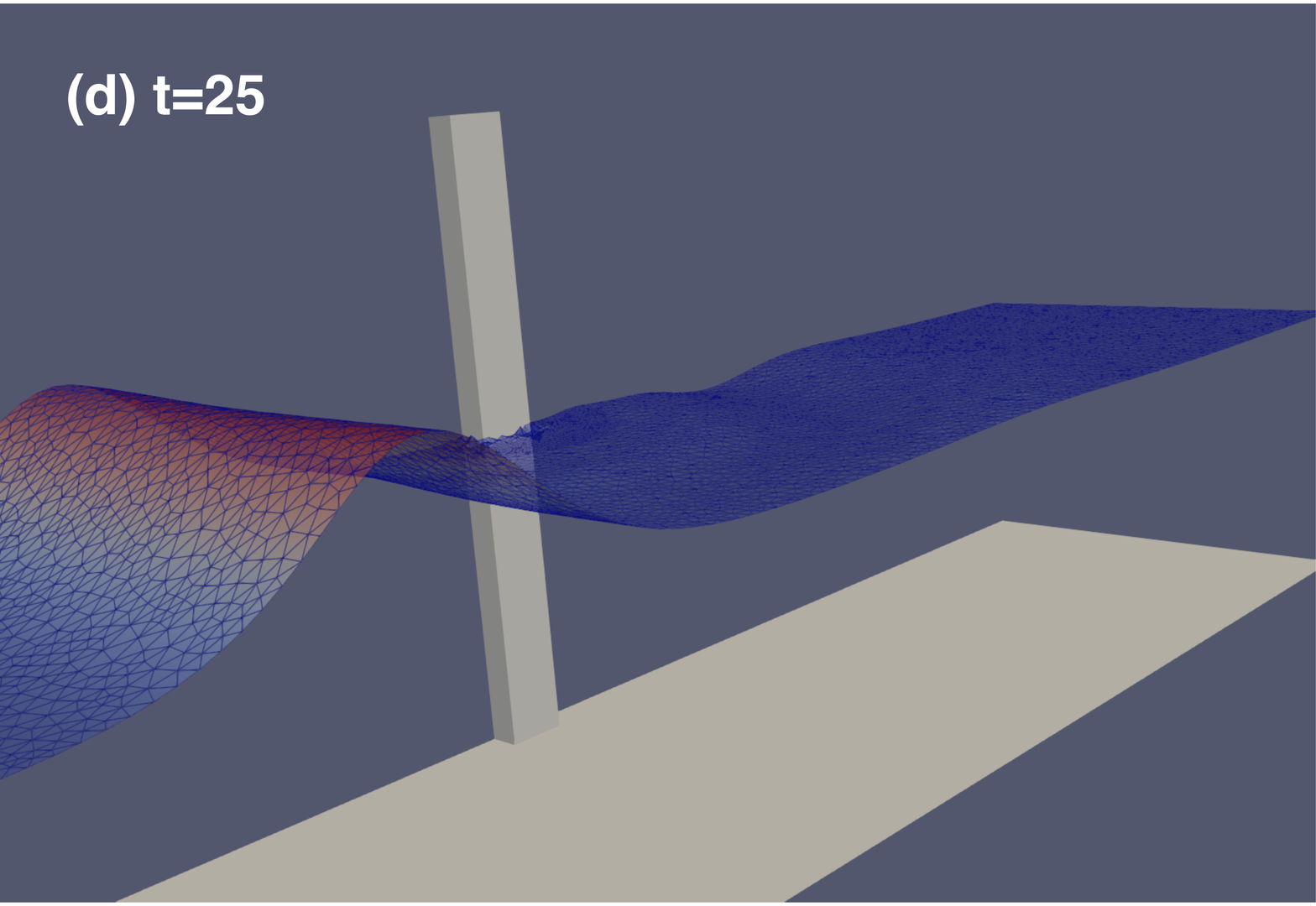}
\caption{Interaction of solitary waves with a rectangular obstacle: Due to symmetry, only upper half computational domain is used for the simulation}
\label{fig:TC4_0}
\end{figure}

\begin{figure} 
\centering 
\includegraphics[trim=1cm 0cm 1.7cm 0cm, clip=true, scale=0.3, angle = 0]{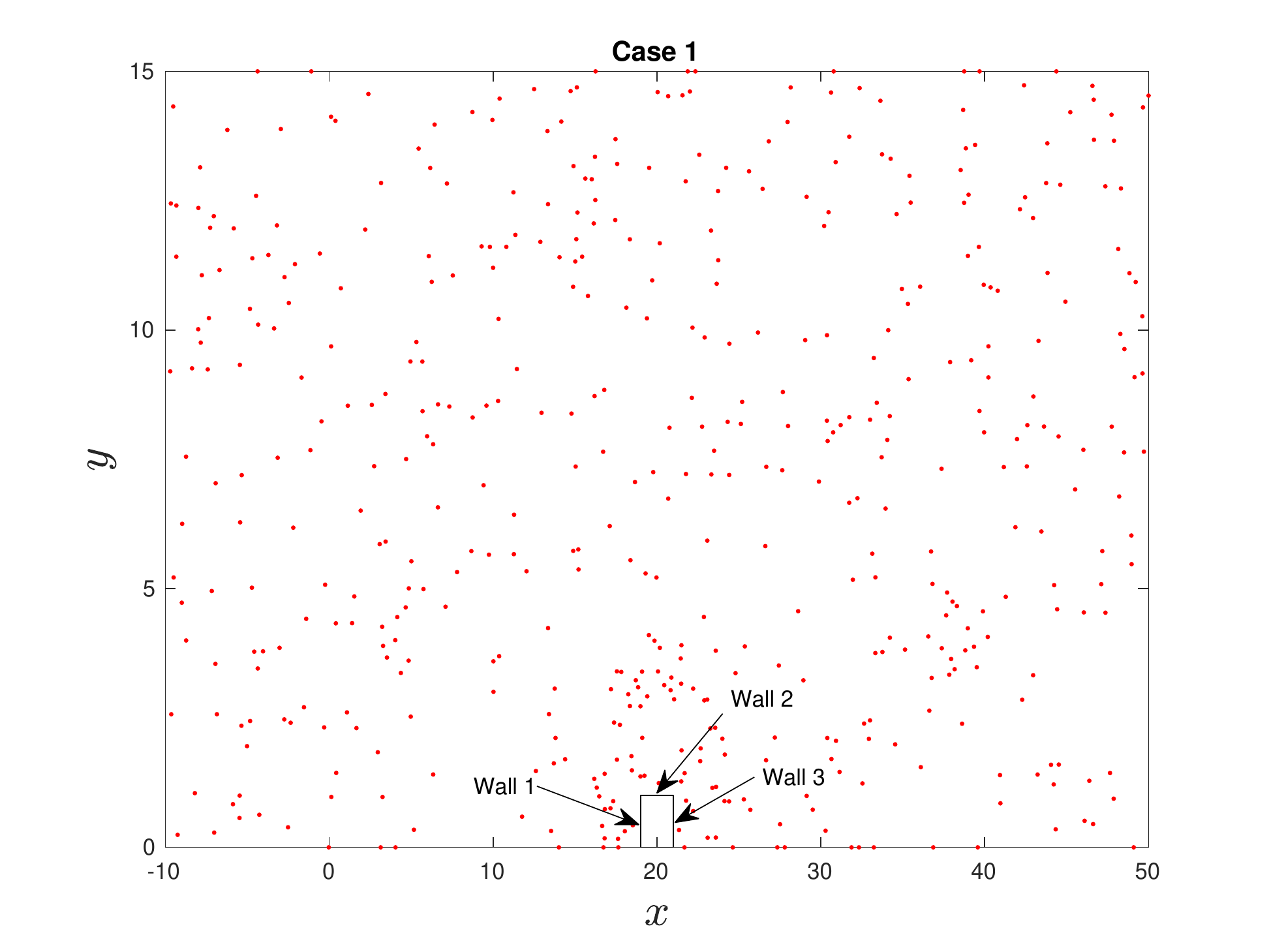}
\includegraphics[trim=1cm 0cm 1.7cm 0cm, clip=true, scale=0.3, angle = 0]{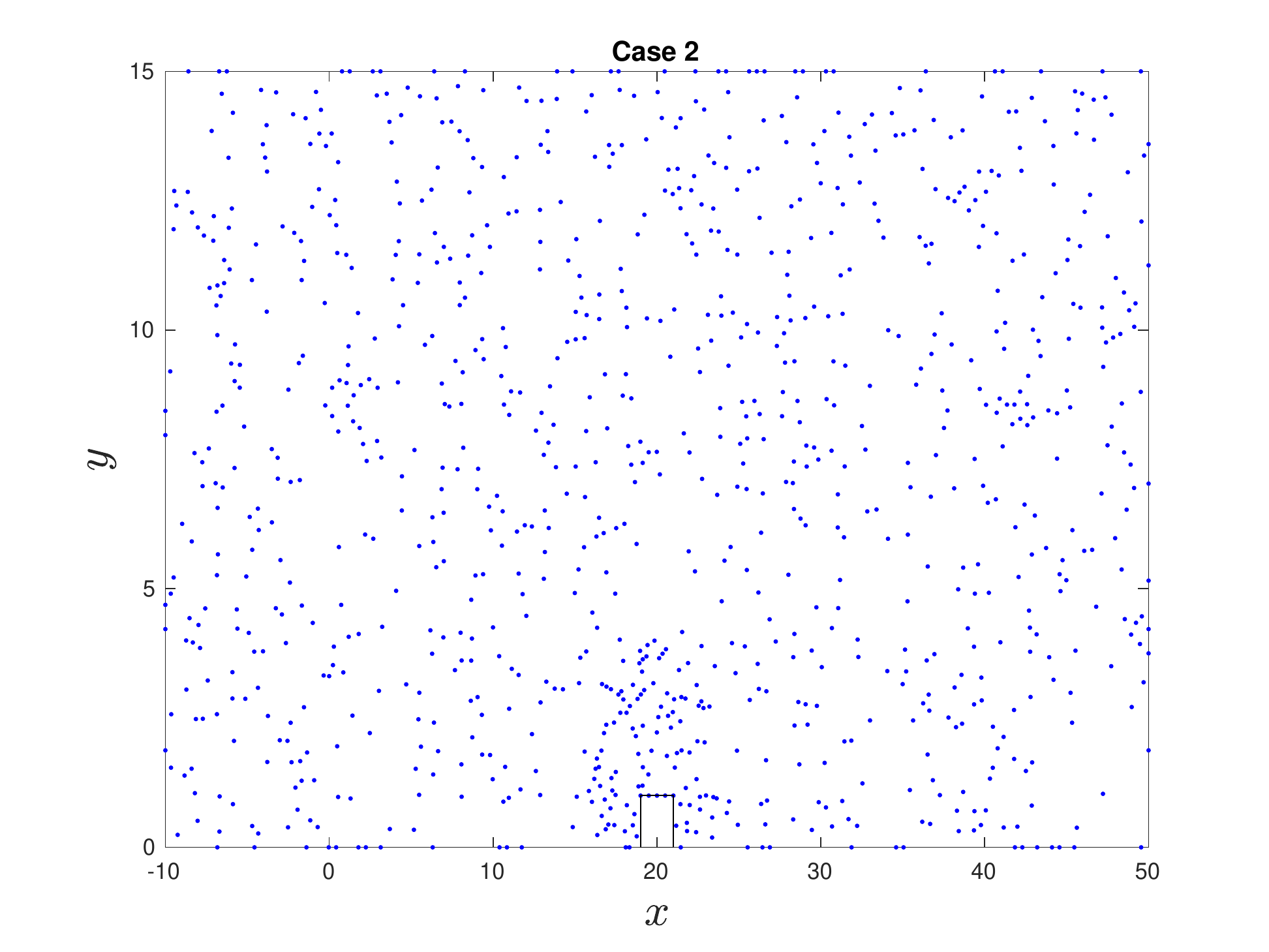}
\includegraphics[trim=1cm 0cm 1.7cm 0cm, clip=true, scale=0.3, angle = 0]{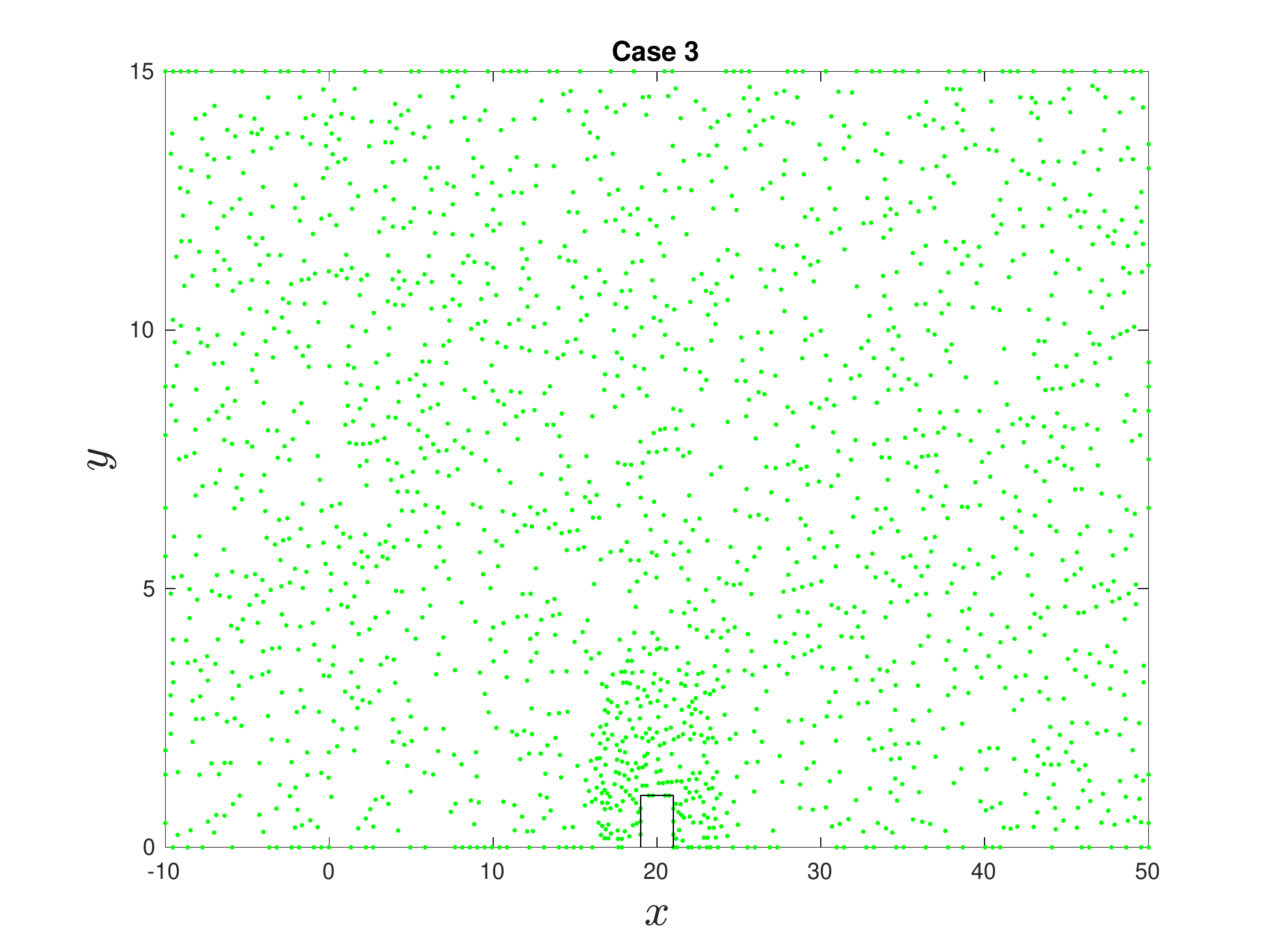}
\caption{Interaction of solitary waves with a rectangular obstacle: Three different cases of spatial locations of synthetic data (course, medium and fine) in the upper half computational domain used in PINNs}
\label{fig:TC4_1}
\end{figure}

In this experiment we consider the Serre-Green-Naghdi equations (\ref{eq:Serre1})--(\ref{eq:Serre2}) written in dimensionless and unscaled form, and we compute the force applied to the walls in a collision of a solitary wave with a rectangular obstacle. The domain consists of a rectangular channel with dimensions $[-10,50]\times[0,15]$, while the obstacle is represented by the rectangular hole $[19,21]\times[0,1]$. Due to symmetry, only the upper half computational domain is used for the simulation as shown in figure \ref{fig:TC4_1}. The solitary wave is described by formula (\ref{eq:solitwav}) with $A=0.1$, and it was translated initially in such a way that its maximum value was achieved on the line $x=8$. To generate synthetic data for this particular problem, we employed the finite element method of \cite{mitsotakis2017b} with general (Delauney) triangulations of two-dimensional domains. Using the slip-wall boundary condition $\bu\cdot\bn=0$ and $10020$ triangles we generated the synthetic data as shown in Figure \ref{fig:TC4_0}. We also computed the force applied on the faces of the three obstacle walls using formula (\ref{eq:iforce2}) at the locations $(19,0)$, $(21,0)$ and $(20,1)$.

\begin{figure} 
\centering 
\includegraphics[trim=0.6cm 0cm 1.7cm 0cm, clip=true, scale=0.3, angle = 0]{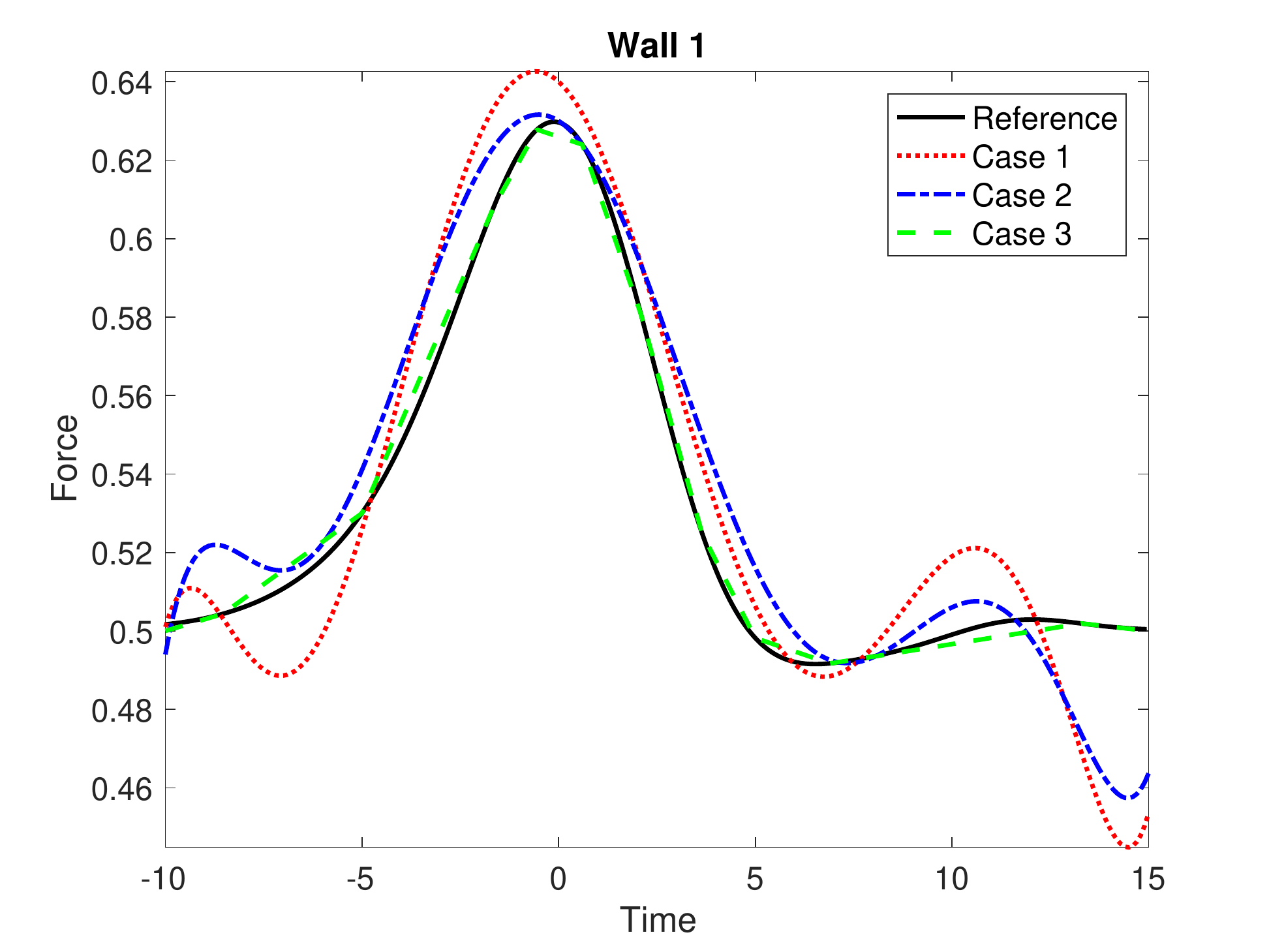}
\includegraphics[trim=0.6cm 0cm 1.7cm 0cm, clip=true, scale=0.3, angle = 0]{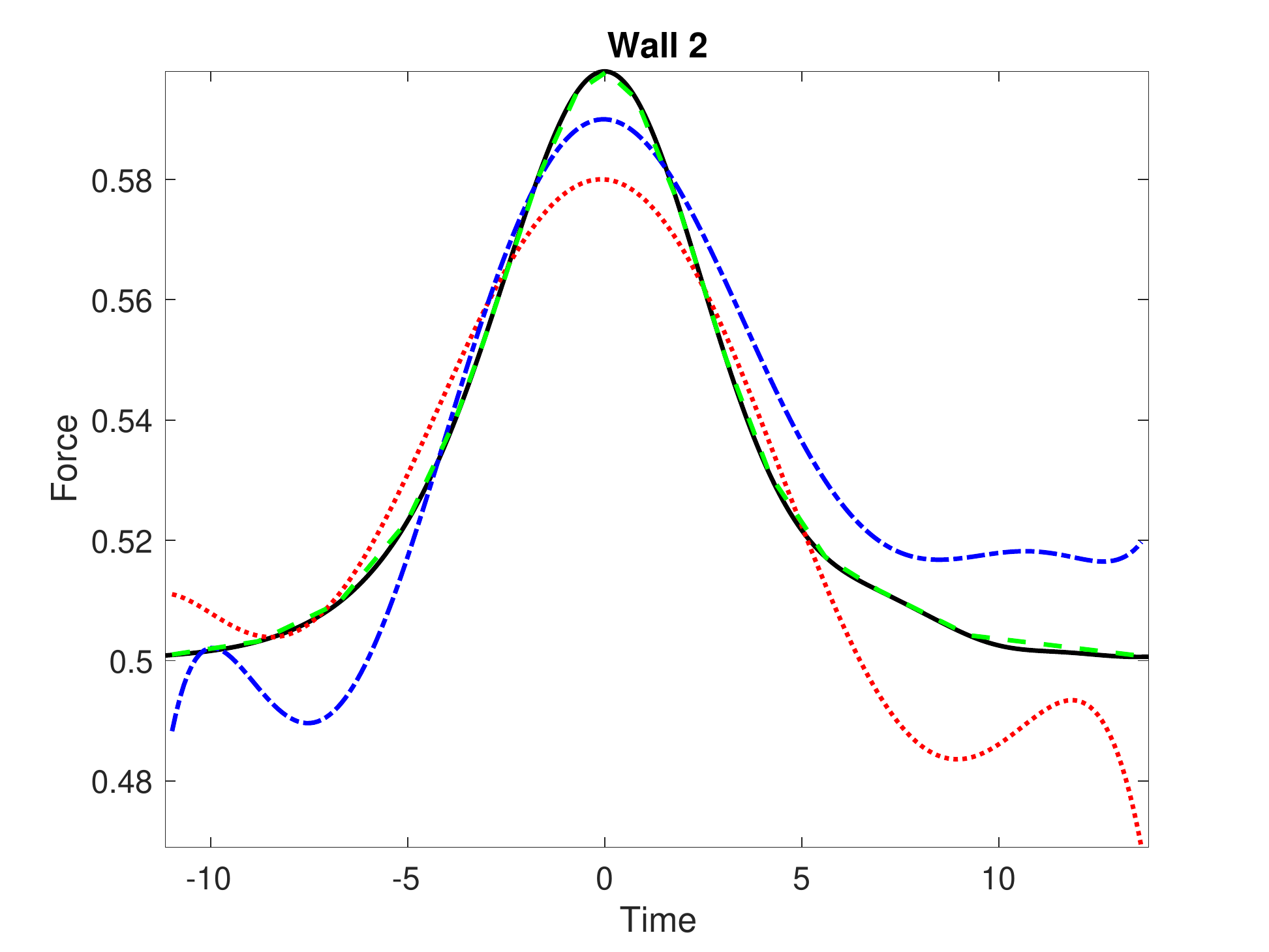}
\includegraphics[trim=0.6cm 0cm 1.7cm 0cm, clip=true, scale=0.3, angle = 0]{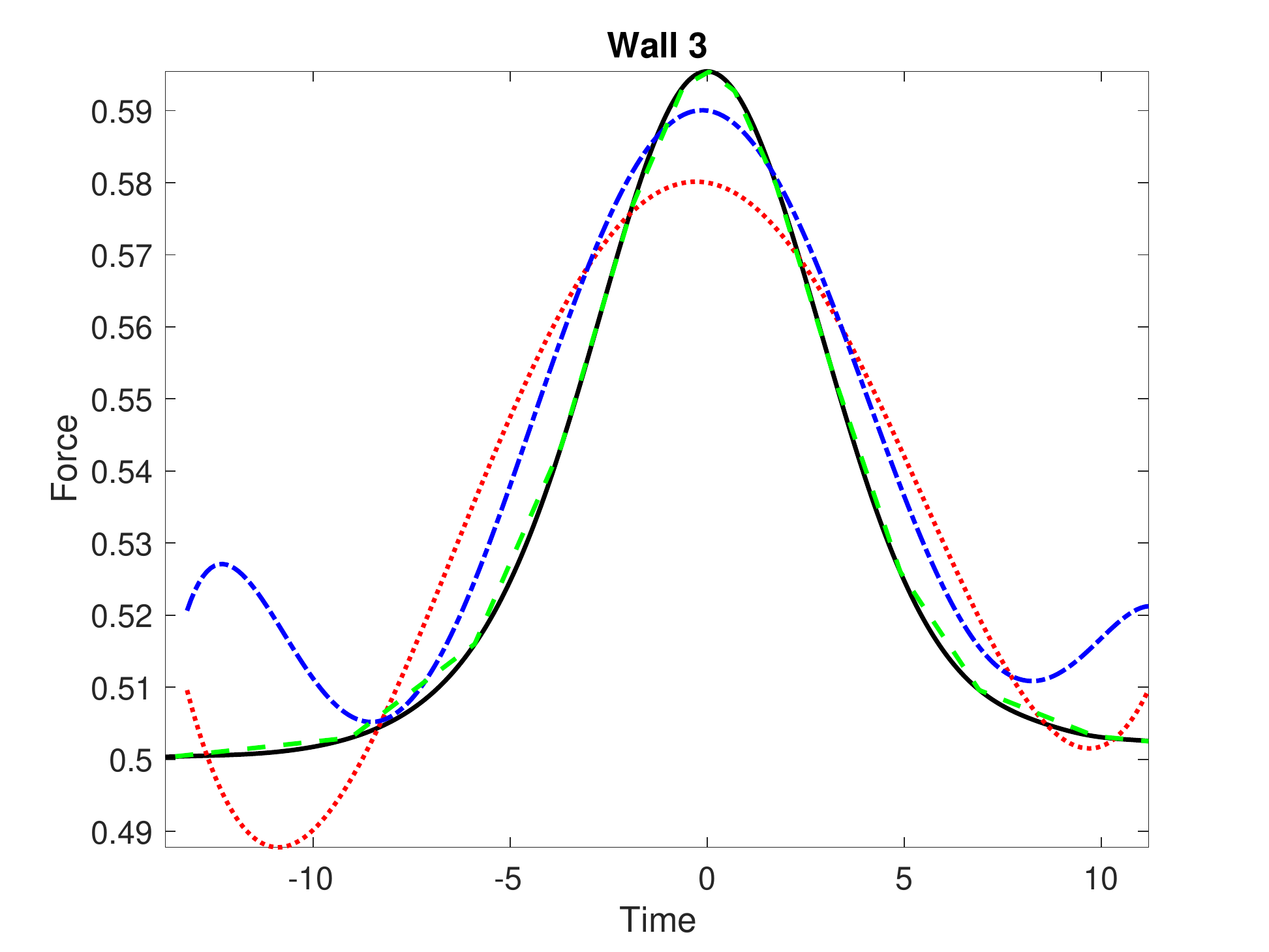}
\caption{Interaction of solitary waves with a rectangular obstacle: The reconstructed normalized force $F_w/\rho g D_0^2$ as a function of the scaled time $t-t_0\sqrt{g/D_0}$ applied by the solitary wave at the three faces of the rectangular obstacle}
\label{fig:TC4_2}
\end{figure}
\begin{figure} 
\centering 
\includegraphics[trim=0cm 0cm 1cm 0cm, clip=true, scale=0.56, angle = 0]{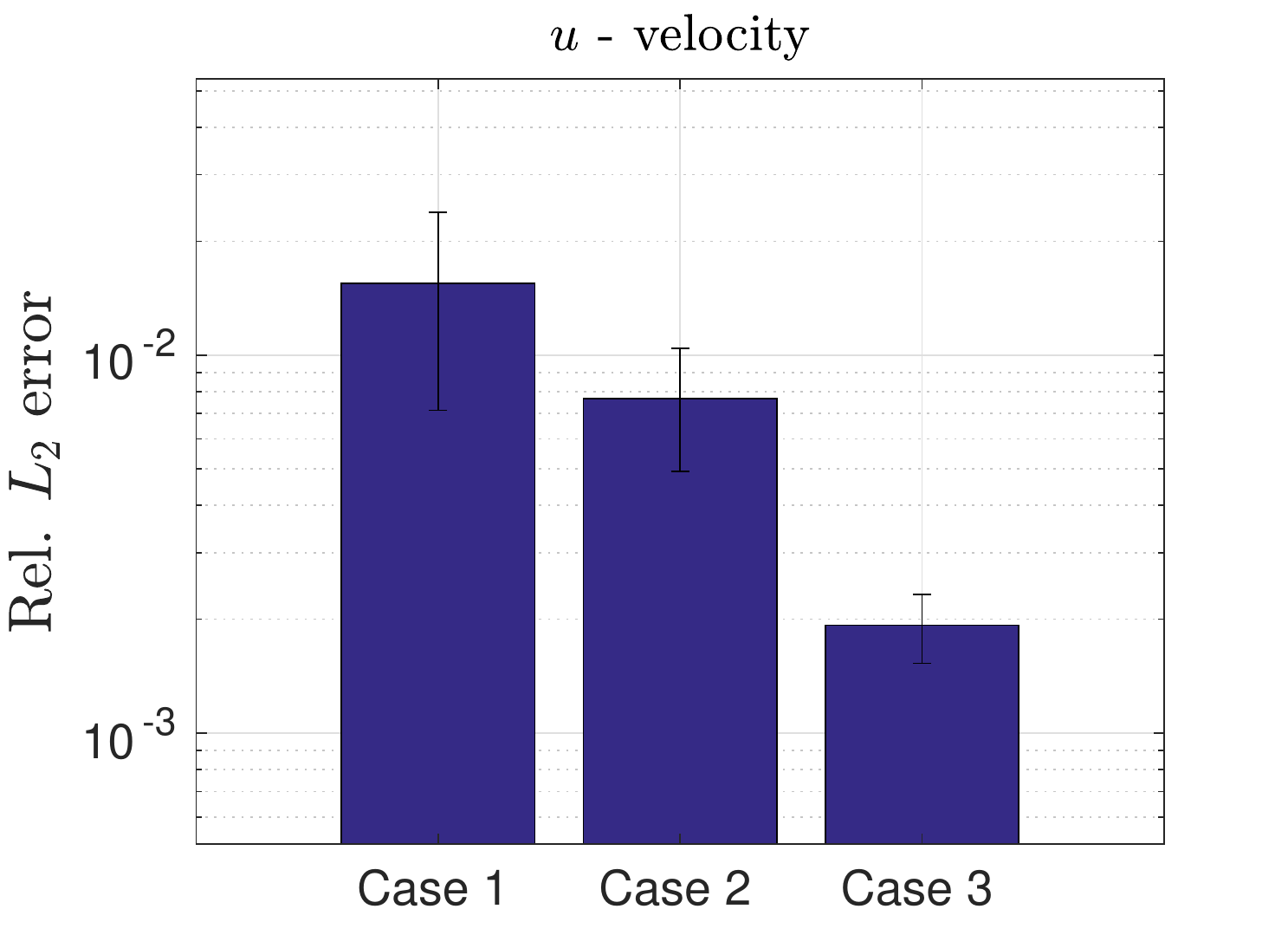}
\includegraphics[trim=0cm 0cm 1cm 0cm, clip=true, scale=0.56, angle = 0]{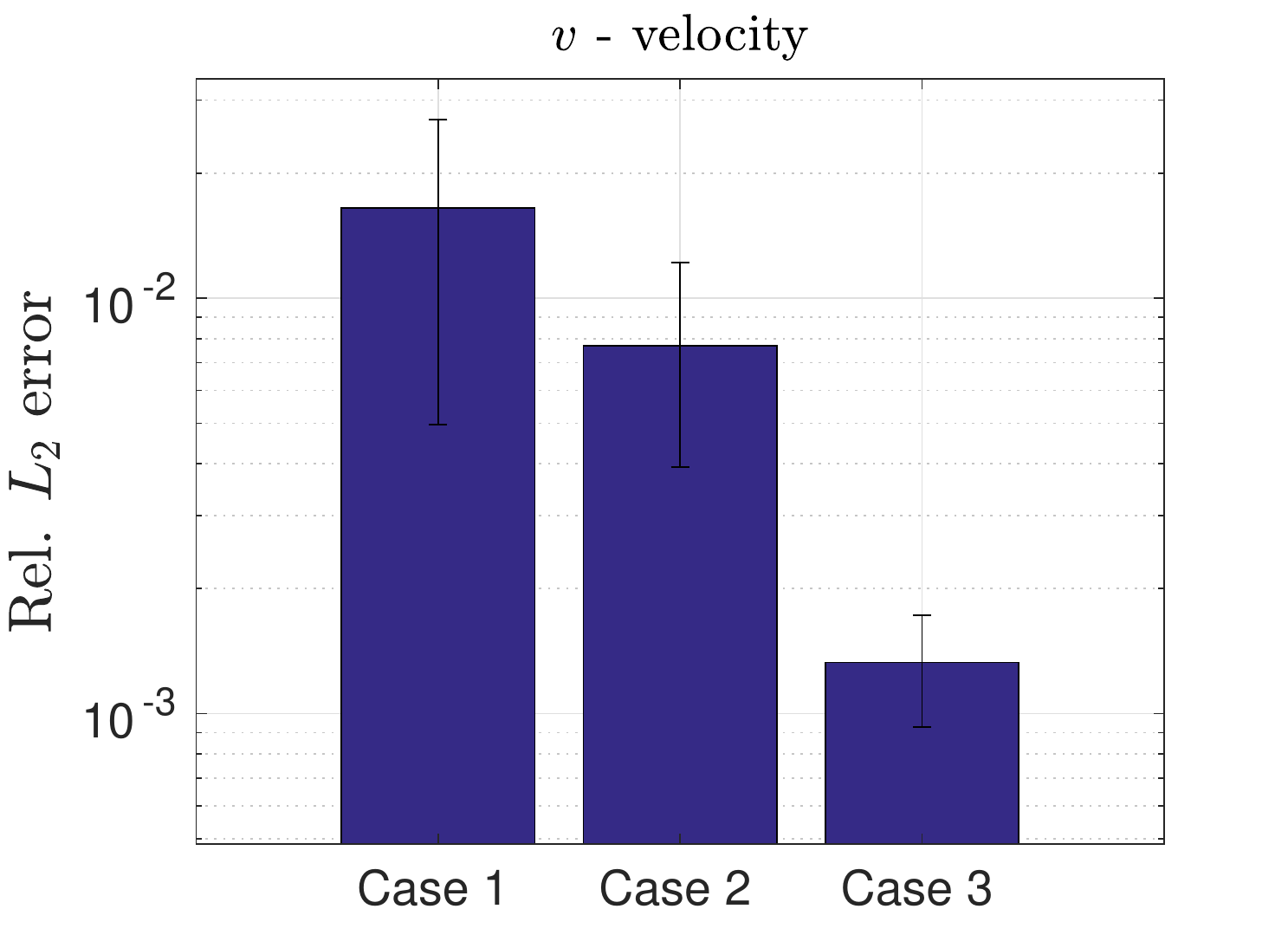}
\caption{Interaction of solitary waves with a rectangular obstacle: Mean and standard deviation of relative $L_2$ error in the velocity field $(u,v)$ for three cases using 10 different initialization of tunable parameters}
\label{fig:TC4_3}
\end{figure}
In this test case we used 5 hidden-layers with 100 neurons in each layer. The activation function is hyperbolic tangent, the learning rate is $8\times 10^{-4}$, and the number of residual points used is $30000$, which are randomly chosen in the entire spatio-temporal domain. The aim is to infer the solitary wave force acting at the midpoint of the obstacle walls. Once again, the only data we used as input in the PINN is the water surface elevation from the gauges as time series data, and the total depth at discrete locations. In order to observe convergence of the computed solution to the reference data we considered three cases, where in each case we used different number of data. Figure \ref{fig:TC4_1} shows the three different cases of spatial locations of synthetic data (coarse, medium and dense) in the computational domain, used for PINN method. The location of the solid obstacle with walls 1, 2 and 3 is also shown in the Figure \ref{fig:TC4_1}. Cases 1,2 and 3 use 500, 1000 and 1500 wave gauges, respectively. The number of data given by the wave gauges in the temporal direction is 80 in each case.  The reason for using such a scattered data over the entire spatio-temporal domain is to accurately infer the velocity field in the entire domain along with the transient forces acting on the walls. As the solitary wave approaches the rectangular obstacle, the magnitude of the force increases until it reaches its peak, and then decreases as the solitary wave passes around the obstacle.
The comparison of the predicted force by PINN on wall 1, 2 and 3 with the reference numerical solution given by all the three cases is presented in Figure \ref{fig:TC4_2}. It can be seen that by increasing the number of data points the predictive accuracy of the PINN solution consistently increases. Similar observation can be made from Figure \ref{fig:TC4_3} where mean and standard deviation of the relative $L_2$ error for the velocity fields are plotted for three different cases using 10 different randomly chosen initializations of all the tunable parameters. With increase in the number of data points, the predictive accuracy of the PINN methodology consistently increases.

\subsubsection{Multi-fidelity Simulation}
\begin{figure} 
\centering 
\includegraphics[trim=0cm 0cm 0cm 0cm, clip=true, scale=0.45, angle = 0]{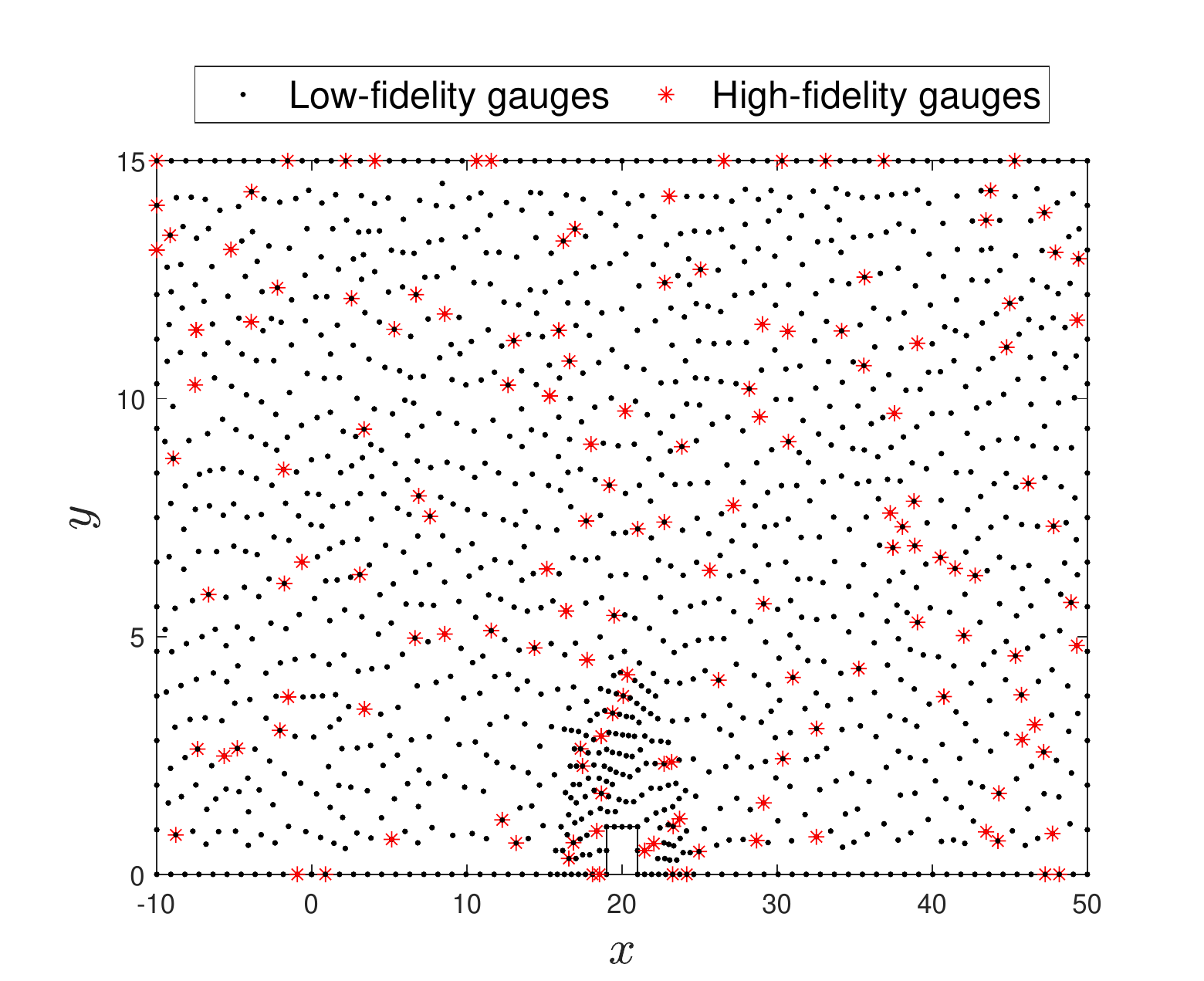}
\includegraphics[trim=0cm 0cm 0cm 0cm,, clip=true, scale=0.462, angle = 0]{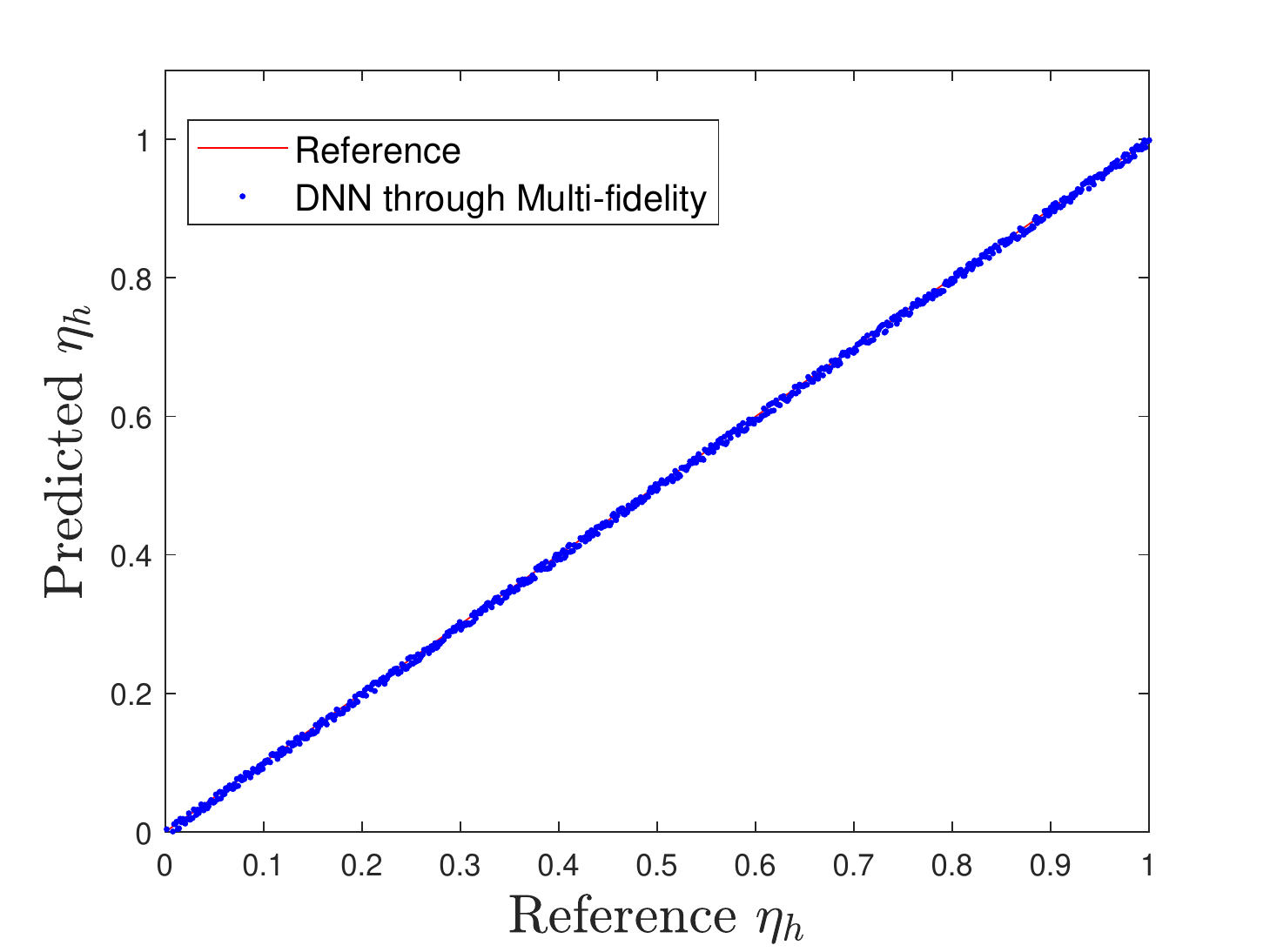}
\caption{Interaction of solitary waves with a rectangular obstacle: (Left) locations of low- and high-fidelity gauges, where 1360 low and 50 high-fidelity gauges are used. (Right) the predicted vs reference normalized high-fidelity $\eta$ data, denoted by $\eta_h$.}
\label{fig:TC4_3}
\end{figure}
\begin{figure} [htbp]
\centering 
\includegraphics[trim=0.5cm 0cm 1.2cm 0cm, clip=true, scale=0.43, angle = 0]{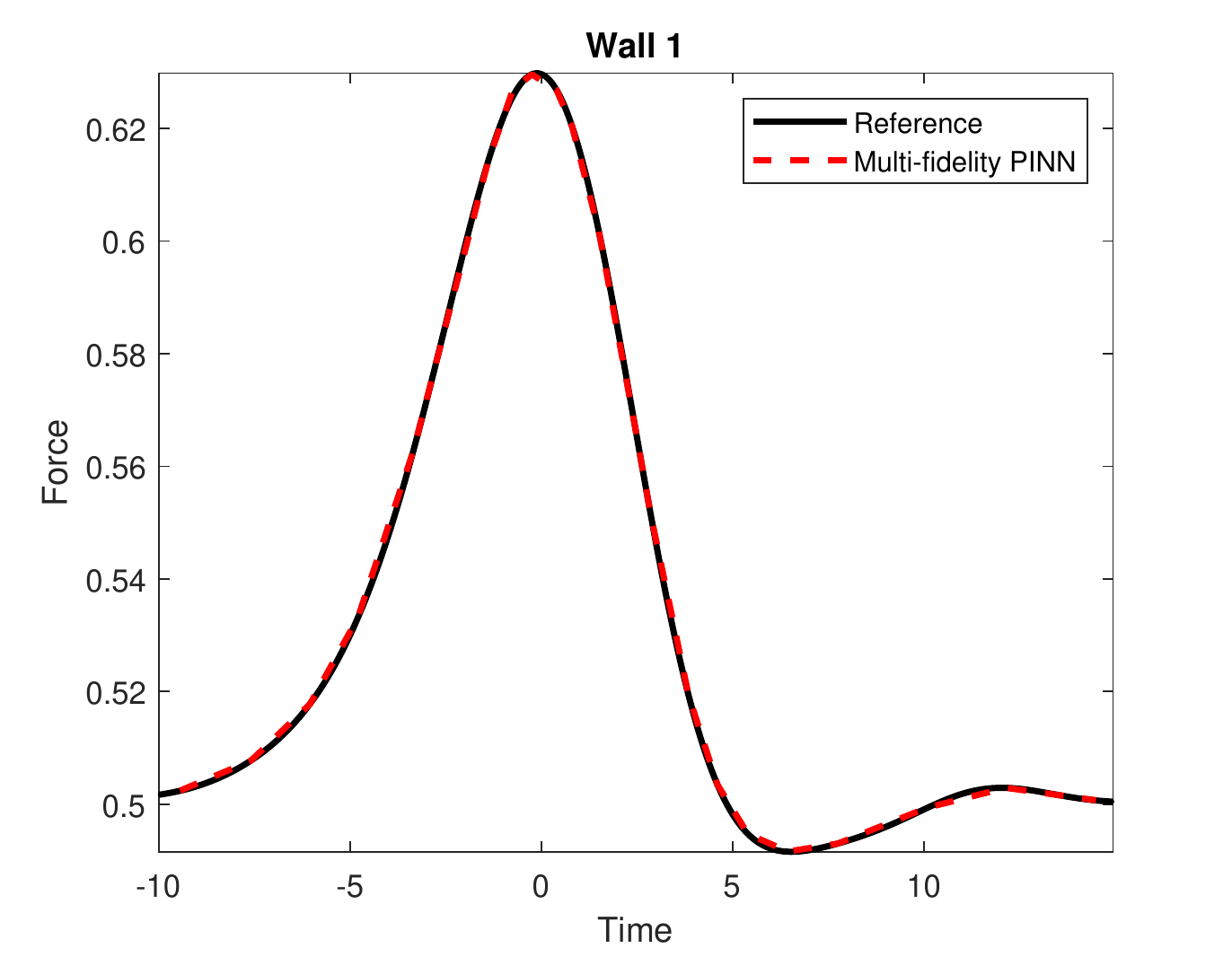}
\includegraphics[trim=0.5cm 0cm 1.2cm 0cm,, clip=true, scale=0.43, angle = 0]{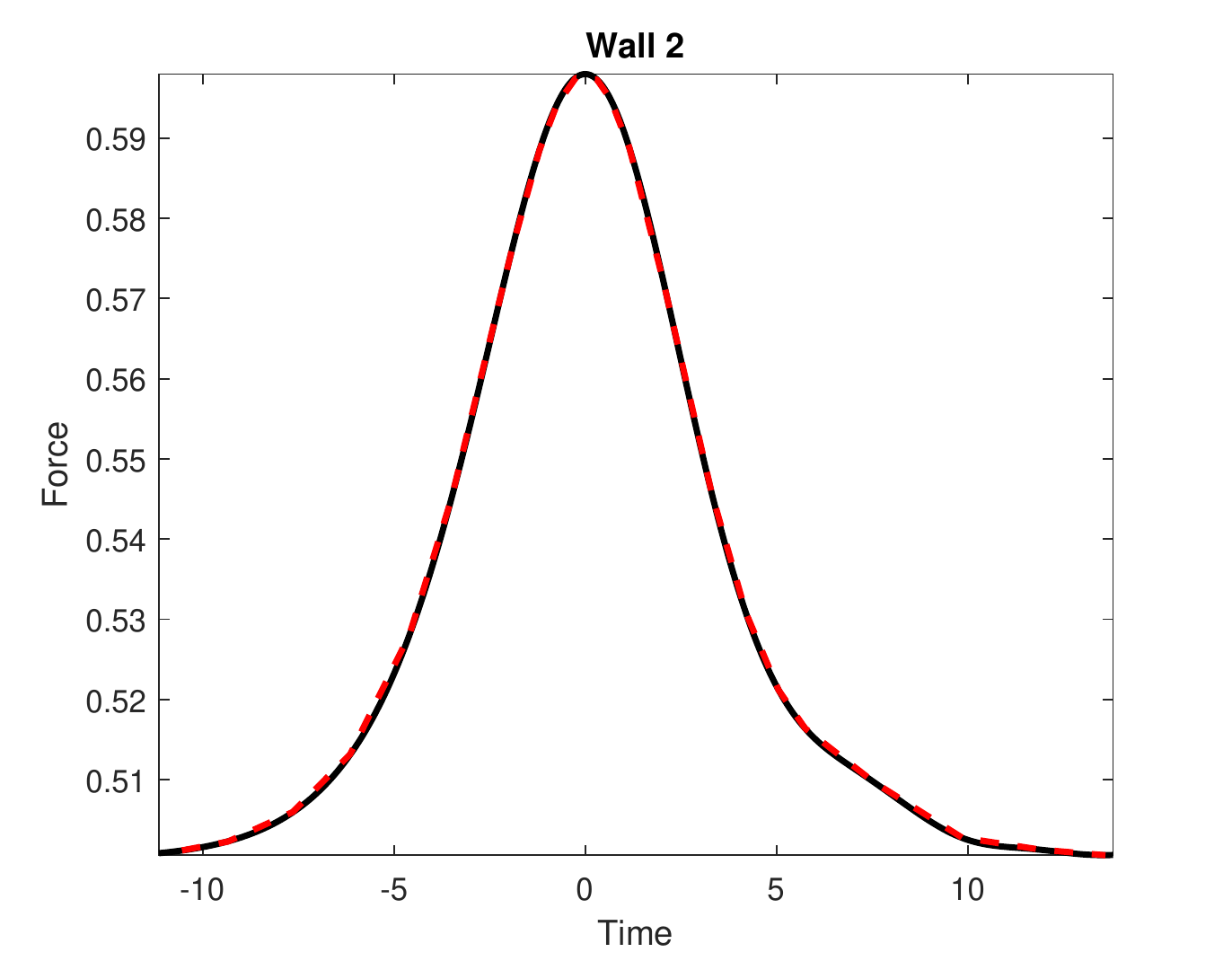}
\includegraphics[trim=0.5cm 0cm 1.2cm 0cm,, clip=true, scale=0.43, angle = 0]{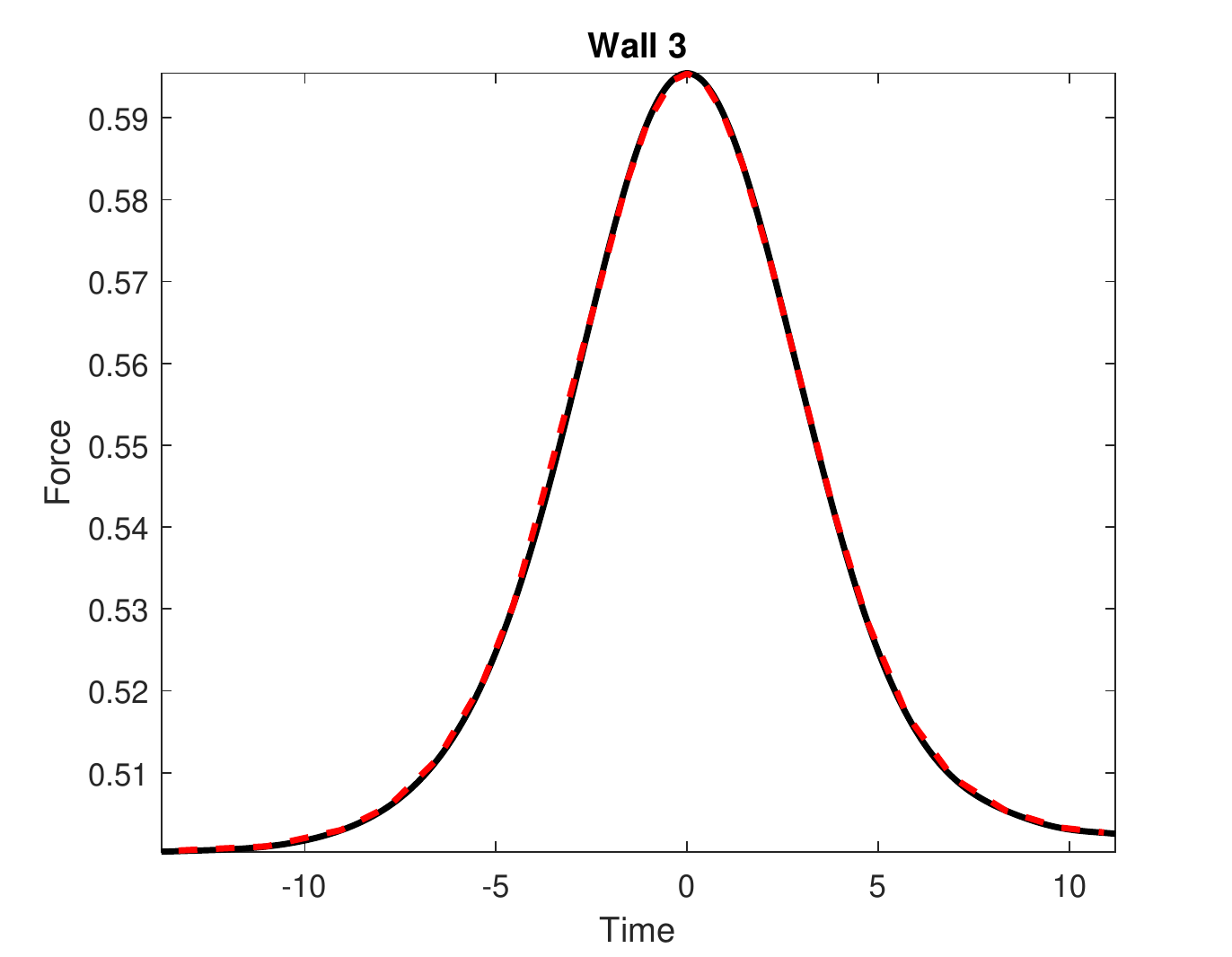}\\
\includegraphics[trim=0.5cm 0cm 1.2cm 0cm, clip=true, scale=0.4, angle = 0]{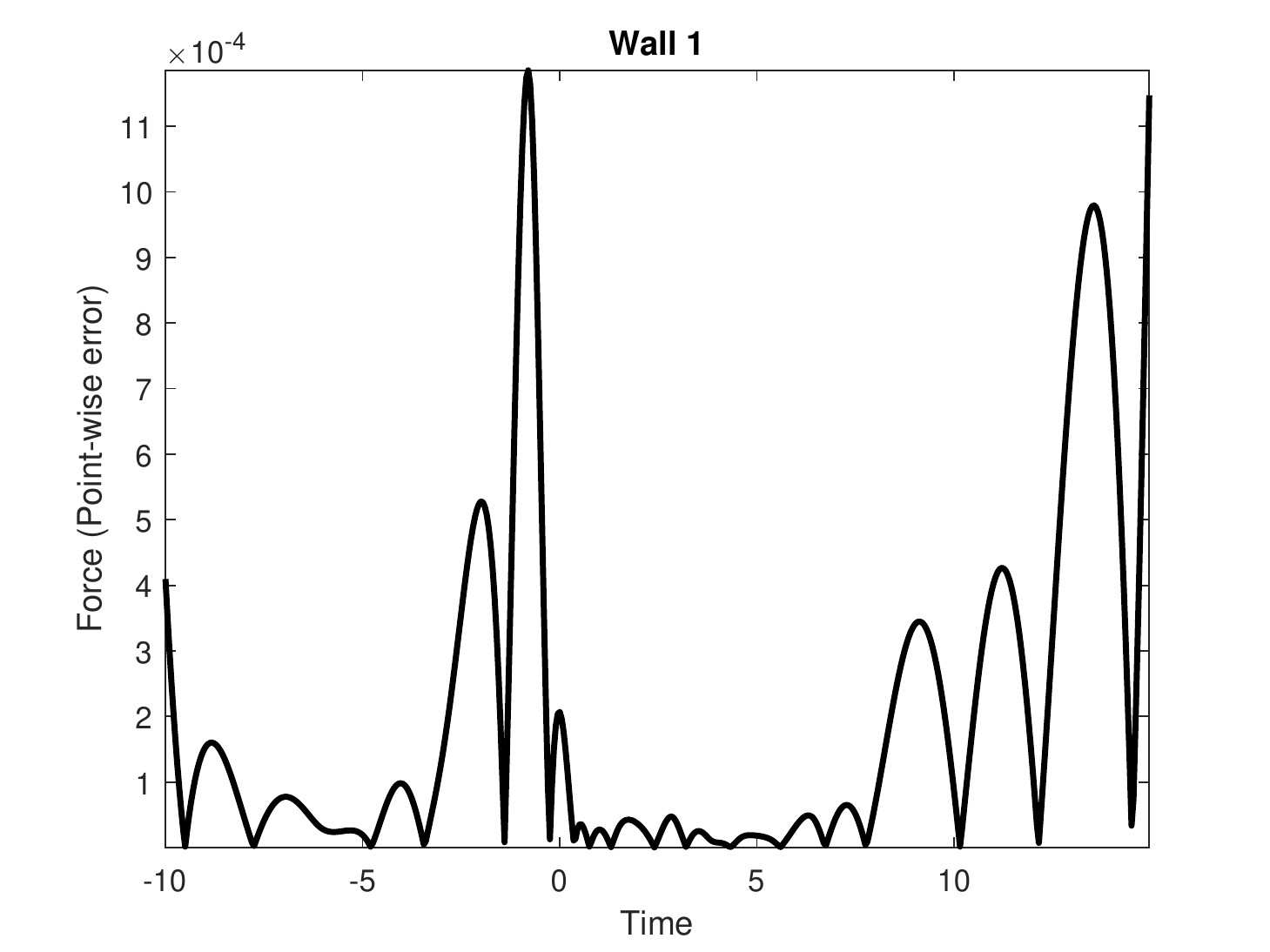}
\includegraphics[trim=0.5cm 0cm 1.2cm 0cm, clip=true, scale=0.4, angle = 0]{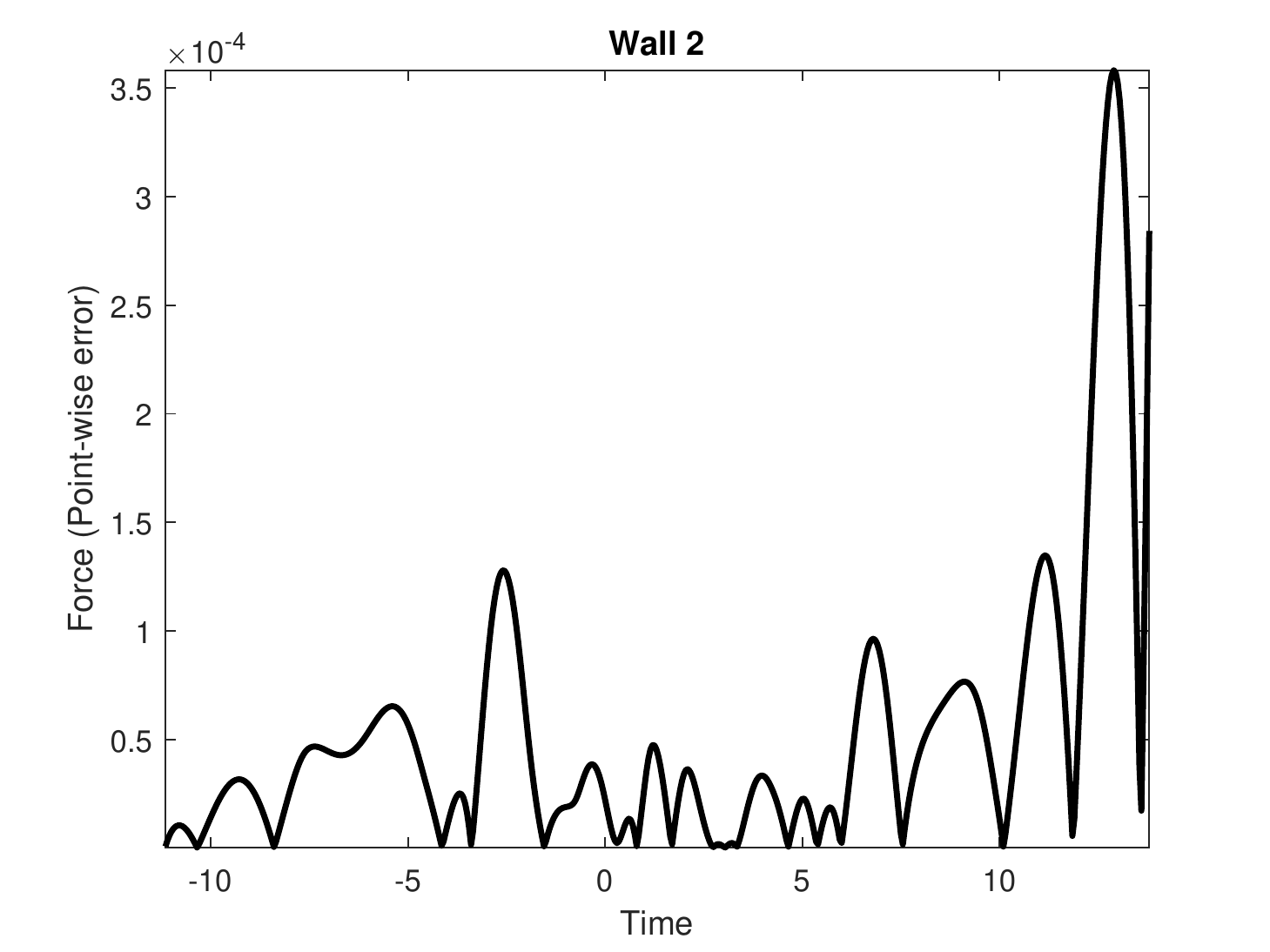}
\includegraphics[trim=0.5cm 0cm 1.2cm 0cm, clip=true, scale=0.4, angle = 0]{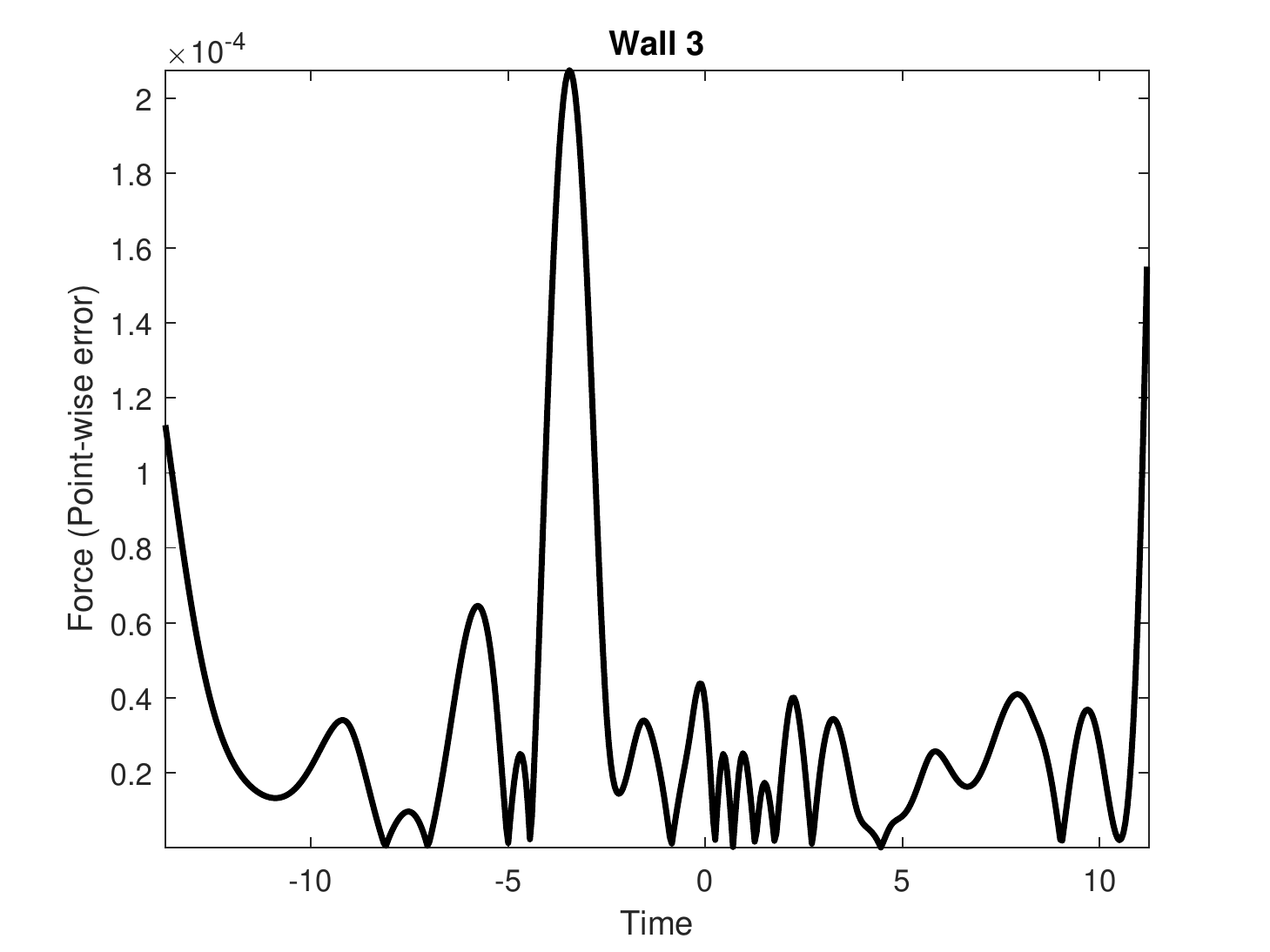}
\caption{Interaction of solitary waves with a rectangular obstacle: (Top row) reconstructed normalized force $F_w/\rho g D_0^2$ as a function of the scaled time $t-t_0\sqrt{g/D_0}$ applied by the solitary wave at the three faces of the rectangular obstacle using multi-fidelity PINNs. (Bottom row) point-wise errors on the three walls.}
\label{fig:TC4_4}
\end{figure}
For real world applications, the number of high-fidelity sensors or gauges that employed to acquire data are often limited due to high cost associated with it. On the other hand, plenty of low-fidelity data can be made available through cheap gauges. Multi-fidelity modeling has been shown to be very efficient in achieving the better predictive accuracy by leveraging the high- as well as low-fidelity data sets. In the literature, Meng \& Karniadakis \cite{meng2020composite} proposed the composite neural network framework that can be used to obtain the nonlinear correlation between the high- and low-fidelity data sets. In a recent study, Chakraborty \cite{chakraborty2021transfer} proposed transfer learning approach for multi-fidelity data. In the present work we used the composite neural networks to map the low-fidelity $\eta$ data onto high-fidelity level, and then further use it in PINNs for solving the inverse Serre-Green-Naghdi problem.  The low- and high-fidelity $\eta$ data are obtained from the low- and high-resolution numerical methods, respectively. Figure \ref{fig:TC4_3} (left) shows the locations of low- and high-fidelity gauges where we used 1360 low  and 50 high-fidelity gauges, shown by black-dots and red asterisk, respectively. The predicted $\eta_h$ (normalized high-fidelity $\eta$ data) versus the reference $\eta_h$ matches very well as shown in Figure \ref{fig:TC4_3} (right). The predicted high-fidelity data is further used to train the PINN with the same hyperparameters used before.
\begin{table}
\begin{center}
\small \begin{tabular}{cccc} \hline 
 &  $u$-velocity & $v$-velocity
 \\ \hline
 \\
Rel. $L_2$ error & 4.2756e-3 & 6.9980e-3   
 \\
 \hline 
 \end{tabular}
\caption{Relative $L_2$ error in the spatio-temporal domain for velocity components}\label{TableTC4}
\end{center}
\end{table}
We trained the network first with Adam optimizer, and then followed by L-BFGS optimizer. Table \ref{TableTC4} shows the relative $L_2$ error for velocity components inferred by the PINN. With the multi-fidelity data, the predictive accuracy of the velocity field is better and the corresponding transient forces (see Figure \ref{fig:TC4_4}) on the three walls are obtained with good accuracy.

\section{Concluding remarks}
In this work we solved highly ill-posed inverse water wave problems using the physics-informed neural networks (PINNs) methodology. The PINN has an ability to incorporate experimental data as well as the governing physical laws, and can accurately and efficiently solve ill-posed problems.
In particular, we solved the
Serre-Green-Naghdi system, which is an asymptotic model for strongly nonlinear and weakly dispersive surface water waves. A derivation of the particular system along with a formula for the instantaneous force of the water is presented using asymptotic techniques. We assumed that the free surface elevation from the wave gauges and the total depth at discrete spatial locations are known, and the aim is to infer the depth integrated averaged horizontal velocity of the water as well as the transient force exerting on the wall due to the water waves. Furthermore, we have also discussed that how to choose the optimal gauge locations for the best predictive accuracy. These are important problems that arise in the construction of off-shore structures such as, for example, oil platforms and wind turbines. Both experimental and synthetic training data are used to train the PINN surrogate model. Moreover, both single-fidelity and multi-fidelity data are used to train the PINN. The problem under consideration is highly ill-posed and can not be solved using the existing traditional methods, which shows the efficacy of PINNs methodology. Both one- and two-dimensional test cases of Serre-Green-Naghdi system are solved using PINNs. 
The predicted velocity field and the corresponding transient force acting on the solid obstacle shows good agreement with the reference solutions. In summary, the PINN method is a robust, accurate, and a promising approach for solving ill-posed water wave problems, which has many real-world applications.









\appendix

\section{Derivation of the Serre-Green-Naghdi equations}\label{sec:derivation}

In this Appendix we present a derivation of the two-dimensional Serre-Green-Naghdi equations with general bottom topography $D(x,t)$ along the lines of \cite{peregrine1967} and \cite{barthelemy2004}. For an alternative derivation we refer to \cite{CiCP2018}.

Let $(\tilde{\bx},\tilde{z})=(\tilde{x},\tilde{y},\tilde{z})$ denote the horizontal and vertical Cartesian coordinates respectively with $\tilde{z}$ measured upwards from the still water level. The water-surface deviation from its rest position at every time $\tilde{t}$ is denoted by $\tilde{\eta}(\tilde{\bx},\tilde{t})$ and the fluid velocity is $(\tilde{\bu}(\tilde{\bx},\tilde{t}),\tilde{w}(\tilde{\bx},\tilde{t}))^T$. The depth of the bottom is assumed to be also time-dependent $\tilde{D}(\tilde{\bx},\tilde{t})$. The Euler equations, which describe the two-dimensional motion of a perfect (inviscid and irrotational) fluid of density $\rho$, \cite{whitham2011}, consist of the momentum conservation equations
\begin{align}
&\tilde{\bu}_{\tilde{t}}+(\tilde{\bu}\cdot\nabla)\tilde{\bu}+\tilde{w}\tilde{\bu}_{\tilde{z}}=-\frac{1}{\rho}\nabla\tilde{p}\ , \label{eq:mom1}\\
&\tilde{w}_{\tilde{t}}+\tilde{\bu}\cdot\nabla\tilde{w}+\tilde{w}\tilde{w}_{\tilde{z}}=-g-\frac{1}{\rho}\tilde{p}_{\tilde{z}}\ , \label{eq:mom2}
\end{align}
the mass conservation equation
\begin{equation} \label{eq:mass}
\nabla\cdot\tilde{\bu}+\tilde{w}_{\tilde{z}}=0\ ,
\end{equation}
and the irrotationality condition
\begin{equation} \label{eq:irrot}
\left\{ \begin{aligned}
&\nabla\times \tilde{\bu}=0\ ,\\
&\tilde{\bu}_z-\nabla \tilde{w}=0\ ,
\end{aligned}
\right.
\end{equation}
where $\tilde{p}=\tilde{p}(\tilde{\bx},\tilde{z},\tilde{t})$ is the pressure field, $g$ is the acceleration due to gravity, $\nabla=\nabla_\bx=(\partial_x,\partial_y)$ and $\tilde{z}\in (-\tilde{D},\tilde{\eta})$.

The Euler equations along with the kinematic boundary condition at the free surface
\begin{equation}\label{eq:bound1}
\tilde{\eta}_{\tilde{t}}+\tilde{\bu}\cdot\nabla\tilde{\eta}-\tilde{w}=0\ , \quad\mbox{ for }\quad \tilde{z}=\tilde{\eta}(\tilde{\bx},\tilde{t})\ ,
\end{equation}
and the kinematic boundary condition at the bottom
\begin{equation}\label{eq:bound2}
\tilde{D}_{\tilde{t}}+\tilde{\bu}\cdot\nabla\tilde{D}+\tilde{w}=0\ , \quad\mbox{ for }\quad \tilde{z}=-\tilde{D}(\tilde{\bx},\tilde{t})\ ,
\end{equation}
form a well-determined system of equation. Without loss of generality, we can assume that the pressure $\tilde{p}(\tilde{\bx},\tilde{\eta},t)=0$ since any other constant is simplified in the course of the derivation of the Serre equations.

In order to derive the Serre equations we consider a characteristic (mean) depth $D_0$ and a typical wavelength $\lambda_0$ and a typical wave height $a_0$ the nondimensional independent variables
\begin{equation}\label{eq:indnondim}
\bx=\frac{\tilde{\bx}}{\lambda_0},\quad z=\frac{\tilde{z}}{D_0}, \quad t=\frac{c_0}{\lambda_0}\tilde{t}\ ,
\end{equation}
and the nondimensional dependent variables
\begin{equation}\label{eq:depnondim}
\bu=\frac{D_0}{a_0c_0}\tilde{\bu}, \quad w=\frac{\lambda_0}{a_0c_0}\tilde{w},\quad \eta=\frac{\tilde{\eta}}{a_0},\quad D=\frac{\tilde{D}}{D_0}\ ,\quad p=\frac{\tilde{p}}{\rho g D_0}\ ,
\end{equation}
where $c_0=\sqrt{gD_0}$ is the linear speed of propagation. Then the governing equation for the fluid motion in nodimensional and scaled variables take the form
\begin{align}
& \varepsilon \bu_t+\varepsilon^2\left((\bu\cdot\nabla)\bu+w\bu_z\right)=-\nabla p \ , \label{eq:6a}\\
& \varepsilon\sigma^2 w_t+\varepsilon^2\sigma^2 \left(\bu\cdot\nabla w+ ww_z\right)=-1-p_z\ , \label{eq:6b}
\end{align}
for $-D<z<\varepsilon\eta$, and with $\varepsilon=a_0/D_0$ and $\sigma=D_0/\lambda_0$. The mass conservation equation is written as
\begin{equation}\label{eq:7a}
\nabla\cdot \bu+w_z=0, \mbox{ for } -D<z<\varepsilon\eta \ ,
\end{equation}
while the irrotationality condition becomes
\begin{equation}\label{eq:7b}
\begin{aligned}
& \nabla\times \bu=0,\\
& \bu_z-\sigma^2 \nabla w=0,
\end{aligned} \mbox{ for } -D<z<\varepsilon\eta \ .
\end{equation}
The boundary conditions at the free-surface and bottom of the fluid domain take the form
\begin{equation}\label{eq:8a}
\eta_t+\varepsilon \bu \cdot\nabla \eta-w= 0 \mbox{ and } p=0\ , \mbox{ on } z=\varepsilon\eta\ ,
\end{equation}
and
\begin{equation}\label{eq:8b}
 D_t+ \varepsilon \bu \cdot \nabla D+ \varepsilon w=0\ , \mbox{ on } z=-D\ .
\end{equation}

In the sequel we denote any depth averaged quantity 
\begin{equation}
\bar{v}=\frac{1}{h}\int_{-D}^{\varepsilon\eta}v~dz\ ,
\end{equation}
where $h=D+\varepsilon\eta$ is the total depth of the water measured from the free-surface $\varepsilon\eta$.

Integrating the mass conservation equation (\ref{eq:7a}) from $-D$ to $\epsilon\eta$, and using the Leibniz rule and the boundary conditions (\ref{eq:8a}) and (\ref{eq:8b}) we obtain the mass conservation equation for the depth-averaged velocity $\bar{u}=\bar{u}(x,t)$
\begin{equation}\label{eq:massc}
h_t+\varepsilon \nabla\cdot [h\bar{\bu}]=0\ .
\end{equation}
The momentum equation (\ref{eq:6a}) after integration from $-D$ to $\epsilon\eta$ and the boundary conditions (\ref{eq:8a}) and (\ref{eq:8b}) can be written as
\begin{equation}\label{eq:monet}
\varepsilon(h\bar{\bu})_t+\varepsilon^2\int_{-D}^{\varepsilon\eta} [(\bu\cdot\nabla)\bu+\bu(\nabla\cdot\bu)]~dz=-\int_{-D}^{\varepsilon\eta}\nabla p~dz\ .
\end{equation}

Using the Leibniz rule and after some simplifications we observe that
\begin{equation}
\int_{-D}^{\varepsilon\eta}[(\bu\cdot\nabla)\bu+\bu(\nabla\cdot\bu)]~dz = \begin{pmatrix}
\partial_x\int_{-D}^{\varepsilon\eta}u^2+\partial_y\int_{-D}^{\varepsilon\eta}uv\\
\partial_y\int_{-D}^{\varepsilon\eta} v^2+\partial_x\int_{-D}^{\varepsilon\eta}uv
\end{pmatrix}\doteq \nabla\cdot\int_{-D}^{\varepsilon\eta} \bu\otimes\bu~dz \ .
\end{equation}
Using in addition the irrotationality condition (\ref{eq:7b}), equation (\ref{eq:monet}) can further be simplified in the form
\begin{equation}\label{eq:star}
\varepsilon h\bar{\bu}_t+\varepsilon h (\bar{\bu}\cdot\nabla)\bar{\bu}+\varepsilon^2\nabla\cdot\int_{-D}^{\varepsilon\eta}\bu\otimes\bu-\bar{\bu}\otimes\bar{\bu}~dz=-\int_{-D}^{\varepsilon\eta} \nabla p ~dz\ .
\end{equation}

The momentum equation (\ref{eq:6b}) is rewritten as
\begin{equation}\label{eq:13}
\varepsilon\sigma^2\Gamma(\bx,z,t)=-1-p_z\ ,
\end{equation}
where 
\begin{equation}\label{eq:14}
\Gamma(\bx,z,t)=w_t+\varepsilon \bu\cdot\nabla w+\varepsilon ww_z\ .
\end{equation}
Integrating (\ref{eq:13}) from $z$ to $\varepsilon\eta$ yields
\begin{equation}
p(\bx,z,t)=(\varepsilon\eta-z)+\varepsilon\sigma^2\int_z^{\varepsilon\eta}\Gamma~dz\ ,
\end{equation}
which after integrating from $-D$ to $\varepsilon\eta$ and multiplying with $h$ becomes
\begin{equation}\label{eq:16}
h\bar{p}=\frac{1}{2}h^2+\varepsilon\sigma^2\int_{-D}^{\varepsilon\eta}\int_{z}^{\varepsilon\eta} \Gamma~dz\ .
\end{equation}
Observing that 
\begin{equation}
\int_{-D}^{\varepsilon\eta}\nabla p~dz=\nabla(h\bar{p})-D_xp(\bx,-D,t)\ , 
\end{equation}
and using (\ref{eq:16}) we write (\ref{eq:star}) in the form
\begin{equation}\label{eq:basic}
\begin{split}
\varepsilon h\bar{\bu}_t&+\varepsilon h (\bar{\bu}\cdot\nabla)\bar{\bu} +\varepsilon^2\nabla\cdot\int_{-D}^{\varepsilon\eta}\bu\otimes\bu-\bar{\bu}\otimes\bar{\bu}~dz=\\
&= \nabla\frac{h^2}{2}-\varepsilon\sigma^2\nabla\int_{-D}^{\varepsilon\eta}\int_{z}^{\varepsilon\eta}\Gamma~d\zeta~dz +\nabla Dh+\varepsilon\sigma^2\nabla D\int_{-D}^{\varepsilon\eta}\Gamma~dz\ .
\end{split}
\end{equation}

In order to approximate the double integral in (\ref{eq:16}) we need to find approximations of the horizontal and vertical velocities. We observe, after integration from $-D$ to $z$ of the irrotationality condition (\ref{eq:7b}), that 
\begin{equation}\label{eq:23}
\bu=\bu_b+\sigma^2\int_{-D}^z \nabla w~dz\ ,
\end{equation}
where $\bu_b=\bu(\bx,-D,t)$ is the horizontal velocity at the bottom $z=-D$.
Integrating over the same depth the mass conservation equation (\ref{eq:7a}) we obtain 
\begin{equation}\label{eq:a27}
w=w_b-\int_{-D}^{\varepsilon\eta} \nabla\cdot \bu~dz\ .
\end{equation}
Using the boundary condition (\ref{eq:8b}) and (\ref{eq:a27}), (\ref{eq:23}) we obtain
\begin{equation}
w=-\frac{1}{\varepsilon}D_t-\bu_b\cdot\nabla D-\int_{-D}^z \nabla\cdot \bu~dz\ ,
\end{equation}
which after using (\ref{eq:23}) is approximated by
\begin{equation}\label{eq:25}
w=-\frac{1}{\varepsilon}D_t\bu_b\cdot \nabla D-(z+D)\nabla\cdot \bu_b+O(\sigma^2)\ .
\end{equation}
Substituting (\ref{eq:25}) into (\ref{eq:23}) we have
\begin{equation}\label{eq:26}
\begin{split}
\bu=\bu_b -\sigma^2\left\{\frac{1}{\varepsilon}\nabla D_{t}+\nabla(\bu_b\cdot\nabla  D)+\nabla\cdot \bu_b \nabla D\right\}(z+D) \\
-\sigma^2\nabla \nabla\cdot \bu_b\frac{(z+D)^2}{2}+O(\sigma^4)\ .
\end{split}
\end{equation}
From (\ref{eq:26}) we observe $\bu_b=\bu+O(\sigma^2)$ and also $\bu_b=\bar{\bu}+O(\sigma^2)$. Thus, we can write $w$ as
\begin{equation}\label{eq:27}
w=-\frac{1}{\varepsilon}D_t-\bar{u}\cdot\nabla D-(z+D)\nabla\cdot \bar{\bu}+O(\sigma^2)\ ,
\end{equation}
and $\bu_b$ as
\begin{equation}\label{eq:28}
\bu_b=\bu+\sigma^2\left\{\frac{1}{\varepsilon}\nabla D_{t}+\nabla(\bar{\bu}\cdot \nabla D)+\nabla\cdot \bar{\bu}\nabla D\right\}(z+D)+\sigma^2\nabla\nabla\cdot \bar{\bu}\frac{(z+D)^2}{2}+O(\sigma^4)\ ,
\end{equation}
which after integration from $-D$ to $\varepsilon\eta$ becomes
\begin{equation}\label{eq:29}
\bu_b=\bar{\bu}+\sigma^2\left\{\frac{1}{\varepsilon}\nabla D_{t}+\nabla(\bar{\bu}\cdot \nabla D)+\nabla D\nabla\cdot \bar{\bu}\right\}\frac{h}{2}+\sigma^2\nabla\nabla\cdot\bar{\bu}\frac{h^2}{6}+O(\sigma^4)\ .
\end{equation}
Substituting approximation (\ref{eq:29}) in (\ref{eq:26}) we obtain the horizontal velocity approximation
\begin{equation}\label{eq:30}
\begin{split}
\bu= \bar{\bu} &+\sigma^2\left\{\frac{1}{\varepsilon}\nabla D_{t}+\nabla D \nabla\cdot \bar{\bu}+\nabla (\bar{\bu}\cdot\nabla D)\right\}\left(\frac{h}{2}-(z+D)\right)\\ & +\sigma^2\nabla\nabla\cdot \bar{\bu}\left(\frac{h^2}{6}-\frac{(z+D)^2}{2}\right)+O(\sigma^4)\ .
\end{split}
\end{equation}
From (\ref{eq:30}) we have
\begin{equation}
\begin{split}
u^2=\bar{u}^2 &+\sigma^2\left\{\frac{1}{\varepsilon}\bar{u}D_{xt}+D_x\bar{u}(\bar{u}_x+\bar{v}_y)+\bar{u}(\bar{u}D_x+\bar{v}D_y)_x\right\}\left(h-2(z+D)\right)\\ &+\sigma^2\bar{u}(\bar{u}_x+\bar{v}_y)_x\left(\frac{h^2}{3}-(z+D)^2\right)+O(\sigma^4)\ ,
\end{split}
\end{equation}
from which we deduce that
\begin{equation}
\int_{-D}^{\varepsilon\eta}u^2-\bar{u}^2~dz=O(\sigma^4)\ .
\end{equation}
Similar results hold true for the terms $v^2-\bar{v}^2$ and $uv-\bar{u}\bar{v}$, and thus we deduce that 
\begin{equation}
\int_{-D}^{\varepsilon\eta}\bu\otimes\bu-\bar{\bu}\otimes\bar{\bu}~dz =O(\sigma^4)\ .
\end{equation}
Substitution of (\ref{eq:27}) and (\ref{eq:28}) in (\ref{eq:14}) yields
\begin{equation}
\begin{split}
\Gamma=-\frac{1}{\varepsilon}D_{tt}-2\bar{\bu}\cdot \nabla D_{t}- [\bar{\bu}_t+\varepsilon(\bar{\bu}\cdot\nabla)\bar{\bu}+\varepsilon\bar{\bu}(\bar{\bu}\cdot\nabla)]\cdot\nabla D \\
-(z+D)[\nabla\cdot\bar{\bu}_{t}+\varepsilon\bar{\bu}\cdot\nabla(\nabla\cdot\bar{\bu})-\varepsilon(\nabla\cdot\bar{\bu})^2]+O(\sigma^2)\ ,
\end{split}
\end{equation}
which leads to the approximations
\begin{equation}\label{eq:36}
\begin{split}
\int_{-D}^{\varepsilon\eta}\int_{z}^{\varepsilon\eta} \Gamma~dz=&
-\frac{h^2}{2}\bigg\{\frac{1}{\varepsilon}D_{tt}+2\bar{\bu}\cdot\nabla D_{t}\\&+[\bar{\bu}_t+\varepsilon(\bar{\bu}\cdot\nabla)\bar{\bu}+\varepsilon\bar{\bu}(\bar{\bu}\cdot\nabla)]\cdot\nabla D \bigg\}\\
&-\frac{h^3}{3}\left(\nabla\cdot\bar{\bu}_t+\varepsilon\bar{\bu}\cdot\nabla(\nabla\cdot\bar{\bu})-\varepsilon(\nabla\cdot\bar{\bu})^2\right)+O(\sigma^2)\ ,
\end{split}
\end{equation}
and 
\begin{equation}\label{eq:37}
\begin{split}
\int_{-D}^{\varepsilon\eta}\Gamma~dz= &-h\left\{\frac{1}{\varepsilon}D_{tt}+2\bar{\bu}\cdot\nabla D_{t}+[\bar{\bu}_t+\varepsilon (\bar{\bu}\cdot\nabla)\bar{\bu}+\varepsilon\bar{\bu}(\bar{\bu}\cdot\nabla)]\cdot\nabla D\right\}\\
&-\frac{h^2}{2}\left(\nabla\cdot\bar{\bu}_{t}+\varepsilon\bar{\bu}\nabla(\nabla\cdot\bar{\bu})-\varepsilon(\nabla\cdot\bar{\bu})^2\right)+O(\sigma^2)\ .
\end{split}
\end{equation}
The momentum conservation equation (\ref{eq:basic}) after substituting (\ref{eq:36}) and (\ref{eq:37}) and dividing by $\varepsilon h$ becomes
\begin{equation}\label{eq:38}
\bar{\bu}_t+\nabla\eta+\varepsilon(\bar{\bu}\cdot\nabla)\bar{\bu}-\frac{\sigma^2}{h}\nabla\left\{h^2\left(\frac{1}{3}\tilde{P}+\frac{1}{2}\tilde{Q} \right)\right\}+\sigma^2\nabla D\left(\frac{1}{2}\tilde{P}+\tilde{Q}\right)=\tilde{F}+O(\sigma^4)\ ,
\end{equation}
where
\begin{align}
\tilde{P}&=h[\nabla\cdot\left(\bar{\bu}_t+\varepsilon\bar{\bu}\nabla\cdot\bar{\bu}\right)-2\varepsilon(\nabla\cdot\bar{\bu})^2]\ , \label{eq:39} \\
\tilde{Q}&=[\bar{\bu}_t+\varepsilon (\bar{\bu}\cdot\nabla)\bar{\bu}+\varepsilon\bar{\bu}(\bar{\bu}\cdot\nabla)]\cdot\nabla D\ , \label{eq:40} \\
\tilde{F}&=\frac{\sigma^2}{2h}\nabla\left(h^2\tilde{G} \right)-\sigma^2\nabla D\cdot\tilde{G}\ , \label{eq:41}\\
\tilde{G}&=\frac{1}{\varepsilon}D_{tt}+2\bar{\bu}\cdot\nabla D_{t}\ . \label{eq:41b}
\end{align} 
Discarding the high-order terms, the Serre-Green-Naghdi equations with time-dependent bottom topography in dimensional variables is written in the form
\begin{align}
& h_t+\nabla\cdot[h\bar{\bu}]=0\ , \label{eq:cserre1a}\\
&\bar{\bu}_t+g\nabla\eta+(\bar{\bu}\cdot\nabla)\bar{\bu}-\frac{1}{h}\nabla\left\{h^2\left(\frac{1}{3}P+\frac{1}{2}Q \right)\right\}+\nabla D\left(\frac{1}{2}P+Q\right)=F\ ,  \label{eq:cserre2a}
\end{align}
where
\begin{align}
P&=h[\nabla\cdot\left(\bar{\bu}_t+\bar{\bu}\nabla\cdot\bar{\bu}\right)-2(\nabla\cdot\bar{\bu})^2]\ , \label{eq:cserre3a} \\
Q&=[\bar{\bu}_t+(\bar{\bu}\cdot\nabla)\bar{\bu}+\bar{\bu}(\bar{\bu}\cdot\nabla)]\cdot\nabla D\ , \label{eq:cserre4a}\\
F&=\frac{1}{2h}\nabla\left(h^2G \right)-\nabla D\cdot G, \label{eq:cserre5}\\
G&=D_{tt}+2\bar{\bu}\cdot\nabla D_{t}\ , \label{eq:cserre6}
\end{align}
and $h=\eta+D$.

\section{Instantaneous force at a point}\label{sec:force}

From Equations (\ref{eq:16}) and (\ref{eq:36}) we derive the formula for the depth averaged pressure

\begin{equation}\label{eq:59}
\begin{split}
h\bar{p}=&\frac{1}{2}h^2-\varepsilon\sigma^2 \frac{h^2}{2}\bigg\{\frac{1}{\varepsilon}D_{tt}+2\bar{\bu}\cdot\nabla D_{t}+[\bar{\bu}_t+\varepsilon(\bar{\bu}\cdot\nabla)\bar{\bu}+\varepsilon\bar{\bu}(\bar{\bu}\cdot\nabla)]\cdot\nabla D \bigg\}\\
&-\varepsilon\sigma^2\frac{h^3}{3}\left(\nabla\cdot\bar{\bu}_t+\varepsilon\bar{\bu}\cdot\nabla(\nabla\cdot\bar{\bu})-\varepsilon(\nabla\cdot\bar{\bu})^2\right)+O(\varepsilon\sigma^4)\ .
\end{split}
\end{equation} 

The particular formula for the pressure is useful for the computation of the instantaneous  applied force at a point located at $\bx=\bx_w$ by the water. The formula of the instantaneous force is given as
\begin{equation*}
F_w(t)=\int_{-D(\bx_w,t)}^{\varepsilon\eta(\bx_w,t)}p(\bx_w,z,t)~dz= h(\bx_w,t)\bar{p}(\bx_w,t)\ ,
\end{equation*}
where we take into account that the horizontal velocity on the wall for all $t\geq 0$. This formula of the instantaneous force in the case of a flat bottom $D(\bx,t)=D_0$ around the wall takes the form
$$
F_w(t)=\left[\frac{h^2}{2}-\varepsilon\sigma^2\frac{h^3}{3}\left(\nabla\cdot\bar{\bu}_t+\varepsilon\bar{\bu}\cdot\nabla(\nabla\cdot\bar{\bu})-\varepsilon(\nabla\cdot\bar{\bu})^2\right)\right]_{\bx=\bx_w}+O(\varepsilon\sigma^4)\ ,
$$
and discarding high-order terms is written in dimensional variables as
\begin{equation}\label{eq:iforce}
F_w(t)=\rho g\left[\frac{h^2}{2}-\frac{h^3}{3g}\left(\nabla\cdot\bar{\bu}_t+\bar{\bu}\cdot\nabla(\nabla\cdot\bar{\bu})-(\nabla\cdot\bar{\bu})^2\right) \right]_{\bx=\bx_w}\ .
\end{equation}

\bibliographystyle{plain}

\end{document}